\documentclass[onecolum]{svjour3}           
\smartqed  
\usepackage{graphicx}
\usepackage{latexsym}
\usepackage{natbib}
\usepackage{epstopdf}
\newcommand{\dap}{\tau_\mathrm{D}/P}
\newcommand{\mutilde}{\tilde{\mu}}
\newcommand{\etac}{\eta_\mathrm{C}}
\newcommand{\taud}{\tau_{\mathrm{D}}}
\newcommand{\kms}{$\rm km ~ s^{-1}$}
\newcommand{\eg}{{\it e.g.,}\/}
\newcommand{\et}{{\it et al.}}
\newcommand{\ie}{{\it i.e.,}\ }

\newcommand\apj{ApJ\ }%
\newcommand\apjl{ApJ\ }%
\newcommand\aap{A\&A\ }%
\newcommand\jgr{J.~Geophys.~Res.\ }%
\newcommand\mnras{MNRAS\ }%
\newcommand\nat{Nature\ }%
\newcommand\pasj{PASJ\ }%
\newcommand\solphys{Sol.~Phys.\ }%
\journalname{Space Science Reviews}
\begin{document}

\title{Physics of Solar Prominences: II - Magnetic Structure and
Dynamics}

\author{D.H. Mackay\footnote{The authors would like to thank Aad van Ballegooijen
for his hard work and dedication without which this review would not exist.} 
\and J.T. Karpen \and J.L. Ballester \and B. Schmieder \and G. Aulanier}

\authorrunning{Mackay et al.} 

\institute{D.H. Mackay \at
              School of Mathematics and Statistics,
              University of St~Andrews,
              North Haugh, St~Andrews, Fife, Scotland, KY16 9SS.
               \email{duncan@mcs.st-and.ac.uk}
           \and
           J.T. Karpen \at
              Code 674,
              NASA Goddard Spaceflight Center,
              Greenbelt, MD 20771,
              USA.
           \and
           J.L. Ballester \at
              Department de F\'{i}sica,
              Universitat de les Illes Balears,
              E-07122 Palma de Mallorca,
              Spain.
           \and
           B. Schmieder \at
              LESIA, Observatoire de Paris,
              92195 Meudon Cedex Principal,
              France.
           \and
           G. Aulanier \at
              LESIA, Observatoire de Paris,
              92195 Meudon Cedex Principal,
              France.
}

\date{Received: date / Accepted: date}

\maketitle

\begin{abstract}
Observations and models of solar prominences are reviewed. We focus on
non-eruptive prominences, and describe recent progress in four areas
of prominence research: (1) magnetic structure deduced from observations
and models, (2) the dynamics of prominence plasmas (formation and flows), (3) 
Magneto-hydrodynamic (MHD) waves in prominences and (4) the formation and large-scale
patterns of the filament channels in which prominences are located. Finally,
several outstanding issues in prominence research are discussed, along with
observations and models required to resolve them.

\end{abstract}

\begin{tableofcontents}

\end{tableofcontents}

\section{Introduction}
\label{sec:1}

Solar prominences consist of relatively cool, dense plasma that is
suspended in the solar corona at heights up to 100 Mm above the
chromosphere. They are observed  as ``filaments'' on the solar disk,
where they are seen in absorption in strong spectral lines (such as
H$\alpha$) and in the Extreme Ultraviolet (EUV) continuum. When described as
``prominences'' they are seen above the solar limb, where they appear as bright
features against the dark background. In this review the terms
``filament'' and ``prominence'' will be used interchangeably.  The
existence of cool, dense plasma suspended in the hot corona has been a
mystery ever since the first observations of filaments and prominences
\citep[reviewed in][] {Hirayama1985, Zirker1989, Tan-Han1995}.
The discovery by \citet{Babcock55} that solar filaments lie between
different polarities of the Sun's magnetic field, provided one piece of the 
puzzle by identifying the primary source of support and constraint for the
prominence mass \citep[][] {Kippenhahn1957}. However, the magnetic
structure of prominences is still not fully understood, with many 
observations and theoretical models differing on the exact nature of
the magnetic field.  As a result the
physical processes governing the origin and subsequent behavior of the
prominence plasma remain a lively topic of debate.

In recent years, high-resolution observations of prominences from the
ground and from space have greatly improved our knowledge of the
structure and dynamics of prominence plasmas. There have also been
renewed efforts to measure the vector magnetic fields in and below
solar prominences using spectro-polarimetric methods \citep[see][]
{Lopez2007}. These observations have inspired a wide variety of
theoretical models describing different aspects of the physics of
prominences. The purpose of this review is to report on the progress 
that has been made in recent years, since the comprehensive review by
\citet{Tan-Han1995}. Both state of the art space and ground based 
observations, along with models in four different areas of prominence 
research will be described. The four areas are chosen to give a broad 
overview of solar filaments, from the smallest structures currently observed, 
to their properties found on their scale size and finally filament properties 
on a global scale.  These specific topics are:
\begin{enumerate}
\item {\bf{Prominence magnetic structure}} : The first topic describes 
the basic structure and morphology of solar filaments, including new high 
resolution observations and magnetic field measurements obtained from ground and space. 
The relationship of filaments to their underlying magnetic field along with 
recently developed static 3D magnetic field models, many of which are derived from 
observations, are described.
\item {\bf{The dynamics of prominence plasmas (formation and flows)}}: The second
topic considers the underlying physics and complex range of models that may describe
the origin and behavior of the filaments mass. Models include those based around
magnetic forces, such as injection or levitation mechanisms, to thermal pressure 
force models of evaporation and condensation. In particular, significant new work
on thermal non-equilibrium models is discussed.
\item {\bf{MHD waves in prominences}}: The third topic considers
the significant advances that have been made in the last 10 years in observing
and modeling the small-scale oscillations observed in filaments and prominences. 
The effect of flows and damping mechanisms on linear-MHD waves is discussed.
\item {\bf{The formation and large-scale patterns of filament channels}}: The final
topic considers the properties of filament channels and filaments
in a global context. In particular H$\alpha$ observations and theoretical models 
for the formation of filaments and filament channels are discussed, along with schemes 
used to categorize filaments. 
\end{enumerate}
Other aspects of prominence research,
including results from space-based spectroscopy of solar prominences,
are considered in the review by \citep[][hereafter Paper~I]
{Labrosse2009}. We focus on the properties of non-eruptive
prominences.


\section{Magnetic Structure of Filaments and Prominences}
\label{sec:2}

In this section we first describe recent observations relevant to
prominence magnetic structure, and then describe related modeling.

\subsection{Observations}
\label{sec:2.1}

\subsubsection{Basic Properties}
\label{sec:2.1.1}

\begin{figure}[t]
\centering\includegraphics[scale=0.5]{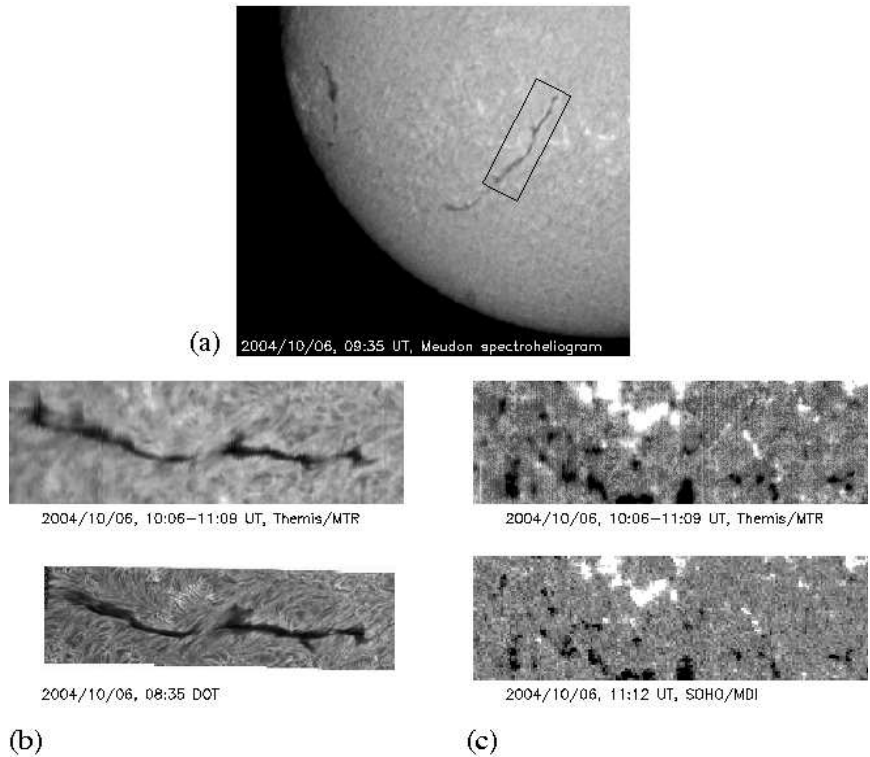}
\caption{H$\alpha$ filament observed on 6$^{th}$ October 2004 from
\cite{Dudik2008}.
(a) Meudon spectroheliograph in H$\alpha$ line core at 09.35 UT. The
box corresponds to the THEMIS/MTR field of view. (b) THEMIS/MTR 
(10:06-11:09 UT) and DOT (08:35 UT) observations in H$\alpha$ line center.
(c)  THEMIS/MTR (10:06-11:09 UT) and SoHO/MDI (11;12 UT) longitudinal magnetograms,
both saturate at $\pm 40$ G.} 
\label{fig:new_fig1}
\end{figure}

Filaments are always located above Polarity Inversion Lines (PILs),
i.e., lines on the photosphere where the radial component $B_r$ of
the magnetic field changes sign.  Filaments can be found above PILs
inside activity nests consisting of multiple bipolar pairs of spots
(``active regions filaments''), 
at the border of active regions
(``intermediate filaments''), and on the quiet Sun (``quiescent
filaments''), including the polar crown.  This nomenclature refers to
their location on the Sun relative to underlying magnetic fields and 
not with respect to their internal plasma motions.
Intermediate and active region filaments are located at sunspot latitude
belts, however quiescent filaments may exist over all latitudes on the Sun.   
On the quiet sun the
magnetic fields are concentrated into discrete network elements that
are well separated from each other, with much weaker fields in between. 
Therefore, a PIL on the quiet sun is actually a zone of mixed
polarity, and the precise location of the PIL can best be defined by 
using spatially smoothed magnetograms \citep[][]{Jones2004}.
An example of a quiescent filament seen in H$\alpha$ through different instruments, 
along with the underlying photospheric magnetic field distribution 
can be seen in Figure \ref{fig:new_fig1}. In the next paragraph we first 
consider the basic structure of filaments as seen on the solar disk. Following this,
the varying structures that are found when prominences are viewed above the limb 
are illustrated.

\begin{figure}[t]
\centering\rotatebox{90}{\includegraphics[scale=0.5]{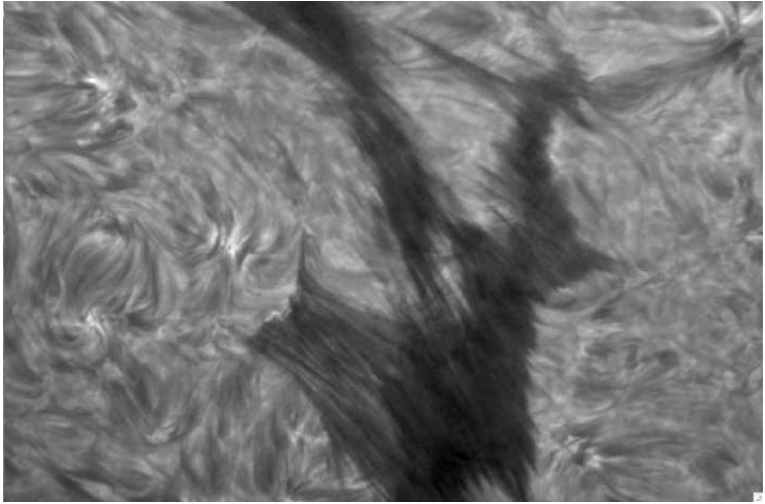}}
\caption{High-resolution H$\alpha$ images obtained from the Swedish Solar Telescope
illustrating thin threads aligned with the local magnetic field (from \cite{Lin2005a})
. } 
\label{fig:new_fig2}
\end{figure}

Filaments typically consists of 3 structural components: a spine, barbs, and 
two extreme ends. The spine runs horizontally along the top of the filament, 
although there may be sections along the filament where the spine is nearly 
invisible. The barbs protrude from the side of the filament (see Figure 1(b) 
bottom panel) and when observed closer to the limb, using standard
instrumentation with moderate spatial resolution, the barbs are seen to
extend down from the spine to the chromosphere below. The barbs, as well 
as the ends of the
filament (also called ``legs'') may be a collection of threads that
appear to terminate at a single point or at multiple points \citep[][]
{Lin2008a}. When viewing a quiescent filament on the solar disk at high resolution,
H$\alpha$ observations indicate that each of these three structural components consist 
of thin thread-like structures \citep[see examples in][] {Malherbe1989, Martin1998,
pec00}. In Figure~\ref{fig:new_fig2} these thin threads may be seen, where
the observations are at the limit of present day resolution. The 
threads are found to have widths of about 200 km \citep[][]{Lin2005a, Lin2005b,
Lin2008a, Lin2008b} and are not necessary aligned with the
structural component of the filament to which they belong. Individual threads have 
lifetimes of only a few
minutes, but the filament as a whole can live for many days. These 
thin threads are thought to be aligned with the local magnetic field.
\citet{Lin2005a} find that for quiescent H$\alpha$ filaments it is
not possible to associate the ends of individual filament threads
with bright points in G-band images (such bright points are proxies
for small kilogauss flux elements in the photosphere). They conclude
that filament threads are located in thin bundles of field lines
that are longer than
the observed threads, i.e., only a fraction of each field line is
filled with absorbing plasma. In a recent paper, \citet{Lin2008a} argue 
that at high resolution, all filaments
ranging from active region to quiescent are made up of 
such thin threads. While short threads are commonly seen in quiescent
filaments, active region filaments seem to be composed of relatively longer
threads. These differences may be related to the average angle made by the magnetic field 
to the axis of the filament spine.

\begin{figure}[t]
\centering\includegraphics[scale=0.5]{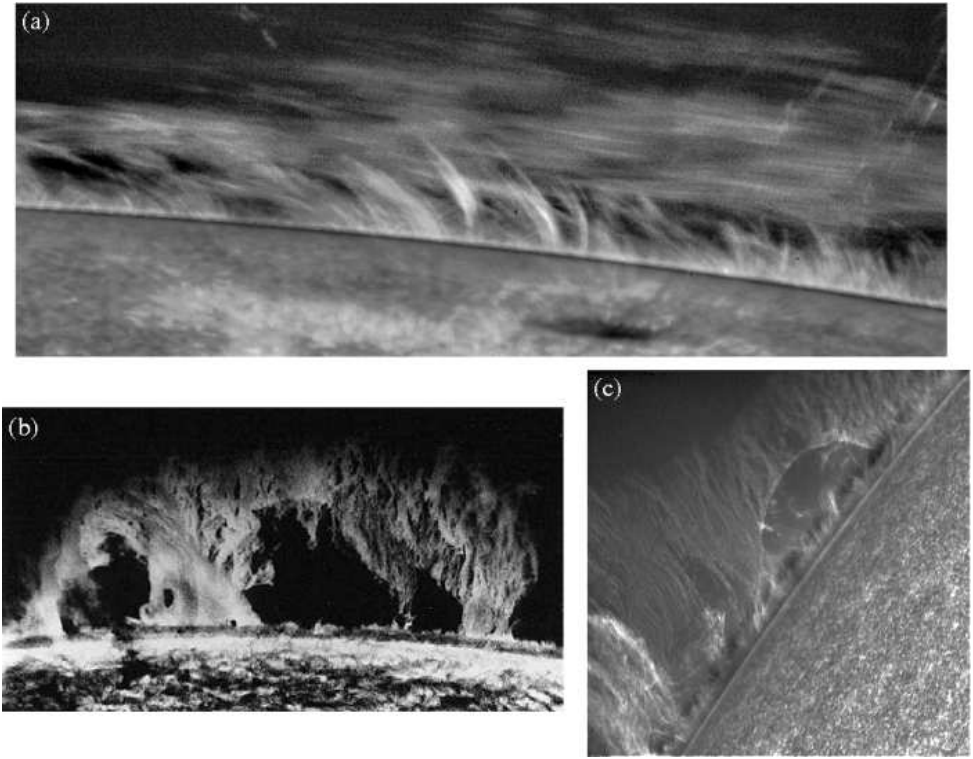}
\caption{Examples of solar prominences at the limb in different wavelengths: 
(a) Ca ~II H HINODE/SOT image from November 9, 2006  (from \citet{Okamoto2007}) 
; (b) BBSO H$\alpha$ image from 1970; (c) Ca II H HINODE/SOT image (courtesy of T. Berger). } 
\label{fig:new_fig3}
\end{figure}

In contrast to the view on the disk, when observed above the solar limb 
different prominences can have 
very different appearance as can be seen from the three examples in
Figure~\ref{fig:new_fig3}. In some cases the prominence
consists of a collection of nearly horizontal threads or elongated
blobs (Figure~\ref{fig:new_fig3}(a)), similar to the features seen in 
filaments on the disk. High-resolution H$\alpha$ and Ca~ II H observations of 
prominences have recently been obtained with the Solar Optical Telescope (SOT)
on the Hinode Satellite. \cite{Okamoto2007} observed horizontal threads in a
prominence near an active region and studied the oscillatory motions of these
threads. A movie of the evolution of these threads may be seen
in \cite{Okamoto2007}. These
horizontal threads are most likely aligned with the local magnetic
field. In other cases, the prominence consists of a collection of quasi-vertical 
threads (Figure~\ref{fig:new_fig3}(b) and (c)). \citet{berg08} observed a 
hedge-row prominence and
found that the prominence sheet is structured by both bright
quasi-vertical threads and dark inclusions. The bright structures are
down-flow streams with velocity of about 10 \kms, and the dark
inclusions are highly dynamic up-flows with velocity of about 20
\kms. It is unclear what drives these up-flows and down-flows. The
down-flow velocities are much less than the free-fall speed, indicating
that the plasma is somehow being supported against gravity
\citep{pec00, Mackay2001b}. How these quasi-vertical threads relate
to the magnetic field in such {\it hedge-row prominences} is not well 
understood. For example, \citet{Schmieder2010} have shown that dopplershift
line-of-sight velocities in such threads can be of the same magnitude as observed
vertical velocities. Assuming field-aligned flows, the authors argue for the existance
of an oblique magnetic field in 3D, that is projected on the plane of the sky.
In addition, it is unclear how the lower altitude quasi-vertical threads 
seen above the limb relate to the barb threads seen in filaments on the disk. A major 
challenge of prominence research is to reconcile the different structures observed
in filaments and prominences when seen on the disk versus limb.

Filaments and prominences can also be observed in the H~I Lyman lines
\citep[e.g.,][]{Korendyke2001, Heinzel_etal2001b, Patsourakos2002,
Schmieder2007, gun07} and in the He~II 304 {\AA} resonance line
\citep[][]{wang99}. These lines likely originate in the
prominence-corona transition region (PCTR). While prominences 
simultaneously observed in He~II 304 {\AA} and H$\alpha$ must lie in 
the same magnetic configuration, they commonly
have different morphologies \citep{Wang1998,Heinzel_etal2001a, aul02b, Dudik2008}.
Differences may include the observed width and whether the spine is visible or not.
Such differences may be explained by different formation mechanisms of these lines (see
sections 8.1 and 8.2 in Paper I) and their optical depths. Another question that
arises from considering different lines in prominences, is whether lines formed 
from ions and neutrals should exhibit the same prominence structure and if so what
role ion-neutral coupling plays in this?.  Observations of prominences with
the SUMER instrument on board the Solar and Heliospheric Observatory
(SoHO) satellite have shown that the Lyman lines have asymmetric
profiles, indicating that (1) there are multiple threads along the
line of sight, and (2) the threads move relative to each other with
velocities of order 10 {\kms} \citep[][]{gun08}. For more detailed
discussion of these observations, see sections 8 and 9 of Paper I.

It should be kept in mind that the ``filament'' plasma seen in
absorption on the disk may not be exactly the same as the
``prominence'' plasma seen above the limb some days earlier (East
limb) or later (West limb). One complication in comparing filaments
and prominences is that the prominence fine structure changes
continually with time. Also, active region filaments, those filaments
lying within the centers of activity complexes, generally lie at
such low heights (less than 10Mm, the approximate height of the spicule 
forest) that they are difficult to see at the
limb. Figure~\ref{fig:new_fig3} (a) and (c) illustrate prominences which
lie at active latitudes, but can  be seen above the spicular forest. They may
be seen, as they are not active region filaments but rather ``intermediate filaments" 
which lie on the borders of active regions and can reach much higher heights. 
Some prominences can be clearly seen in He~II 304 {\AA} above
the limb, but may not be so easily visible on the disk because of
temperature and density effects (see section 8.2 of Paper I).
\citet{Schmieder2003} and \citet{Schwartz2006} show that cool plasma
($\sim 10^4$ K) with very low density could be present in the vicinity
of filaments. This plasma is not visible in H$\alpha$ but is
detectable in EUV due to absorption of UV line radiation by the Lyman
continuum (see Paper I).
This may explain why prominences do not have the same aspect as
filaments when crossing the limb.

\subsubsection{Filament Channels}
\label{sec:2.1.2}

\begin{figure}[t]
\centering\includegraphics[scale=0.7]{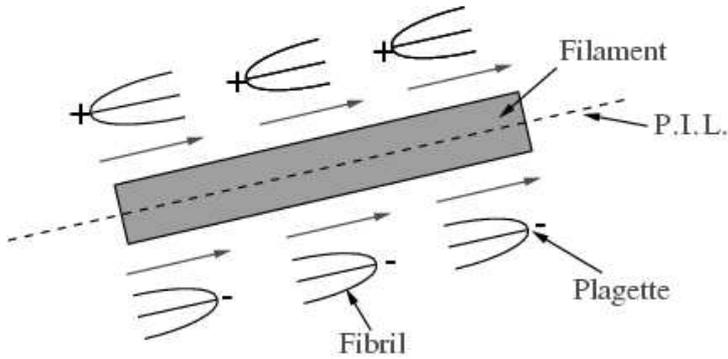}
\caption{Schematic of a filament channel with fibrils which lie (1)
anti-parallel to one-another on either side of the PIL and (2) nearly
parallel to the path of the PIL. The anti-parallel alignment indicates
that the magnetic field (arrows) is dominantly horizontal and points
in the same direction on either side of the channel.}
\label{fig:mackay_fig1}
\end{figure}

\begin{figure}[t]
\centering \includegraphics[scale=0.3]{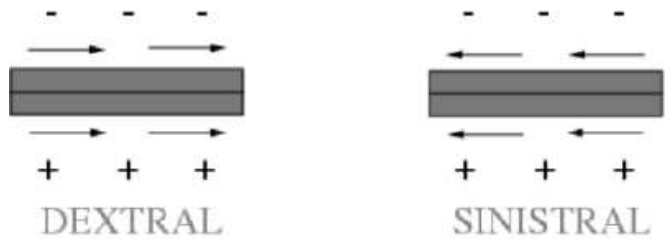}
\caption{The {\it chirality} of a filament channel is defined in terms
of the direction of the magnetic field along the channel (denoted by
arrows) when viewed by an observer on the positive polarity side of
the channel. For a dextral (sinistral) channel, the magnetic field
points to the right (left).}
\label{fig:mackay_fig4}
\end{figure}

Filaments are located in {\it filament channels}, described as regions
in the chromosphere surrounding a PIL where the chromospheric fibrils
are aligned with the PIL \citep{Martres1966, Gaiz1998}. These fibrils
are interpreted as giving the direction of the magnetic field in the chromosphere.
\citet{Foukal1971a, Foukal1971b} noted that the fibrils emanating from
magnetic elements in the channel show a streaming pattern that is
opposite on the two sides of the channel (see
Figure~\ref{fig:mackay_fig1}). Given the magnetic polarity of these
elements, Foukal deduced that the horizontal component of magnetic
field must point in the same direction on the two sides of the channel
\citep[also see][]{Martin1992, Martin1994}. This magnetic field along
the PIL is believed to extend some height into the corona, and the
filament is embedded in this field. The existence of this horizontal
field within filaments was confirmed by direct measurements using the
Zeeman and Hanle effects \citep[][]{Hyder1965, Rust1967, Leroy1983,
Leroy1989, Zirker1998b}.
\citet{Martin1992} introduced the concept of {\it chirality} of
filament channels. They classified filament channels as either
``dextral'' or ``sinistral'' depending on the direction of the axial
component of the field as seen by an observer standing on the
positive-polarity side of the channel (Figure~\ref{fig:mackay_fig4}).
To determine the chirality of filament channels, which do not
necessarily contain a filament, high resolution H$\alpha$ images (to resolve
individual chromospheric fibrils) are required. In strong field
regions usually only H$\alpha$ images are required. In contrast, for weak field 
regions where fibril patterns are not strong, magnetograms may also
be used to aid the determination of chirality, by using them to determine the 
polarity of the magnetic elements from which the fibrils extend from or go into.  
\citet{Martin1994} showed that channels in the northern
hemisphere are predominantly dextral, while those in the south are
predominantly sinistral \citep[also see][] {Leroy1983, Zirker1997}.
The origin of this hemispheric pattern will be further
discussed in Sections \ref{sec:5.1.2} and \ref{sec:5.5}.

As magnetic fields in filament channels are believed to extend
to higher heights and the filament embedded in this field. Filaments may also
be classified as being of ``dextral'' and ``sinistral type''. The chirality of
filaments may be deduced, either, indirectly from that of the channel or
directly from magnetic field measurements \citep[also see][] {Leroy1983}. To
date no simultaneous studies comparing the chirality in filaments determined
both indirectly and directly has taken place.

Spectral diagnostics of the magnetic field orientation in a
prominence observed with SUMER can also be used. Lyman lines
have reversed/non-reversed profiles when the magnetic field of the
prominence is perpendicular/parallel to the line of sight \citep[][]
{Heinzel2005}. This has been tested by \citet{Schmieder2007} with
a round shaped filament crossing the limb. In the EUV, filament
channels are much broader than H$\alpha$ filament channels \citep[][]
{Heinzel_etal2001a, aul02b}. \citet{Heinzel_etal2001a} explain this
enhanced width by absorption of EUV line radiation by the H~I Lyman
continuum; lower emission in the channel (``emissivity blocking'')
is another mechanism. \citet{aul02b} used their magnetic flux tube
models to deduce that magnetic field lines with concavities 
may be present
in EUV filament channels where plasma is cool.
The existence of concave up magnetic field lines has been tested
recently by polarimetric measurements, in filament channels
\citep[][]{lites05} and close to the feet of filaments \citep[][]
{lop06}.

\subsubsection{Filament Barbs}
\label{sec:2.1.3}

When observed from above (i.e., at disk center), the filament barbs
are seen to protrude at an acute angle with respect to the long axis
of the filament, like ramps off an elevated highway \citep[][]
{Martin1992}. Therefore, filaments can be classified as either
{\it right-bearing} or {\it left-bearing} depending on the directions
of the barbs as seen from above. Martin and collaborators found a
one-to-one correlation between the right/left-bearing structure of the
barbs and the chirality of the filament channel: filaments in dextral
channels have right-bearing barbs, and those in sinistral channels
have left-bearing barbs. An important goal of filament modeling is to
explain the cause of this relationship. \citet{Bernasconi2005} have
developed software for automated detection of filament barbs and the
chirality of filament channels.

When observed away from disk center, 
the barbs are seen to be inclined structures that extend from the filament 
spine to the chromosphere below \citep[e.g., SST observations of][]{Lin2008a}. 
The vertical extent of a barb is much
larger than the gravitational scale height of the prominence plasma
(about 200 km), so the barbs cannot be static field-aligned
structures; the plasma must somehow be supported against gravity.
Therefore, a key issue is the orientation of the magnetic field in the
barbs. This issue will be further discussed in section \ref{sec:2.2}.

The relationship between the filament barbs and the underlying
photospheric magnetic field has been discussed by many authors.
\citet{Martin1994} argue that the ends of the barbs of filaments on
the disk are connected to weak magnetic fields in between the network
elements. \citet{me94} propose that the filament barbs are anchored
in parasitic (minority) polarity elements, i.e., weak magnetic fields
with opposite polarity compared to the dominant (majority) polarity
elements on the side of the filament where the barb is located.
\citet{aul98} show that the barbs move in accordance with the changes
of parasitic polarities during one day of observations. \citet{wang99,
wang01}, \citet{chae05} and \citet{Lin2005a} find that the ends of the
barbs are located very close to small-scale PILs between majority and
minority polarities on the side of the filament.

\citet{wang01} and \citet{Wang2007} proposed that flux cancellation
between the parasitic polarity and the neighboring dominant polarity
plays a key role in the formation of filament barbs. Flux cancellation
refers to the disappearance of photospheric magnetic flux by mutual
interaction of opposite polarity fields in the photosphere \citep[][]
{Livi1985, Martin1985}. It occurs everywhere on the Sun, but is
particularly common at large-scale PILs on the quiet Sun where
opposite polarity elements intermix and a zone of mixed polarity is
created.  At the edge of such a mixed-polarity zone, parasitic flux
elements are likely to cancel against dominant polarity elements.
\citet{wang99, wang01} argued that magnetic reconnection accompanying
photospheric flux cancellation is the dominant mechanism for injecting
mass into quiescent prominences. In contrast, \citet{Martens2001}
propose that downward-inclined barbs, linked to parasitic polarity
elements, arise as a result of failed cancellation, i.e., the motion
of flux elements across the PIL without cancellation. Changes of
minor polarities in filament channels lead to strong changes of
H$\alpha$ filaments and prove directly the relationship between minor
polarities and barbs/ends of filaments.  The disappearance and
reappearance of a part of a filament has been directly related to
emergence of magnetic flux, followed by canceling flux \citep[][]
{Schmieder2006}.

\citet{Zirker1998a} observed counter-streaming of plasma in a quiescent
filament (see section 4, Paper I). Using observations at three wavelengths 
in the H$\alpha$
line, they found evidence that matter flows along the spine of the
filament in both directions. Similar counter-streaming occurs within
the barbs: in one wing of the spectral line they observe up-flows from
the barbs into the spine, while in the other wing they see down-flows.
It is unclear what drives these flows, but they appear to be parallel
to the magnetic field.
Therefore, the observed flow direction can be used to infer the
direction of the magnetic field, which may help resolve the question
whether the magnetic fields in barbs are inclined or horizontal.
\citet{Zirker1998a} propose that a filament barb consists of a set of
closely spaced flow channels that are highly inclined with respect to
the vertical direction, with some channels having up-flows and others
having down-flow. The velocities involved are 5 to 10 km/s, much less
than the free-fall velocity corresponding to the filament
height. \citet{Deng2002} found counter-streaming in an active region
filament. \citet{Lin2003} and \citet{Schmieder2008} observed such
flows in large polar crown filaments two days before filament
eruptions, so the counter-streaming may be a signature of the slow
rise of a filament before eruption. With new high-resolution
observations such as those by Hinode, many more examples of 
counterstreaming are expected to be found. With this it will be determined
whether counterstreaming is a property common to all filaments and prominences,
or whether it is a signature of the early stages of the onset of eruption.

\begin{figure}[t]
\centering \includegraphics[scale=0.4]{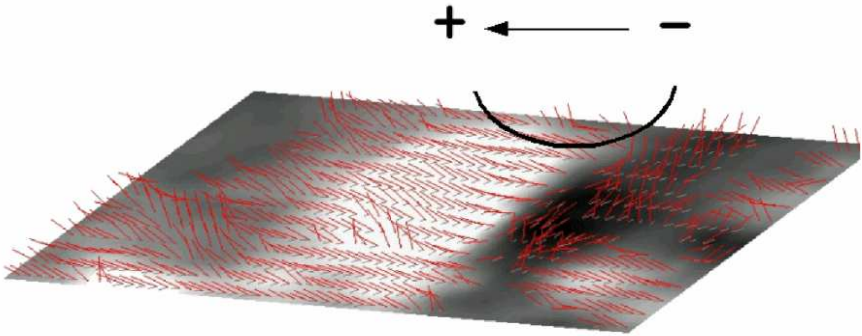}
\caption{
Vector magnetic field in a filament channel as observed with
TH\'{E}MIS. From \citet{lop06}.}
\label{fig:lopez_fig1}
\end{figure}

To investigate the nature of magnetic fields in the photosphere 
below and around barbs, polarimetric measurements were recently made by 
\citet{lop06} using TH\'{E}MIS/MTR (T\'{e}lescope H\'{e}liographique 
pour l'Etude du Magn\'{e}tisme et des Instabilit\'{e} Solaires). In 
Figure~\ref{fig:lopez_fig1} an illustration of the 3D field can be seen,
where white/black denotes the positive/negative normal field component and the
red arrows, the direction of the horizontal field. Through using filament channel
chirality rules to resolve the 180$^\circ$ ambiguity, the authors deduce that the
field is of inverse polarity at the level of the photosphere. This type
of photospheric configuration, with a collection of high altitude magnetic dips crossing
the local PIL, as hinted from Figure~\ref{fig:lopez_fig1}, may fit many different structures 
of the coronal field
lying above. An alternative interpretation of the coronal structure associated with
the measured photospheric magnetic field is a vertical current sheet with no connections across the
PIL. However such a structure is not stable to any form of non-ideal MHD perturbation and
is likely to be short lived. Non-ideal MHD perturbations would first result in the tearing 
of the field \citep{furth63} and then the  formation of an X-point structure  
(when viewed in a 2D projection along the PIL). Subsequent reconnection would then produce 
connections across 
the PIL. Depending on the force balance of the field and the nature of the extended current 
systems formed in the relaxation process, either a coronal arcade (of normal polarity) or 
bald patch (of inverse polarity) may result. As the long term existence of such a current 
sheet is unlikely, and since the photospheric field is of inverse polarity, the authors deduce that bald patches, 
i.e. magnetic dips that touch the photosphere, are present. These magnetic dips or hammocks are 
deduced to support the plasma in the barbs. 
Such observations are a major step in understanding the magnetic field structure in filaments 
and barbs. However to fully understand the magnetic field in filaments, new high resolution 
polarimetric measurements in many filaments are required.

\subsubsection{Measurements of Prominence Magnetic Field}
\label{sec:2.1.4}

While observations of prominence fine structure provide some clues to
the magnetic topology, the most direct determination of the prominence
magnetic field comes from the inversion of spectro-polarimetric data
\citep[see reviews by][] {Paletou2003, Paletou2008, Lopez2007}.
Only a few spectral lines in the optical spectrum of prominences are
suitable for magnetic field measurements using Hanle-effect diagnostic
(the Hanle effect is the change in the polarization
state of resonantly scattered radiation due to presence of a magnetic
field). The H$\alpha$ line is strong and has interesting polarization
profiles \citep[][]{Lopez2005, Bianda2006, Ramelli2006}, but the line
is generally optically thick and the polarization transfer in this
line is not understood. The lines most often used for measuring
prominence magnetic field are the He~I multiples at 5876 {\AA} in the
visible (He~$\rm D_3$) and at 10830 {\AA} in the near-infrared. For
recent attempts in modeling these lines see \citet{Leger2009}. The
spatial resolution of such measurements is typically a few arc seconds,
i.e., they cannot yet resolve the fine-scale structures seen in
high-resolution images of prominences such as those obtained in
SOT.

A comprehensive effort to measure prominence magnetic fields was
conducted in the 1970's and early 1980's, using the facilities at
{\it Pic du Midi} (France) and Sacramento Peak Observatory (USA).
The results of this work are reviewed by \citet{Leroy1989}, and can be
summarized as follows \citep[][]{Paletou2003}. The magnetic field in
quiescent prominences has a strength of 3-15 G. The field is mostly
horizontal and makes an acute angle of about $40^\circ$ with respect to the
long axis of the prominence \citep[][]{Bommier1998}. The field strength
increases slightly with height, indicating the presence of dipped
field lines. Most prominences have {\it inverse} polarity, i.e., the
component of magnetic field perpendicular to the prominence has a
direction opposite to that of the potential field. The component along
the prominence axis obeys a large-scale pattern \citep[][] {Hyder1965,
Rust1967, Leroy1983}, which is now known as the chirality pattern
\citep[][]{Martin1994, Zirker1997} and will be further discussed in
Sections \ref{sec:5.1.2} and \ref{sec:5.5}. In the following discussion
unless otherwise stated, angles given for the magnetic field in prominences
are measured as an acute angle relative to the long axis of the prominence. Many 
features may effect these measurements, such as the position of the 
filament on the disk and the spatial resolution of the images. For each set of 
observations the reader should refer back to the original papers for such
information.

\citet{HLin1998} presented the first full-Stokes
spectro-polarimetric observations of a filament on the disk in He~I
10830 {\AA}. \citet{Trujillo2002} demonstrated the importance of
lower-level atomic polarization in such measurements (the lower level
of the 10830 {\AA} line is the ground level of the triplet system of
neutral helium). They showed that this atomic polarization creates
linear polarization of radiation in forward scattering at disk
center. This provides a new technique for measuring the azimuth of
magnetic fields in filaments. \citet{Collados2003} applied this method
to observations of two filaments in the Northern hemisphere. They find
that the magnetic field is nearly horizontal, and that the magnetic
vector is rotated with respect to the filament axis by an angle in the
range 15 to 25 degrees.

\citet{Paletou2001} reported full-Stokes observations of a prominence
in He~$\rm D_3$. Besides linear polarization signals due to the
Hanle effect, they also found circular polarization signals and they
derived magnetic field strengths of 30-45 G in an active part of the
prominence. \citet{Casini2003} published the first map of the vector
field in a prominence. They found that the average magnetic field in
prominences is consistent with earlier studies \citep[see][]
{Leroy1989}: a mostly horizontal field with a strength of about 20 G
and with the magnetic vector pointing $20^\circ$ to $30^\circ$ off the
prominence axis. In addition, the map shows magnetic fields
significantly stronger than average (up to 80 G) in clearly organized
patches within the prominence \citep[also see][] {Wiehr2003,
Lopez2002, Lopez2003, Lopez2007}. It is unclear whether these
structures are related to the fine structures of quiescent prominences
(e.g., vertical threads).

\citet{Casini2005} discuss various aspects of Stokes profile
inversion, including the problem of finite optical depth of the
He~$\rm D_3$ line in the prominence, and the effects of line-of-sight
integration of the polarization signals from different magnetic
configurations. They confirm the presence of magnetic fields
significantly stronger than average in limited regions of the
prominence. They also emphasize the importance of full Stokes
inversion for a correct diagnostic of magnetic fields in prominences.
For certain values of magnetic field strength and inclination of the
field with respect to the vertical direction, the Van Vleck effect
produces a $90^\circ$ ambiguity in the position angle of the magnetic
field as projected onto the plane of the sky (in addition to the
well-known $180^\circ$ ambiguity). This $90^\circ$ ambiguity makes
it difficult to distinguish between horizontal and vertical
fields in certain cases, and including Stokes V in the analysis helps
disentangle this ambiguity. However, the results from the earlier
studies based only on linear polarization measurements are thought
not to be strongly affected by this ambiguity (L\'{o}pez Ariste,
private communication).

Recently, \citet{Merenda2006} observed He~I 10830 {\AA} in a polar
crown prominence above the limb, and found evidence for fields of
about 30 G that are oriented only $25^\circ$ from the vertical
direction. The reason for the different orientation of the magnetic
field in this prominence (compared to the more horizontal fields found
in other studies) is unclear. One possibility is the $90^\circ$
ambiguity due to the Van Vleck effect, but the authors claim their
particular measurement is unambiguous. 

\subsubsection{Magnetic Fields and Flows in the Photosphere Below
the Prominence}
\label{sec:2.1.5}

An indirect method for probing the prominence magnetic field is to
measure the vector field in the photosphere below the prominence.
\citet{lop06} deduced the presence of magnetic dips in the
photosphere below a filament after having resolved the 180-degree
ambiguity in photospheric vector-field measurements using the chirality
rules for filaments (see Figure \ref{fig:lopez_fig1}). These so-called
{\it bald patches} are sites on the PIL where the horizontal field
perpendicular to the PIL has ``inverse'' polarity, i.e., the field
points from negative to positive polarity, opposite to the direction
expected for a potential field \citep[see also][]{lites05}.
\citet{lop06} argue that the dips are consistent with the
presence of a weakly twisted flux rope in the corona above the
PIL. If this is the case, the filament is thought to be supported 
by similar dips in the
helical field lines at larger heights.

Flows in the photosphere below the prominence may play an important
role in the evolution and stability of the magnetic field supporting the
prominence. Observations of shear flows, highlighted by maps of the
vorticity, have been found below filaments, but mostly during their
eruption \citep{Bala98,Keil99}. Until recently, observations of
organised photospheric flows below quiescent filaments were very rare.
\citet{Rondi2007} measured meso- and super-granular flows
in the vicinity and below a filament before and during its eruption.
They found that the disappearance of the filament starts where both
parasitic and normal magnetic polarities are swept by a continuously
diverging flow corresponding to supergranular diffusion. They also
observed the interaction of opposite polarity elements, which is
another candidate for triggering the filament destabilization as it
implies a reorganization of the overlying coronal magnetic field.

\citet{Roudier2008} studied large-scale horizontal flows in the
photosphere below a filament, using series of full-disk Dopplergrams
and magnetograms obtained with the Michelson Doppler Imager (MDI) on
SoHO. In Dopplergrams, the supergranular pattern is tracked using
local correlation tracking (LCT), while in magnetograms LCT is used to
obtain horizontal velocities. The topology of the observed large-scale
flow-field changes during the eruptive phase suggests a coupling
between the surface flows and the coronal magnetic field. An unusually
fast north-south flow stream of amplitude 30 to 40 $\rm m ~ s^{-1}$
was found to be compatible with the rotation of a part of the
filament. This behavior suggests that the filament footpoints have
been moved by the surface flows. The influence of the flow stream was
enhanced by differential rotation. The authors measured an
increase of the zonal shear in the region below the starting point of
the eruption, and a decrease after the eruption.  These results
indicate that, even if large-scale flows do not lead to large-scale
changes of the photospheric magnetic field, they nevertheless might be
able to perturb the coronal field to an extent that filament eruptions
and coronal mass ejections can be triggered.

\subsubsection{Coronal Structures Above the Filament Channel}
\label{sec:2.1.6}

Quiescent prominences are located in cavities at the base of coronal
helmet streamers \citep[][]{Engvold1989}. In white-light observations
above the limb, the cavity is darker than its surroundings and the
height and width of the cavity is roughly twice the height of the
prominence \citep[][]{sai73}. Some streamers contain concentric arch
systems \cite[][]{sai68}, suggesting the presence of an arcade
of coronal loops overlying the cavity. \citet{Low1995} argue that the
cavity represents a magnetic flux rope.

In EUV and X-ray observations of filament channels on the solar disk,
the coronal loops overlying a prominence often cannot be clearly
identified, perhaps because they are too faint compared with the
lower-altitude structures. However, X-ray arcades can often be seen
in active regions, and after eruptive events on the quiet Sun
\citep[][]{Martin1996}. The coronal loops in these arcades do not
cross the PIL at right angles, but are skewed in the direction along
the PIL (``sigmoids''). The loops can be right-bearing or left-bearing
with respect to the direction along the PIL (as seen from above). The
observations indicate that dextral (sinistral) channels have left
(right)-skewed coronal arcades, opposite to the right (left)-bearing
structure of the filament barbs \citep[][] {Martin1996, Martin1998,
McAllister2002, Schmieder2004}. Taking into account the polarity of
the dominant flux on either side of the filament, this implies that
the axial component of magnetic field in the arcade is the same as
that in the filament channel. Therefore, this implies that
the filament channel observed
in the chromosphere is part of a larger structure that extends to
heights well above the prominence.

\subsection{Models of Prominence Magnetic Structure}
\label{sec:2.2}

The above-mentioned observations show that prominences are embedded in
magnetic fields that are highly non-potential. As the dominant
component of the field lies along the filaments long axis, such non-potential
fields exhibit strong magnetic shear. However, the detailed
structure of these fields and the associated electric currents are not
well understood. In the following we review various models for
prominence magnetic structure. To begin with we focus on models that
describe the structure as it exists at one instant of time, and we
in general will ignore for the moment the question of how the field came 
to be that way. In Section~\ref{sec:3} we will see that the mass in 
prominences is infact dynamic. However since the flow speed is generally
much less than the Alfven speed and within filaments the plasma-$\beta$
is low, such motions are often expected to be along field lines and 
not to strongly effect the field. Therefore, under this approximation, static 
modeling is useful to describe the large-scale magnetic field structure 
of the prominence. This field is expected to evolve on times-scales much longer 
than that of the observed flows. Issues related 
to the formation of the prominence magnetic field and its evolution will be 
addressed in Sect.~\ref{sec:5}.

An important question in prominence modeling concerns the role of
magnetic dips in the structure and dynamics of the prominence plasma.
\citet{Kippenhahn1957} showed how cool plasma can be supported
against gravity by a magnetic field containing {\it dips} in the field
lines, i.e., sites where the field lines are locally horizontal and
curved upward. For example, dips can occur in a potential field when
the underlying photospheric magnetic sources have a quadrupolar structure
(``quadrupolar dips''). However, the magnetic fields in filament channels
are not potential, and contain significant electric currents. Dips in
non-potential magnetic fields can occur for several reasons:
\begin{enumerate}
\item
The structure of the magnetic field is significantly affected by the
{\it weight} of the prominence plasma, distorting the field in such a
way as to create magnetic dips.
Two examples are the isothermal \citet{Kippenhahn1957} model and the
non-isothermal \citet{Hood1990} extension. In both the shape of
the dipped field lines is determined by the balance between magnetic
and gravitational forces. Many authors have constructed
magneto-hydrostatic (MHS) models of prominence threads based on the
Kippenhahn-Schl\"{u}ter model \cite[e.g.,][]{Malherbe1983, hein01,
Heinzel2005, Low2005}. Another example is the ``normal polarity'' flux
rope model by \citet{Low2002}. In this model the sense of magnetic
twist in the flux rope is opposite to that in the overlying coronal
arcade, and the force of gravity plays a crucial role in the existence
of dipped field lines. In Paper I, Section 9.2 the Kippenhahn-Schl\"{u}ter 
model is used to compute the hydrogen spectrum in non-LTE modeling. 
\item
Another possibility is that the dips exist in the absence of
prominence plasma, i.e., the dips are not caused by the weight
of the prominence. According to such models, the filament plasma
is likely to be located near the dips of the field lines (see,
however, Section~\ref{sec:3}). If the plasma $\beta < 1$, these
currents must flow nearly parallel to the field lines, so that the
magnetic field is approximately force-free ($\beta$ is the ratio of
gas pressure and magnetic pressure).
Such {\it force-free fields} can contain dips in the field
lines even when the underlying photospheric field is dipolar.
In fact, many authors have suggested that filaments are supported by
nearly force-free {\it flux ropes} that lie horizontally above the PIL
\citep[e.g.,][]{Kuperus1974, Pneuman1983, Balle1989, Priest1989,
Rust1994, Aulanier1998, chae01b, Gibson2006}. An alternative
force-free configuration with dips, the {\it sheared arcade}, has
also been studied as a description of the filament channel magnetic
structure \citep{Antiochos1994, DeVore2000, Aulanier2006}.
\end{enumerate}

In quiescent prominences with low field strengths ($B < 10$ G) we may
have a mixture of these two cases: the existence of the dips is due
to the presence of a large-scale flux rope or sheared arcade (as in
case 2), but the magnetic field near the dips is significantly
distorted by the weight of the prominence plasma ($\beta > 1$).
In this case there are significant perpendicular currents near the
dips, and the flux rope or sheared arcade is not force free.

In the remainder of this section we consider the second category of
models which describe the 3D structure of the magnetic field in the
prominence, under the approximation that the weight of the prominence 
plasma does not play an
essential role in the large-scale structure of the magnetic field.
These models differ in terms of the requirement of dips for the
formation and persistence of filaments. While the models differ
in this sense, one common feature is that the magnetic field
has a strong axial component along the PIL. Such a field may be described
as strongly ``sheared". Magnetic shear either in the form of an arcade or
a flux rope may build up through a number of process. Examples 
include: shearing motions \citep{Antiochos1994}, emergence of the upper part 
of a flux rope \citep{man04,arch08b,DeVore2000,fan09} or shearing motions 
followed by convergence and
cancellation of photospheric flux \citep{Balle1990,Litvinenko2005}. The
origin of the axial fields in filaments will be fully discussed in 
Section~\ref{sec:5.1} where a wide range of models and mechanisms for 
producing magnetic shear are discussed. Returning to the
topic of static configurations, Section~\ref{sec:2.2.1} discusses the 
\citet{me94} Empirical Wire Model, which suggests that 
dips are not essential. Following this we describe two models in which 
dips play increasingly important roles. These are the sheared arcade model 
(Section~\ref{sec:2.2.2}) and finally the linear and non-linear flux rope 
models (Section~\ref{sec:2.2.3}).   

\subsubsection{Empirical ``Wire Model"}
\label{sec:2.2.1}

Through following a time series of H$\alpha$ observations of a 
quiescent filament, along with photospheric magnetic field distributions, 
\citet{me94}, proposed an empirical ``wire model" for the magnetic field 
geometry in a filament. A key assumption of their model is that the fine 
scale structure observed in H$\alpha$ lies parallel to the magnetic field. 
In the ``wire-model'' the filament plasma is located on magnetic arches 
that are highly sheared in the direction along the PIL. Although they are
highly sheared, the arches do not contain dips. This feature 
distinguishes this model from others discussed in later sections.
Some field lines run along the entire length of the filament and outline
the filaments ``spine". Other shorter ones run partially along the spine,
but spread out from it and connect down to minority polarity elements on 
either side of the PIL. These shorter structures represent the filament barbs.
As neither the spine or barb field lines contain dips, the model requires
the existence of other (non-magnetic) forces to act parallel to the
inclined field lines to supported the cool dense plasma against
gravity. Without such forces, the
cool plasma would fall down to the chromosphere in a matter of
minutes, which is not observed. The model however does not address this
question. In Section~\ref{sec:3} possible candidates for such forces 
will be discussed. While the \citet{me94} model does not consider the origin 
of the sheared arcade, such a structure may be formed by shearing motions 
of the photospheric footpoints or by the emergence of the upper part of a 
flux rope as will be discussed in Section~\ref{sec:2.2.2}.

\begin{figure}[t]
\centering \includegraphics[scale=0.5]{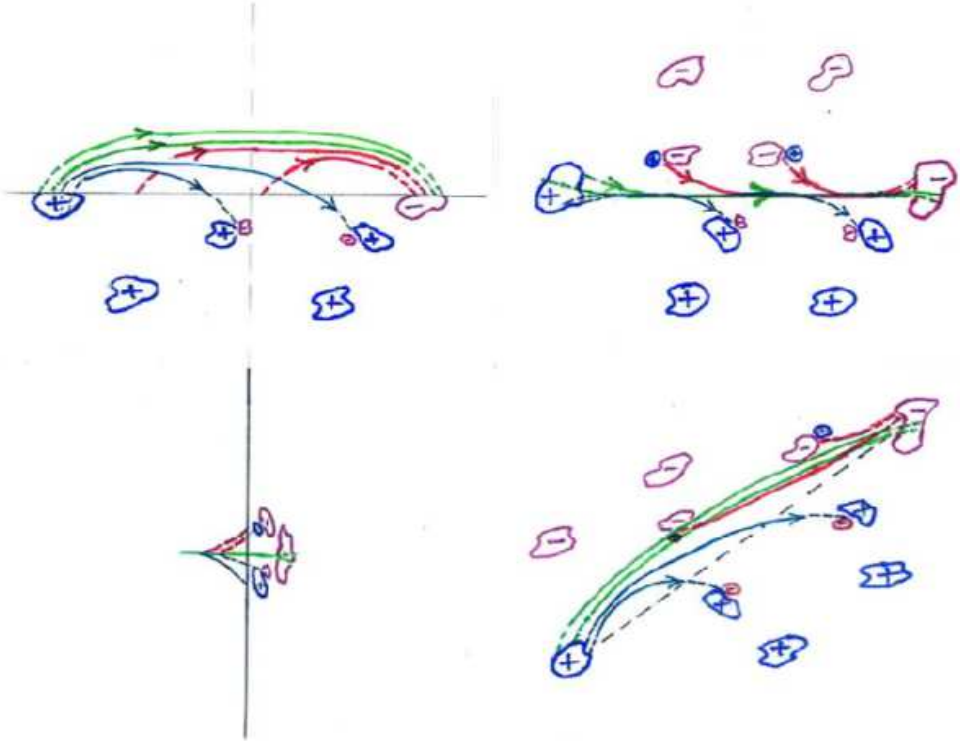}
\caption{
The \citet{me94} model of filament magnetic structure. The four
panels show a dextral filament as seen from different perspectives.
The {\it green} field lines show the filament spine, and the {\it
blue} and {\it red} field show the barbs. From \citet{Lin2008a}.}
\label{fig:ylin_fig}
\end{figure}

The Martin \& Echols model was further discussed and developed by 
\citet{Lin2008a}, who presented the diagram shown in Figure
\ref{fig:ylin_fig}. The panels illustrate the magnetic field
geometry deduced from four high resolution H$\alpha$ images of filaments
taken from different angles and including different classifications 
(active region-top left, intermediate - top right and 
two quiescent filaments - bottom). In each image the spine is represented 
by the green field lines and the barbs by blue and red field lines. Finally the end of 
the filament is denoted by the dashed lines entering the network fields at 
either end. From high resolution $H\alpha$ images the authors deduce that 
thread like structures are common to all classifications of filaments and that
in some examples the barbs consist of long thin threads that are inclined with 
respect to the horizontal. The authors conclude that such thin threads as the basic 
building blocks of all filaments no matter their classification. Subsequently the only 
difference between classifications may be the relative importance of the spine, barbs and 
ends. 

A key assumption of the Martin \& Echols model that is illustrated by \citet{Lin2008a} 
is that as the barbs, which are inclined field lines, spread out from the main body 
and only connect to parasitic polarities in the photosphere. If this is the 
only allowed case it follows that dextral channels produce filaments with right-bearing
barbs, and sinistral channels produce left-bearing barbs, consistent
with observations. However, as the authors consider barbs as inclined field lines,
the model does not explain {\it why} these inclined field lines
cannot extend upwards from network fields on either side and join into the spine, thus producing the
opposite bearing orientation of barbs on either side. Therefore, the model only provides 
a partial explanation for the observed correlation between the right/left-bearing structure 
of the barbs and the chirality of the filament channel. Later when discussing flux 
rope models (Section~\ref{sec:2.2.3}) we will see that these models can provide a possible 
explanation for this structural asymmetry when magnetic dips are considered.

\subsubsection{Sheared Arcade Model}
\label{sec:2.2.2}

\begin{figure}
\includegraphics[width=4.8in]{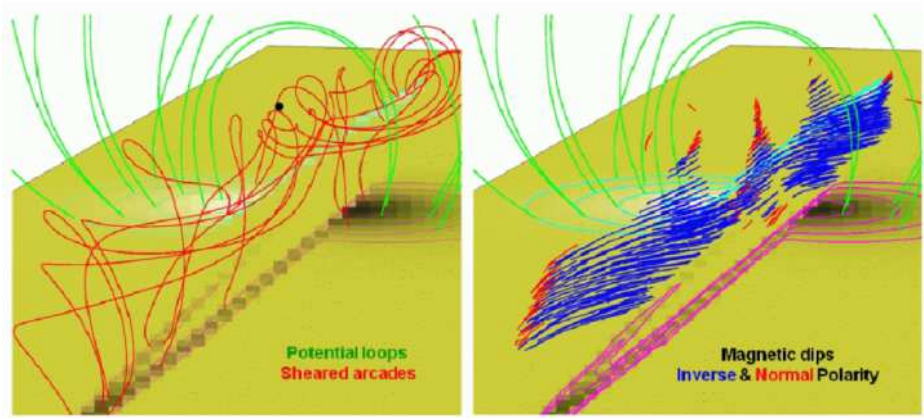}
\caption{
Sheared arcade model for magnetic field in a filament channel:
(a) Sheared field lines ({\it red}) with overlying potential loops
({\it green}). (b) Distribution of normal polarity ({\it red}) and
inverse polarity ({\it dark blue}) magnetic dips. Each field line is
plotted up to a height of 300 km above the dip to simulate the effect
of the pressure scale height of the prominence plasma. Adapted from
\citet{aul02a}.}
\label{fig:aulanier_fig1}
\end{figure}
A ``sheared magnetic arcade'' may be formed for example by shearing motions 
of the photospheric footpoints localized around the PIL  
\citep{Antiochos1994, DeVore2000, aul02a}, by the emergence of the upper 
part of a flux rope or by large-scale vortical motions.
Numerical studies of this filament-channel formation
mechanism began with a simple bipole embedded in a larger scale
background dipolar field, subjected to strong footpoint motions,
parallel to the PIL and confined to a narrow zone on either side
\citep{DeVore2000, DeVore2005}. These oppositely directed flows drag
the innermost portion of the bipole into a zone of weaker overlying
field, yielding elongated, low-lying field lines that bulge upward at
their less constrained ends and hence become dipped. As this stressed
system relaxes, a variety of field-line paths are formed: some dipped
field lines lean across the PIL, producing inverse polarity compared
with the initial bipole, while others reconnect in the corona,
producing weakly twisted field lines.  Consequently the resulting
filament channel is neither fully normal nor fully inverse
polarity. Fig.~\ref{fig:aulanier_fig1} illustrates the resulting
magnetic structure.

\subsubsection{Flux Rope Models}
\label{sec:2.2.3}

In the remainder of this section we consider models in which the weight
of the prominence plasma does not play an essential role in the
large-scale structure of the magnetic field. Such models may be split into
two categories: weakly twisted flux ropes and highly twisted flux ropes
(e.g \citet{Rust1994,Okamoto2008}).
While two categories exist, we focus on models
containing weakly twisted flux ropes with dips in the magnetic field
lines; the prominence plasma is assumed to be located at these dips.
Weakly twisted flux ropes are only discussed as when non-erupting
filaments or prominences are observed, they do not appear to be
highly twisted. This is commonly observed in H$\alpha$ images and through flows
which are mainly horizontal. In Section~\ref{sec:5} when considering models for
filament formation a wider range of flux rope models will be discussed including
those with strongly twisted flux ropes. Weakly twisted flux ropes are 
similar to sheared arcades, in the sense that the magnetic field is dominated 
by the axial component.
One constraint on such models is that the flux rope must be
approximately in force balance with its surroundings, i.e., the
configuration must be close to magneto-static equilibrium, and this
equilibrium must be stable. Therefore, a key feature of such models
is the existence of a {\it coronal arcade} overlying the flux rope.
The coronal arcade provides the magnetic tension forces necessary to
hold down the flux rope in the low corona. Although as previously stated, weakly
twisted flux rope models are similar to sheared arcades, there is one key difference.
For flux ropes models the flux rope and arcade are independent flux systems with a
separatrix surface between them. In contrast, for a sheared arcade only a single flux
system exists.

The sense of twist within the flux rope is generally assumed to be
consistent with the direction of the magnetic field in the overlying
coronal arcade. Since the prominence plasma is located in the dips of
the helical windings, the flux rope model predicts that the magnetic
field at the prominence has {\it inverse} polarity with respect to the
surrounding photospheric fields, consistent with Hanle measurements
\citep[][]{Leroy1984}. Furthermore, the models predict that filament
channels with dextral orientation of the axial field have left-helical
flux ropes and left-skewed coronal arcades, and sinistral channels
have right-helical flux ropes and right-skewed arcades. This is
consistent with the observation that dextral (sinistral) channels have
left (right)-skewed arcades \citep[][] {Martin1996, McAllister2002,
Schmieder2004}.

If the effects of gas pressure and gravity are ignored, the magnetic
field ${\bf B} ({\bf r})$ must be more or less {\it force-free} with
electric currents flowing parallel or anti-parallel to the field
lines:
\begin{equation}
\nabla \times {\bf B} = \alpha {\bf B} .
\end{equation}
Here $\alpha ({\bf r})$ is the so-called torsion parameter, which is
constant along the field lines (${\bf B} \cdot \nabla \alpha = 0$).
The degree of twist of the flux rope is a matter of debate. In some
models the magnetic field lines inside the flux rope are highly
twisted, but in others the central part of the flux rope is only
weakly twisted and the field lines inside this core are nearly
parallel to the flux rope axis. The key difference between such models
is the spatial distribution of $\alpha ({\bf r})$. In models with
strong twist, $\alpha ({\bf r})$ has its peak on the axis of the flux
rope, whereas in models with weak twist $\alpha ({\bf r})$ has a
hollow core distribution with the peak value of $\alpha ({\bf r})$
occurring at the interface between the flux rope and its local
surroundings \citep[e.g.,][] {bob08, Su2009}. The observed
field-aligned flows may be difficult to explain with highly twisted
flux ropes.

\subsubsection{Linear Magneto-hydrostatic Models}
\label{sec:2.2.4}

Aulanier and collaborators have developed 3D magnetic models of
filaments by extrapolating photospheric magnetograms into the corona,
assuming either linear force free fields \citep[LFFF,][]{Aulanier1998}
or linear magneto-hydrostatic fields \citep[LMHS,][]{aul98,
Aulanier_etal1999, Aulanier2000, Aulanier2003, Dudik2008}. Such
models produce helical flux ropes overlying the PIL, and the filament
plasma is assumed to be located at dips in the helical field
lines. The flux rope may be perturbed by the presence of magnetic elements
in the photosphere. In areas where parasitic polarity elements are
located close to the flux rope, the dips extend away from the main
body of the filament, consistent with the observation that filament
barbs are located near parasitic polarities \citep[][]{me94,
Martin1998, wang01}. A key feature of the flux rope model is that
the dips that represent the main body of the filament are all of inverse
polarity. Therefore only
minority polarity elements may perturb them to produce extended
dips away from the main flux rope that represent barbs. Due to the direction
of the axial field in the flux rope, combined with the inverse polarity nature, such a 
perturbation always results in a right/left-bearing structure of barbs 
in dextral/sinistral orientations of the axial field
\citep[][]{Martin1992, Martin1994}. Therefore the flux rope model, along with
any other model that exhibits these same properties, provides a natural 
explanation for the observed orientation of barbs.
In addition the flux rope model resolves the problem of the opposite orientation 
of filament fine structures and the overlying loops, which is a real problem for the
Martin \& Echol's model.

The modeling by Aulanier and collaborators predicts that the lowermost
dips form bald patches on a part of the PIL separating the
parasitic polarity from the surrounding dominant flux. Therefore,
the ``ends'' of the barbs are located at the PIL, and the field lines
passing through the barb are not actually rooted in the parasitic
polarity. The model agrees with recent observations of bald patches in
the photospheric vector field in areas where filament barbs are located
\citep[][]{lop06}. This means that the ends of the filament barbs
would not be rooted directly in the photosphere as proposed by
\citet{wang01} and \citet{Lin2005a}.
The observations of magnetic dips are in contradiction with the
idea of barbs rooted in parasitic polarities \citep[][]{Lin2005a} and
with the interpretation of the observed counter-streaming motions as
plasma flows along fixed, inclined field lines \citep{Zirker1998a}.
More polarimetric observations should be done jointly with H$\alpha$
high spatial resolution observations of fine structures in quiescent
filaments to distinguish between the different magnetic models for the
barbs. For active filaments and dynamic fibrils the situation may be
different, and the fine structures probably end in the dominant
polarities \citep[][] {Lin2008b}.

\begin{figure}[t]
\centering \includegraphics[scale=0.45]{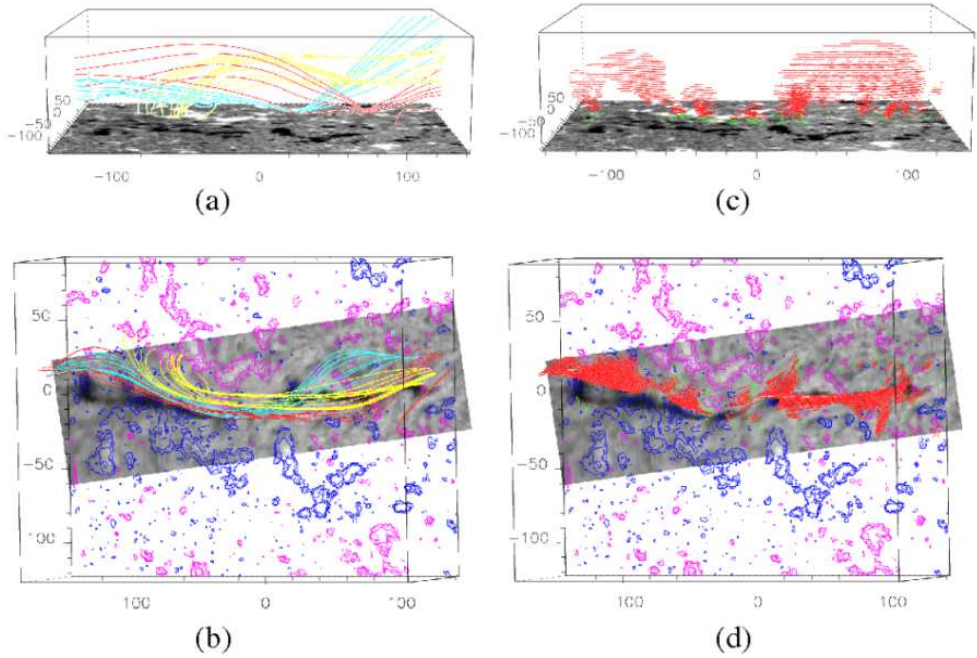}
\caption{
Linear magneto-hydrostatic model of a filament observed with
TH\'{E}MIS \citep[][]{Dudik2008}. 
The {\it left} panels show selected magnetic field lines as seen
(a) from the side of the computational domain, (b) by an observer on
Earth. The {\it yellow} field lines are rooted within the domain,
while the {\it red} and {\it blue} lines pass through the left and
right boundaries. The {\it right} panels show the locations of dips in
the field lines as seen from the same perspectives. Each field line is
plotted up to a height of 300 km above the dip to simulate the effect
of the pressure scale height of the prominence plasma. All panels show
the vertical magnetic field on the photosphere, either as a greyscale
image (upper panels) or as {\it blue} and {\it purple} contours (lower
panels). Panels (b) and (d) also show the co-aligned H$\alpha$ image
from TH\'{E}MIS.}
\label{fig:dudik_fig}
\end{figure}

\citet{Dudik2008} used TH\'{E}MIS/MTR observations of the photospheric
vector field to construct a LMHS model of a filament. Figure
\ref{fig:dudik_fig} shows selected magnetic field lines and a
comparison of the modeled dips with H$\alpha$ observations.
The authors find significant departures from translational invariance
along the PIL. They show that the flux rope is split into two parts:
one part is rooted in a network element (P2) on the positive polarity
side of the filament, while the other part is present along the full
length of the computational domain. This shows that flux ropes can
have a complex magnetic structure, and may consist of multiple strands
that are twisted together. The axial magnetic flux may vary along the
length of the filament.

\subsubsection{Non-Linear Force-Free Field Models}
\label{sec:2.2.5}

Several authors have developed non-linear force-free field (NLFFF)
models that include the local environment of the flux rope. This is
important because the overlying coronal arcade plays a key role in
the equilibrium and stability of the flux rope. For example,
\citet{Regnier2007} construct NLFFF models by ``extrapolating''
observed photospheric vector fields into the corona \citep[also
see][]{Regnier2002, Regnier2004, Regnier2006, Canou2009}. They
consider two active regions: one a decaying active region with strong
currents (AR 8151), the other a newly emerged active region with weak
currents (AR 8210). For the old decaying active region the
connectivity and geometry of the NLFFF model include strong twist and
strong shear, and are very different from the potential field model.
The twisted flux bundles store magnetic energy and helicity high in
the corona (about 50 Mm). The newly emerged region has a complex
topology and the departure from a potential field is small, but the
excess magnetic energy is stored in the low corona and is enough to
trigger powerful flares.

An alternative method for constructing NLFFF models of filaments was
developed by \citet{vanb04}. The technique involves inserting a flux
rope into a potential field based on an observed photospheric
magnetogram, and then evolving the field in time to an equilibrium
state using magneto-frictional relaxation. The advantage of this
method is that the axial flux $\Phi_{axi}$ and poloidal flux $F_{pol}$
of the flux rope can be specified, and these parameters can be varied
until a good fit to the observations is obtained. If the axial and
poloidal fluxes are not too large, the relaxation results in a NLFFF
describing both the flux rope and its surroundings. The method was
applied to a filament observed at the Swedish Vacuum Solar Telescope
(SVST) in June 1998.  This U-shaped filament had a prominent barb that
exhibited interesting fine structure and internal motions. It was
found that the dipped field lines in the NLFFF model reproduce some of
the observed features of the filament barb, but not all. The barb is
attributed to the presence of a magnetic element directly below the
filament, so it is difficult to decide whether this element is
``dominant'' or ``parasitic.''

\citet{bob08} developed NLFFF models of two active regions using the
above flux-rope insertion method. The models are constrained by
H$\alpha$ observations of the filaments and by TRACE observations of
coronal loops in the overlying corona. The best fit to the
observations is obtained for a model in which the flux rope is only
weakly twisted; the peak value of $\alpha$ occurs at the boundary
between the flux rope and its surroundings, not at the flux rope
axis. Bobra {\et} find that the axial fluxes of the flux ropes are
close to the upper limit for stability of the force-free equilibrium;
however, no major eruptions occurred in these active regions either
several days before or after these observations. This suggests that
active regions can release magnetic free energy gradually, and that
the build-up of free energy does not necessarily lead to large flares
or coronal mass ejections. In contrast, \citet{Su2009} modeled a
different active region and found that the flux rope present in that
region has an axial flux that is well below the threshold for
eruption. Clearly, observations of filaments can provide important
constraints on the structure and stability of magnetic fields in
active regions. For filaments between active regions and on the polar
crown it is not known how close they are to the limit of stability.

In more recent modeling of NLFFFs, \citet{Mackay2009} consider how the
structure of dips in a filament may vary as a single bipole polarity
is advected towards the main body of the filament. The authors
demonstrate how the bipole may result in the break up of the filament
and a significant change in vertical extent of the dips. In contrast
to linear force-free models they show that through considering a
time-series of evolved non-linear force-free fields that barbs may form when
either the minority or dominant polarity is advected towards the PIL.

In the previous discussion the observed structure and morphology of solar 
filaments has been discussed along with recent developments in constructing
static 3D models of their magnetic structure. In the next section the review
considers the origin of the mass within filaments, along with the properties 
of observed flows.


\section{Plasma Structure and Dynamics}
\label{sec:3}

Although the Sun's magnetic field clearly is the primary source
of support and constraint for the prominence mass, the underlying physics governing the
origin and subsequent behavior of this material remains a lively topic
of debate. The  observed characteristics of prominence plasmas are
discussed in depth in Paper I (sections 3 and 4) with a brief summary given below. 
The key observational constraints that
must be satisfied by any successful plasma formation
model are as follows:
\begin{enumerate}
\item Active region prominences tend to be shorter (of order 10 Mm),
short-lived (minutes to hours), and lower ($<$10 Mm), while those formed
outside active regions can be hundreds of Mm long, persist through
multiple solar rotations, and extend as high as 100 Mm.
\item Prominence plasma is predominantly dynamic, exhibiting
horizontal and/or vertical motions of order 10-70 km s$^{-1}$
\citep{kub86, schm91, Zirker1998a, kuc03, Lin2003, Okamoto2007,
berg08, chae08}.
Quiescent filaments can go through phases of enhanced internal motion
and activity \citep[][]{mar73, Tan-Han1995, kuc06}, altering the
amount of cool material in the filament \citep[][]{Kilper2009}.
\item Prominence plasma frequently appears {\it in situ} high in the
hot corona \citep{mcm38,sch01}, but also is observed flowing up from
the chromosphere \citep{chae00, liu05, Schwartz2006}.
\item Prominences appear to be composed of multiple knots and threads
with widths down to the resolution limit ($\sim$100 km) and typical
lengths of 3 - 20 Mm \citep{Lin2003, Lin2005a, hein06, gun07, gun08,
berg08}.
\item Barbs --- lateral extensions from the prominence spine toward
patches of parasitic polarity with vertical extents much greater than
the local gravitational scale height --- apparently are connected to
the chromosphere \citep{kiep53, Malherbe1989, Martin1998, chae05}.
\end{enumerate}
As is discussed below, no single model can explain the entire complex
range of characteristics at present.

It was recognized over 30 years ago that the large prominence mass
must come from the chromosphere, because there is not enough plasma
residing in the ambient corona \citep{pik71,sai73,zir94}.  Sufficient
mass to explain prominences must be extracted from the chromosphere,
either through magnetic forces, which inject or lift cool plasma
directly into the corona, or through thermal pressure forces, which
evaporate heated plasma that subsequently condenses into prominence
knots or threads. Here we summarize the current status of these models
for the prominence plasma, and discuss one well-studied example of an
evaporation-based mechanism --- the thermal non-equilibrium model ---
in greater detail.

\subsection{Injection Models}
\label{sec:3.1}

\begin{figure}
\includegraphics[width=4.5in]{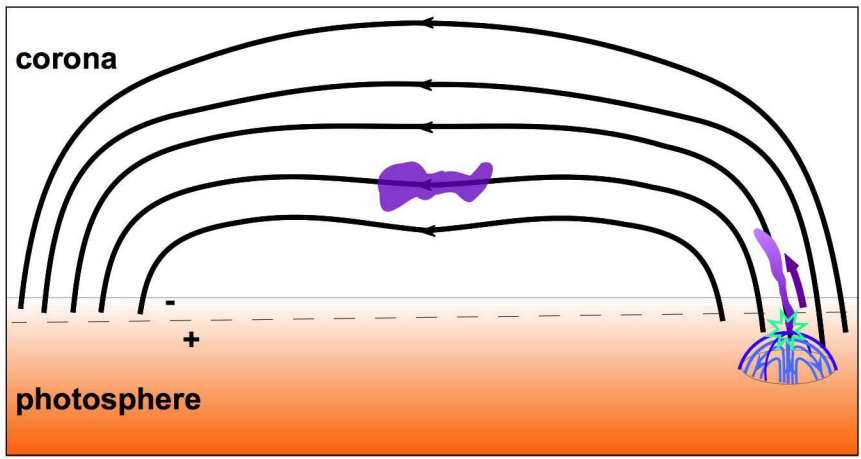}
\caption{Illustration of {\it injection} of prominence plasma (purple)
by reconnection between minority-polarity bipole (blue lines)  and
preexisting filament-channel field (black lines). The polarity
inversion line is dashed.}
\label{fig:karpen_fig1} 
\end{figure}

According to the injection models, cool plasma is forced upward in
filament-channel flux tubes with sufficient force to reach the
observed heights of prominence plasma.  Most injection models (see
Fig.~\ref{fig:karpen_fig1}) generally invoke reconnection low in the
solar atmosphere as the pivotal cause of mass at prominence
temperatures being expelled into the corona \citep{wang99, chae01a}.
This conjecture is largely motivated by the well-observed, but poorly
understood, connection between flux cancellation and filament channel
formation \citep{Balle1989, Martin1998, Wang2007}. Because injection
necessarily produces
unidirectional flows in each flux tube, counter-streaming could arise
either through reconnection occurring at the bases of different flux
tubes on opposite sides of the PIL or through sequential reconnection
at each footpoint.  Some injection models propose that the
reconnection sites are at the PIL \citep[\eg][]{chae03}, while others
suggest that the jets originate at minority-polarity intrusions offset
from the PIL \citep[\eg][]{wang99}. Many
active-region prominences provide strong evidence for the type of
jetting expected from reconnection events, and may well be explained
by injection \citep{chae03}. However, active region prominences are
short-lived, low-lying and highly variable, and are
sufficiently distinct from quiet-Sun prominences that different
physical mechanisms may be involved. Prominence barbs also might be
explained by the injection of cool plasma by reconnection between the
filament channel field and small bipoles emerging in the channel
\citep{lit00,wang01,chae05}. Both large- and small-scale up-flows have
been recorded in quiet-Sun hedge-row prominences \citep{zir94,berg08},
although it is difficult to determine whether these flows originate at
the photosphere due to spicules and other obscuring chromospheric
activity. Recent high-resolution observations by SOT reveal
strong jetting throughout the chromospheric network \citep{depont07};
it is possible that this activity occurs within filament channels on
field lines capable of hosting prominence material.

On the other hand, it is unclear whether injection can account for all
aspects of the observed dynamic evolution of quiet-Sun prominences,
such as the frequency with which cool plasma appears suddenly in the
corona and the predominantly horizontal, fine-scale counter-streaming
flows. Furthermore, the extent to which reconnection can drive cool
filament-channel material as high as 100 Mm in the corona, without
also heating this plasma, has not been demonstrated in a realistic
three-dimensional (3D) geometry with applicable energy sources and
sinks. In simulations of Yohkoh X-ray jets by \citet{yok95},
reconnection produces hot jets directly, while the associated (but not
co-spatial) cool jets are produced by compression of nearby open
field. A similar mechanism might explain the association between flux
cancellation and injection of cool plasma. Another unresolved issue is
whether cancellation reconnection preferentially occurs in or below
the chromosphere; if the interacting flux systems reconnect instead in
the low corona (see below), then cool, dense plasma will not be lifted
or injected directly.  As is well-known, the fact that cancellation is
observed in the photosphere or chromosphere does not necessarily
indicate that the associated reconnection site is at the same level
\citep{zwa87,har99}. The temperature minimum is thought by some to be
the most favorable location for reconnection, because the Spitzer
resistivity is highest there \citep{roum93,stur99,lit07} or because of
enhanced plasma turbulence \citep{chae02}. In summary, the essential
distinguishing features predicted by the injection model are that
photospheric or chromospheric mass is injected with substantial speed
at locations of flux cancellation/reconnection, either at or away from
the PIL, and that this mass rises into the corona at or near its
original cool temperature.

\subsection{Levitation Models}
\label{sec:3.2}

\begin{figure}
\includegraphics[width=4.5in]{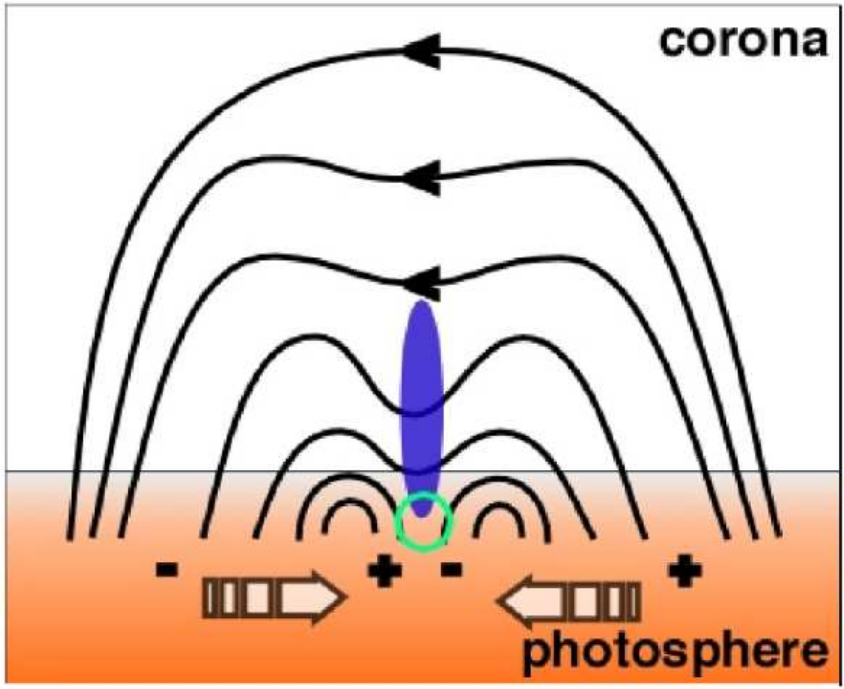}
\caption{Illustration of {\it levitation} of prominence plasma
(purple) by U-loop emergence or post-reconnection relaxation
associated with flux cancellation.}
\label{fig:karpen_fig2}
\end{figure}

Levitation models propose that cool plasma is lifted by rising
magnetic fields at the PIL and transported transverse to the magnetic
field (see Fig.~\ref{fig:karpen_fig2}).
In one class of levitation models, the
filament-channel magnetic structure is a highly twisted flux rope that
brings up cool plasma as the axis and lower portions of the rope
emerge above the photosphere; thus, the prominence plasma should
reside in the upward concavities of the helical field
\citep{Rust1994}. However, such concave-upward formations have been
observed only rarely in non-erupting prominences \citep[see, \eg][]
{lites05, lop06, Okamoto2007}.  An alternative lifting mechanism
is the relaxation of magnetic fields during emergence of U loops
\citep{deng00} or after reconnection associated with flux cancellation
\citep{Balle1989, Priest1996, Oliver1999, lit99, Galsgaard1999,
Litvinenko2005, Welsch2005}. Simulations of the
reconnection-levitation mechanism have verified that rising field
lines are indeed produced by reconnection between bipolar systems, but
significant work remains to prove that this mechanism is responsible
for the observed motions and properties of prominence plasma. Detailed
calculations of reconnecting bipoles \citep{Galsgaard1999,vonr08} predict
significantly different plasma properties, depending on the
dimensionality, the initial conditions, and the terms included in the
energy equation. For example, \citet{Galsgaard1999} found a 20\% density
contrast between the elevated material and the background corona, much
less than the 2 orders of magnitude deduced from prominence
observations, whereas \citet{vonr08} predict that the uplifted plasma
can exhibit X-ray-emitting coronal characteristics, filament-like
photospheric properties, or a mixture of both (note, however, that the
latter model is intended primarily to explain coronal bright points).
The difference between the two results is that \citet{Galsgaard1999}
due to computational reasons only considered a simple hydrostatic
atmosphere with a uniform temperature profile and did not include a
transition region.
One common result is that most of the uplifted cool plasma drains
along the rising field lines onto the chromosphere. Quiet-Sun
prominences that resemble a central pillar from which cool material
streams outward and downward are the best candidates for supporting
this scenario.  The extent to which this process can lift
photospheric or chromospheric material as high as 100 Mm into the
corona also has not been demonstrated.  To date, none of these
levitation models
has allowed flux to retract beneath the photosphere during
reconnection, thus preferentially favoring upward motions. Flux emergence 
simulations have shown that emerging U-loops and the associated photospheric
plasma do not readily break through the photosphere nor rise to coronal heights 
characteristic of quiet-Sun and intermediate prominences, even without the inhibiting
presence of a preexisting coronal field
\citep{fan01,arch04,man04,mag06,gals07,mag08,arch08b}. Although the
U-loops of the emerging tube do not rise to coronal heights, the process 
of flux emergence may still produce a coronal flux rope with dips. A flux rope
may form through the reconfiguration of emerged sheared field lines that 
lie above the emerging tubes axis. The flux rope may be formed either by 
magnetic reconnection \citep{man04,mag06,arch08b} or through helicity injection 
by torsional Alfven waves \citep{fan09}. Once the tube is formed above the 
photosphere, it may rise to coronal heights, in principle dragging cool photospheric 
or chromospheric plasma with it. Such a process acting during 
flux emergence, is unlikely to explain quiescent or intermediate prominences 
but would be of importance to active region prominences. This will be further discussed
in Section~\ref{sec:5.6}. 
One effect not
considered thus far in 3D flux-emergence calculations, the partial
ionization of portions of the chromosphere and photosphere, might
enable more rapid emergence as well as the emergence of more flux (and
frozen-in plasma) into the corona \citep{lea06}. As discussed above
for the injection models, it is unclear where reconnection between
such flux systems would occur in the solar atmosphere; for example,
the reconnection in the \citet{Galsgaard1999} study begins in the corona, at
the apex of a null-null separator, so the bulk of the mass lifted at
that location is coronal. An alternative mechanism proposed to lift
and support prominence plasma in the corona is upward-propagating,
weakly damped MHD waves \citep{pec00}; recent observational evidence
for significant Alfvenic perturbations in chromospheric structures
\citep{depont07} makes this an intriguing suggestion, but further
quantitative work is needed to evaluate this mechanism rigorously. In
general, the key features of levitation models overlap significantly
with those of injection models (see above); however, the levitated
mass does not travel as far or as fast as injected mass would, and
typically is predicted to be located above the PIL.

\subsection{Evaporation-Condensation Models}
\label{sec:3.3}

\begin{figure}
\includegraphics[width=4.5in]{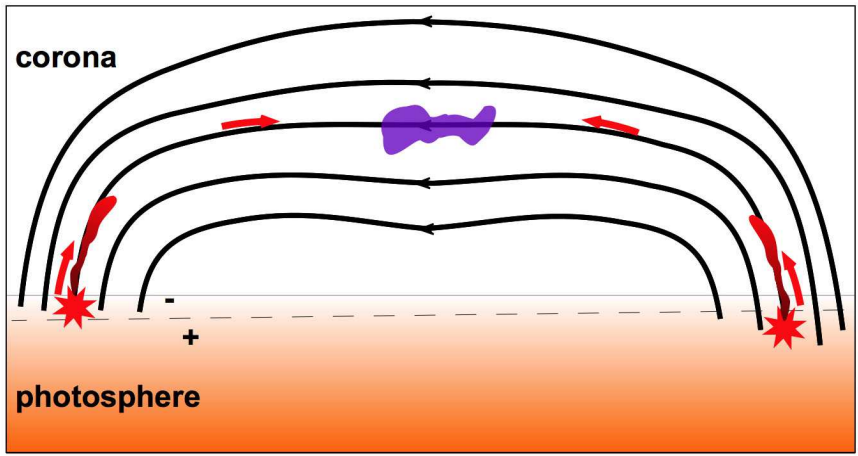}
\caption{Illustration of an {\it evaporation-condensation} model: hot
up-flows (red) driven by heating localized above the footpoints
evaporate chromospheric plasma that ultimately condenses in the corona
as cool prominence material (purple).}
\label{fig:karpen_fig3}
\end{figure}

\begin{figure}
\includegraphics[width=4.5in]{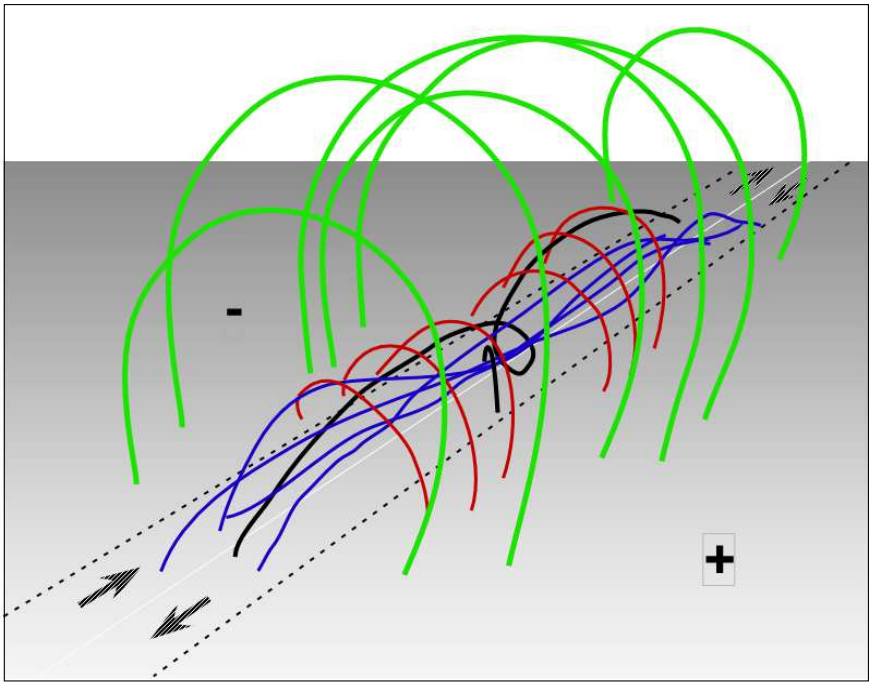}
\caption{
Prominence locations predicted by the thermal non-equilibrium model,
assuming that the filament-channel magnetic structure is a sheared
arcade. The white line on the grey photosphere is the PIL, and the
boundaries of the surrounding filament channel are denoted by dashed
lines. Only the {\it blue} field lines match the requirements for
thermal non-equilibrium to produce dynamic and stationary condensations
consistent with observed prominence structure.}
\label{fig:karpen_fig4}
\end{figure}

Evaporation-condensation models are based on the fact that adding heat
to a coronal loop increases the density of the corona while decreasing
slightly the chromospheric mass (Fig.~\ref{fig:karpen_fig3}).
Different temporal combinations of
heating and cooling were studied, without successfully reproducing the
basic properties of a prominence \citep{eng77,an85,pol86}.  Subsequent
research found that concentrating coronal heating near the footpoints
of a loop should produce a cool condensation at or near the apex
\citep{serio81, mok90, ska91, rbd98}. Although recent analysis of
TRACE observations of some coronal loops independently indicated that
the heating is concentrated near the loop footpoints
\citep{asa01,wine02}, the observational evidence for the spatial
distribution of coronal heating is inconclusive at best. Confirmation
of the basic principle behind this evaporation-condensation model for
prominence plasma formation required the use of adaptive-mesh
numerical simulations, to handle the rapid birth and subsequent
evolution of a new, thin transition region at each interface between
the loop and the cool condensation \citep{ska99}. The central concept
of this model is that, if the heating scale $\lambda$ is small
compared to the length of a coronal loop and localized near the
chromospheric footpoints, then the plasma in the midsection of the
tube, where the heating is negligible, must undergo a
radiatively-driven thermal collapse to low temperatures.  A runaway
situation develops in the coronal plasma because the radiative losses
increase with decreasing temperature $T$ for $T \ge 10^5$ K as well as
with the square of the density: once the local plasma has cooled to
this critical transition-region temperature, it must cool all the way
to chromospheric temperatures to regain equilibrium
\citep{Hood1988}. The ratio of the
heating scale to the loop length is a crucial factor because the total
radiative losses from the loop increase linearly with length, but the
thermal conduction and other energy transport or loss terms either
decrease or remain constant with loop length. Clearly, the radiative
losses will dominate for lengths above a threshold value that is
approximately an order of magnitude greater than the heating scale
\citep{mok90}. As discussed in section \ref{sec:2.2}, the magnetic
structure containing the quiet-Sun prominence material is most likely
to be a sheared arcade \citep{me94, DeVore2000} or weakly twisted flux
rope \citep{Martens2001, bob08}, in which many of the loops nearly
aligned with the PIL are much longer than typical coronal loops.
Therefore, for a given heating scale, condensations are more likely
to form in these elongated loops than in shorter loops rooted outside
the filament channel (see Fig.~\ref{fig:karpen_fig4}).

A series of computational investigations of this
evaporation-condensation process, denoted {\it thermal
non-equilibrium}, has systematically explored the dynamics and
energetics of the plasma within individual elongated flux tubes heated
near the footpoints \citep{ska99, ska00, Karpen2001, Karpen2003,
Karpen2005, Karpen2006, Karpen2008a}. While the nature of the
footpoint heating is not well understood, this work has established
constraints on the conditions
favorable to condensation formation, the magnetic structure of
prominences, and the nature of the associated coronal heating. The key
factors in determining the likelihood and behavior of condensations
are the flux tube geometry and the localized heating properties. In
long, low-lying flux tubes with shallow arches or dips, unequal
quasi-steady heating yields a repetitive cycle of condensation
formation, motion along the tube, and destruction by falling onto the
nearest chromosphere \citep{ska00, Karpen2001, Karpen2006}.
It is important to
note that this process does not require the presence of dipped flux
tubes, in contrast to the assumptions of many magnetic-structure
models \citep[\eg][]{anz03}, although shallow dips facilitate the
collection and retention of cool plasma. On the other hand, if the
host flux tube has a deep dip, as would occur in the outer portions of
a highly twisted flux rope, thermal non-equilibrium produces
condensations that rapidly fall to the lowest part of the dip and
remain there, stationary but growing as long as the heating remains
quasi-steady \citep{Karpen2003, Karpen2005}. Condensations also form
when the energy input is impulsive in nature and randomly distributed
in time, as in nanoflare models of coronal heating
\citep{Klimchuk2006}, so long as
the average interval between energy bursts is shorter than the
radiative loss time in the ambient corona \citep{Karpen2008a}. The
calculated condensation speeds, counter-streaming, lifetimes, and sizes
are consistent with observations of many quiet-Sun
prominences. However, this model does not provide a satisfactory
explanation of active-region prominences, which are too short to
support the thermal non-equilibrium process with typical values of the
heating scale, or of the vertical structure and dynamics of hedge-row
prominences. Barbs could be consistent with thermal non-equilibrium if
they are composed of vertically aligned dips in otherwise horizontal
flux tubes \citep{hein01}, as long as the dips are deep enough to trap
the condensed matter: for example, in photospheric ``bald patches''
near parasitic polarity sites \citep{aul98,aul02b,vanb04,lop06}.

This process could occur in any favorable magnetic structure, but thus
far has been considered systematically only within the context of the
sheared arcade model (section \ref{sec:2.2}). The color coding in
Fig.~\ref{fig:karpen_fig4} illustrates the implications of the thermal
non-equilibrium studies summarized above for the location and
evolution of prominence material within a filament channel formed by
the sheared arcade mechanism.  Only the blue field lines match the
requirements for thermal non-equilibrium to produce dynamic and
stationary condensations consistent with observed prominence
structure. Red field lines are too short; the helical black field line
can host only a condensation sitting at the bottom of the dip; and
green field lines are capable of hosting small, intermittent,
short-lived condensations known as coronal rain \citep{mul03,mul05}.
Three-dimensional MHD simulations of both the magnetic and the plasma
structure with localized heating, radiation, and thermal conduction
are needed to determine the full scope and applicability of thermal
non-equilibrium to solar prominence plasmas. 
For a discussion of energy balance considerations in solar prominences
see section 10 of Paper I.
Evaporation-condensation
models, as represented by thermal non-equilibrium, uniquely predict
both stationary and highly dynamic prominence threads that condense
{\it in situ} in the corona and trace the supporting flux tubes; an
increase in coronal density precedes each condensation episode, while
the collapse of the condensations reduces the ambient coronal density
and generates waves and shocks.

As discussed above, flows are commonly observed in solar filaments
and prominences. In addition to this a wide variety
of oscillations are observed. The properties of these oscillations, along with
theoretical models used to describe them is now discussed.


\section{Magneto-hydrodynamic Waves in Solar Prominences}
\label{sec:4}

\subsection{Overview}
\label{sec:4.1}

Theoretical studies of small-amplitude oscillations and MHD waves in
the solar corona started long before clear observational evidence
about the presence of these phenomena in the solar corona was
available.  The interest of their study lies in their potential
relationship with the coronal heating problem and with the possibility
to perform local seismology of the solar corona.  Thanks to ground-
and space-based observations, evidence about the presence of
oscillations in coronal structures (loops, prominences, plumes, etc)
is now widely available.

Solar prominences are subject to various types of oscillatory motion
(section 4, Paper I).
In the early days of filament observations, all the data collected on
oscillations were related with motions induced by disturbances coming
from a nearby flare \citep[\eg][]{Ramsey1966}.  These disturbances
produce large-amplitude oscillations with velocity amplitudes of 20
{\kms} or higher.  Observations of large-amplitude oscillations in
filaments are rare, although in recent years with the help of new
observational capabilities more detections have been reported
\citep[][]{Eto2002, Okamoto2004, Jing2003, Jing2006, Gilbert2008}.
Also, theoretical modeling of such events is lacking.  On the other
hand, from observations performed with ground-based telescopes it is
well known that most quiescent prominences and filaments display
small-amplitude oscillations.  The motions are mainly detected through
the periodic Doppler shifts of spectral lines
\citep[\eg][]{Tsubaki1988}, and the observations have shown that (1)
the oscillations are of local nature; (2) simultaneously flowing and
oscillating features are present; (3) the oscillations seem to be
strongly damped in time, and (4) there are spatially coherent
oscillations over large regions of prominences/filaments.  A key
aspect of prominence oscillations is the knowledge of its triggering
mechanisms.  While for large-amplitude oscillations they are well
known, in the case of small-amplitude oscillations, the excitation
process still remains a mystery, in spite of the available
observational information.

In this section, we will focus on small-amplitude oscillations in
prominences, which can be interpreted as linear magneto-hydrodynamic
(MHD) waves, and which constitute an important tool for understanding
the physical properties (sections 2 and 3 of Paper I) and the internal 
structure of
prominences. Traditionally, the study of prominence oscillations has
been based in the determination of the normal modes of oscillation of
different equilibrium configurations, such as slabs or cylindrical
flux tubes, without including any dissipative mechanism.  Recently,
the study of the damping of prominence oscillations has been the
subject of intense theoretical modeling by considering different
dissipative mechanisms.  Prominence seismology seeks to obtain
information about prominence physical conditions from a comparison
between observations and theoretical models of oscillations.  Attempts
to determine the Alfv\'{e}n speed, the magnetic field, the shear
angle, etc.~have already been carried out.  In the following we first
provide a summary of the observations of prominence oscillations, and
then discuss various aspects of the theoretical modeling.

\subsection{Observational Background}
\label{sec:4.2}

\subsubsection{Ground-Based Observations}
\label{sec:4.2.1}

Prominences above the limb exhibit oscillations with velocity
amplitudes in the range 2 to 10 \kms. The oscillations appear to be
local in nature, disturbing only some regions of prominences.  Most
information about this type of oscillation has been obtained from
Doppler velocity data, although there are also detections coming from
other spectral indicators such as line intensity and line width 
(section 2.1 and 3.4, Paper I).
Only rarely are the oscillations detected in several of these spectral
indicators at the same time and with the same period, which
constitutes one of the puzzling features of prominence oscillations.
Earlier, the first observational detections of oscillations using
one-dimensional spectroscopic observations led to a classification in
terms of short- and long-period oscillations. However, when the number
of available observations increased, it was clear that a wide range of
periods, between 1 and 90 minutes, were present. Detailed information 
on the oscillatory periods detected in limb prominences is available in 
\citet {Tsubaki1988}, \citet{Oliver1999r} and \citet{Oliver2002}. 
Although the classification in terms of short- and long-period oscillations
is still in use, it does not cast any light nor gives any help with
regard to the nature, origin or exciter of oscillations.

On the other hand, far more interesting results can be obtained from
two-dimensional, high-resolution observations of prominences
\citep[][]{Molowny1999, Terradas2002} which allow the construction of
maps of the Doppler velocity, wave period, damping time and
wave vector, and the extraction of interesting information about
oscillations in prominences.  For instance, \citet{Terradas2002}
reported the existence of large regions with periodic Doppler (line
of sight) velocity oscillations having similar periods, around 75
minutes, noticing also that the oscillatory amplitude tends to
decrease in time in such a way that the periodicity totally disappears
after a few periods.  Reliable values for the damping time,
$\tau_{D}$, have been derived from different Doppler velocity time
series by \citet{Molowny1999} and \citet{Terradas2002}.  The values of
$\tau_{D}$ thus obtained are usually between 1 and 4 times the
corresponding period, and large regions of the prominence display
similar damping times. Also, \citet{Terradas2002} reported the
presence, along two selected paths in the prominence region, of plane
propagating waves as well as a standing wave.  The plane waves
propagate in opposite directions with wavelengths of 67,500 and 50,000
km and phase speeds of 15 {\kms} and 12 \kms, respectively, while in
the case of the standing wave the estimated wavelength is 44,000 km
and a phase speed is 12 \kms.  Furthermore, the analysis has
identified the existence of a wave generating region, indicating that
oscillations are locally excited.

For the case of filaments on the disk, old two-dimensional observations
of filament oscillations \citep[][]{Thompson1991, Yi1991, Yi_etal1991}
reveal that the Doppler signals of filaments form fibril-like structures.  
\citet {Yi_etal1991} concluded that the fibrils form an angle of about $25^{\circ}$ 
with the filament long axis, that individual fibrils or groups of fibrils may
oscillate independently with their own periods,  and that  these fibril 
structures represent mass motions in magnetic flux tubes. Thanks to the
improvement in spatial resolution provided by modern solar telescopes,
high-resolution observations have given us key information about the
internal structure and features of filaments, which is useful for the
study of filament oscillations.  For instance, observations of
filaments made by \citet{Lin2005a, Lin2005b} suggest that the
thickness of thin threads forming the filaments is $\leq 250$ km,
that only a small portion of filament threads are filled with cool
plasma, and that this absorbing plasma is continuously flowing along
the thread structure with velocities $15 \pm 10$ km s$^{-1}$.
\citet{Lin2004} analyzed the Doppler signals obtained from two
different regions of a polar crown filament, showing the presence of
oscillation that are coherent over each observed region (size about 25 Mm)
and are strongly damped after a few periods.
Furthermore, in the middle part of the filament 49 H$\alpha$
moving features display periodic variations in Doppler velocity.
Finally, \citet{Lin2007} have found evidence for traveling waves
in the threaded structure of a filament and, in some cases it seems
that the propagating waves move in the same direction as mass
flows. They also have determined the wavelength of a propagating wave,
obtaining a value of around 3,000 km. In Table~\ref{spatial_distribution}
the periods found from ground based observations are summarized.
\begin{table}[t]
\center
\caption{\em Periods detected in prominence fine structure oscillations.}
\begin{tabular}{ll} 
\hline
\hline
Reference & Period (min)  \\ \hline
Thompson and Schmieder (1991)    &4.4   \\
Yi, Engvold, and Keil (1991)    &5.3, 8.6, 15.8   \\
 Lin (2004)    &26 \\
 Lin et al. (2007)    &3 - 9 \\
\hline 
\end{tabular}
\label{spatial_distribution}
\end{table}

\subsubsection{Space-Based Observations}
\label{sec:4.2.2}

Prominence oscillations have also been observed with space-based
instruments. \citet{Blanco1999} used SUMER to study the behavior
of different lines of Si~IV and O~IV in a limb prominence.  The
results show that a large amount of energy is contained in
oscillations with periods between one and six minutes, corresponding
to characteristic periods of the chromosphere and photosphere.
Furthermore, for all the considered wavelengths, the energy versus
the slit position varies in such a way that minima and
maxima for each wavelength are coincident. Later on, oscillations in
an active-region filament were reported by \citet{Regnier2001}.  Using
SUMER they observed the He~I line, obtaining a time series of the
filament with a duration of 7 h 30 min and with a temporal resolution
of 30 s.  A Doppler velocity time series was derived, and by Fourier
analysis significant power was found at periods between 5 and 65 min
suggesting the presence of oscillations. \citet{Foullon2004} used
SoHO/EIT 195 {\AA} with a time cadence of 12 min to observe an EUV
filament during its crossing of the solar disk.  They reported
intensity variations of long period, 8 - 27 h, with a dominant period
of about 12.1 h, while the amplitude of the intensity variations
reached 10\% of the background intensity.  Furthermore, the
periodic intensity variations seem to be correlated along the
filament, and the most pronounced oscillations were detected during
$6$ days. \citet{Pouget2006} used SoHO/CDS to observe two
filaments, during different days, in the optically thick He~I line 
(section 8.2, Paper I). The duration of
the observations was 15 - 16 h with a time resolution of 20 s. Before
performing the Fourier analysis of Doppler velocities, the mean
velocity averaged over the width of the filament was computed and the
obtained power spectrum points out the existence of a wide range of
periods arriving up to 5 - 6 h.  The results derived from these
observations must be taken with care. First, it should be noted
that the He~I line may be formed either by a 20,000K
plasma or by scattering in a cooler 8,000K plasma. Therefore the emission
may originate in the outer parts of the prominence,
whereas the authors assume that the detected motions are
propagated from the prominence core. Moreover, the averaging of
Doppler signals coming from each pixel means that spatial information
is lost and, furthermore, it is difficult to be sure that the same
region of the filament is tracked when long-time observations are
performed, and that there is no time evolution of the filament
(brightenings, fadings, small eruptions, etc).  Recently,
\citet{Okamoto2007} using SOT and the Ca~II~H line have
obtained about 1 h of continuous images of an active region.  The
Ca~II~H line movie shows continuous horizontal motions along the
prominence.  Some of the flows display constant velocities while
others accelerate.  Hinode movies also show that the threads suffer
synchronous vertical oscillatory motions.  Finally, recent
high-resolution ($0.2''$) observations of limb prominences made by
SOT \citep[][]{berg08} reveal very complex dynamics with
vertical filamentary down-flows and vortices, as well as episodic,
vertical up-flows.  Furthermore, using horizontal time slices taken at
different heights within the prominence, they suggest the presence of
large-scale oscillations with periods between 20 and 40 minutes,
lasting one or two periods, and with a vertical phase speed of 10
\kms. Finally, \citet{Ning2009} have analysed the oscillatory behaviour  of a 
quiescent prominence observed with Hinode. They find that prominence threads
exhibit vertical and horizontal oscillatory motions. In some parts of the 
prominence, the
threads seem to oscillate independently of each other, and the oscillations seem to be 
strongly damped. The periods reported are very short, with the dominant one 
appearing at 5 minutes. The range of periods determined from space based observations
are summarised in Table~\ref{space}.

\begin{table}[t]
\center
\caption{\em Periods detected from space-based observations
of prominence oscillations.}
\begin{tabular}{ll} 
\hline
\hline
Reference & Period (min)  \\ \hline
Blanco et al. (1999)    &1 - 6  \\
R\'egnier et al. (2001)    & 5 - 65   \\
 Foullon et al. (2004)    &720 \\
 Pouget et al.  (2006)    &up to 360 \\
 Okamoto et al. (2007)    &2 - 4.5  \\
 Berger et al. (2008)    &20 - 40 \\
 Ning et al. (2009)    &3.5 - 6 \\
\hline 
\end{tabular}
\label{space}
\end{table}

In summary, from the available observational information some
characteristic features of prominence oscillations can be highlighted:
The oscillations only affect parts of the prominence, and in the case
of filaments are confined to the thread structure. The presence of
simultaneously flowing and oscillating features is a common feature.
The oscillations seem to be strongly damped in time. There are
spatially coherent oscillations over large regions of prominences and
filaments. More detailed information about observations of small
amplitude oscillations in filaments and prominences can be found in
\citet{Engvold2001, Engvold2004, Engvold2008}, \citet{Oliver2002},
and \citet{Banerjee2007}.

\subsection{Theory of Small-Amplitude Oscillations in Prominences}
\label{sec:4.3}

Theoretical interpretations of prominence oscillations are mostly
based on linear MHD waves. An important ingredient in the modeling
of MHD waves in prominences is the chosen equilibrium configuration.
In the past, slab models were used as a representation of limb 
prominences. However, due to present day high-resolution observations, 
models now pay more attention to equilibria based on thin cartesian or 
cylindrical threads. These threads represent  thin magnetic flux tubes, 
partially filled with cold plasma. Such structures seem to be the building 
blocks of filaments when observed on the disc.  Furthermore, attempts have been also made to
reproduce the observed damping of prominence oscillations.  The
damping involves various thermal and non-thermal mechanisms. The
thermal processes include damping due to optically thin radiative
losses (section 8.1 and 10 of Paper I) and thermal conduction. The non-thermal processes 
include ion-neutral collisions in a partially ionized plasma (section 2.3, Paper I) and resonant
absorption of wave energy in the inhomogeneous prominence plasma. 
In the following we first consider undamped, adiabatic MHD waves in
prominence fine structures, and then discuss various damping mechanisms.
Flows seem to be a characteristic feature of prominences (see section
\ref{sec:3}), and the combined influence of flows and non-adiabatic waves on oscillations will be considered
in section \ref{sec:4.3.5}.

\subsubsection{Adiabatic MHD Waves in Prominence Fine Structures}
\label{sec:4.3.1}

\begin{figure}
\centering
{\resizebox{10cm}{!}{\includegraphics{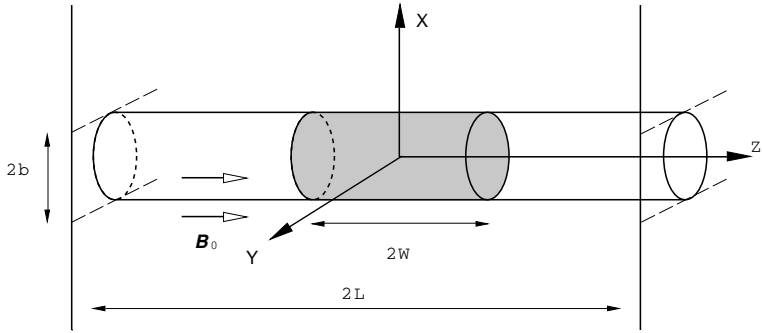}}}
\caption{Sketch of the equilibrium configuration for a cylindrical
model of a prominence thread. The grey zone represents the cold part
of the loop, \ie the prominence. The density in the prominence region
is $\rho_{p}$, in the evacuated (coronal) part of the loop,
$\rho_{e}$, and in the coronal environment, $\rho_{c}$. The magnetic
field is uniform and parallel to the $z$-axis, and the whole
configuration is invariant in the $\varphi$-direction. Adapted from
\citet{Diaz2002}.}
\label{fig:jlb_fig1}
\end{figure}

Cylindrical geometry seems to be the most suitable to model prominence
threads. \citet{Diaz2002} considered a straight cylindrical flux tube
with a cool region representing the prominence thread, which is confined by
two symmetric hot regions (Fig.~\ref{fig:jlb_fig1}). In this case, the
fundamental sausage mode ($m = 0$) and its harmonics are always leaky.
However, for all other modes ($m > 0$), at least the fundamental mode
lies below the cut-off frequency. Regarding the spatial structure of
perturbations, in cylindrical geometry the modes are always confined
in the dense part of the flux tube.  Therefore, an oscillating
cylindrical fibril is less likely to induce oscillations in its
neighboring fibrils, unless they are very close.

\begin{figure}[t]
\centering{\resizebox{10cm}{!}{\includegraphics{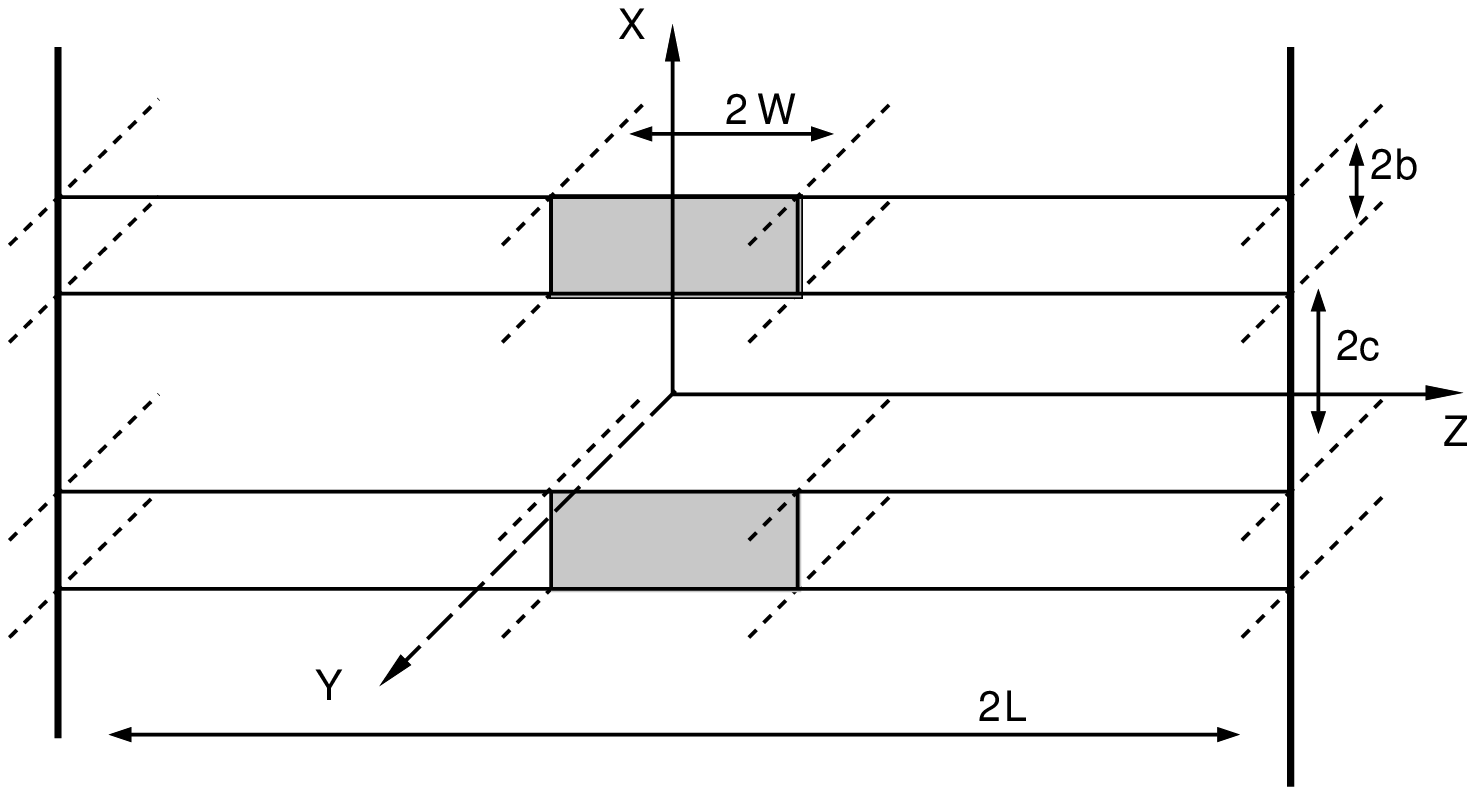}}}
\caption{Sketch of the equilibrium configuration for a collection
of fibrils. The grey zone represents the cold part of the loop, \ie
the prominence. The density in the prominence region is
$\rho_\mathrm{p}$, in the evacuated (coronal) part of the loop,
$\rho_\mathrm{e}$, and in the coronal environment, $\rho_\mathrm{c}$.
The magnetic field is uniform and parallel to the $z$-axis, and the
whole configuration is invariant in the $y$-direction. Adapted from
\citet{Diaz2005}.}
\label{fig:jlb_fig2}
\end{figure}

\citet{Diaz2005} studied multithread systems in Cartesian geometry.
In this case the equilibrium configuration consists of a collection
of threads separated by a distance $2c$ (see Fig.~\ref{fig:jlb_fig2}),
and the individual threads are modeled as slabs \citep[][]{Diaz2001}.
Following this approach, inhomogeneous filaments
were constructed with different thread density ratios representing the
inhomogeneity in density of a real prominence, and the separation
between threads has been chosen randomly within the realistic range.
When the separation between threads is small, there is a strong
interaction between them since the perturbation can easily overcome
the separation. As a result, only one even non-leaky mode can cause
all the threads to oscillate in phase.  The dependence of the
oscillation frequency on the thread separation $c$ can be studied:
in the limit $c \rightarrow \infty$, the structure and the frequencies
of each thread
oscillating alone are recovered, while for small values of $c$ only
the mode described before remains having a slightly smaller frequency
than in the case of the single dominant thread mode.  Therefore, for
realistic values of the separation between threads, the multithread
system would oscillate in phase, with similar amplitudes and the same
frequency.  These results, although derived in Cartesian geometry,
strongly agree with the observations of filament thread oscillations
reported by \citet{Lin2005a}.
However, in the case of cylindrical geometry, the spatial 
structure of perturbations indicates that the modes are always 
confined to the dense part of the flux tube.  Therefore, an oscillating
cylindrical fibril is less likely to induce oscillations in its
neighboring fibrils, unless they are very close.

\subsubsection{Thermal Damping Mechanisms}
\label{sec:4.3.2}

\begin{figure}
\centering
\includegraphics[width=8cm]{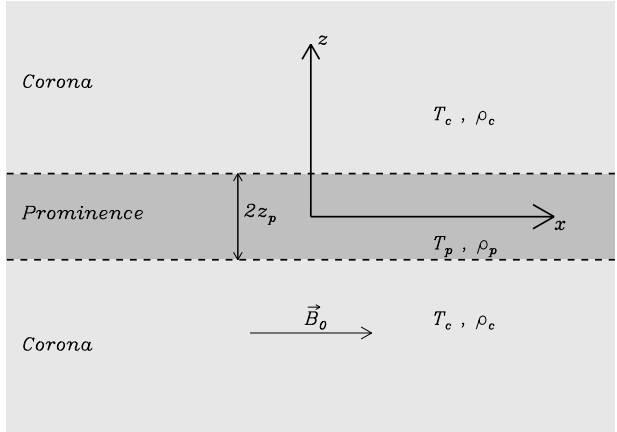}
\caption{
Sketch of the equilibrium configuration for studying the damping of
prominence oscillations in a single thread. Adapted from
\citet{Soler2008a}.}
\label{fig:jlb_fig3}
\end{figure}

\begin{figure}[t]
\centering
\includegraphics[width=12cm]{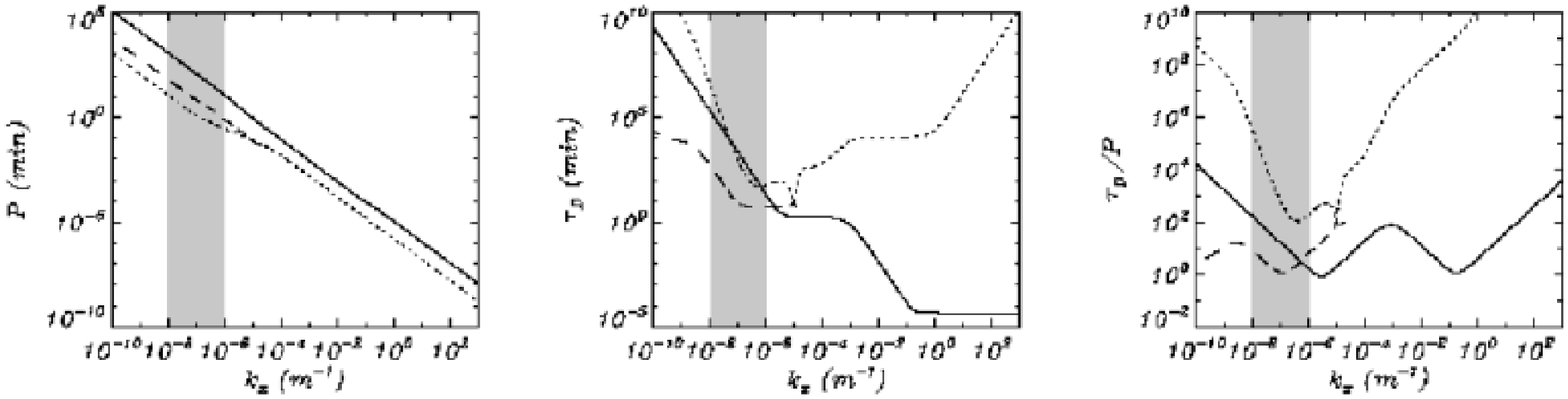}
\caption{
Period (left), damping time (center) and ratio of the damping time to
the period (right) versus the longitudinal wavenumber for the
fundamental kink oscillatory modes: internal slow (solid line), fast
(dotted line) and external slow (dashed line).  The shaded zones
correspond to those wavelengths typically observed in prominence
oscillations. Adapted from \citet{Soler2008a}.}
\label{fig:jlb_fig4}
\end{figure}

\citet{Soler2008a} studied the oscillatory modes of an equilibrium
configuration representing a prominence thread
(Fig.~\ref{fig:jlb_fig3}) which is made of a homogeneous plasma layer
with prominence conditions embedded in an unbounded corona.  Thermal
conduction parallel to the magnetic field, optically thin radiative
losses, and heating have been considered as non-ideal, damping
mechanisms. Figure~\ref{fig:jlb_fig4} shows the period $P$, the
damping time $\taud$, and their ratio, versus the longitudinal
wavenumber, for the fundamental kink modes.  Taking into account the
results in the range of wavelengths typically observed in prominences,
we observe that the internal slow mode (responsible for longitudinal
motions) produces periods compatible with intermediate- and
large-period oscillations, whereas the fast mode (responsible for
transverse motions) could be associated with short-period
oscillations.  On the other hand, the external slow mode mainly
disturbs the surrounding corona, the amplitude of its motions within
the prominence fibril being very small; hence it could be rather
difficult to observe.  Regarding the damping time, both internal and
external slow modes are efficiency attenuated, with damping times of
the order of their periods.  However, again, the fast wave is much
less attenuated since its damping time is between 2 and 6 orders of
magnitude larger than its period.


In order to assess the relative importance of each
non-adiabatic damping mechanism, a comparison was made between the
damping time obtained when considering all non-adiabatic effects
(displayed in the middle column of Fig.~\ref{fig:jlb_fig4}) and the
results obtained when a specific mechanism is removed. The results of
these computations suggest that (1) the internal slow mode is only
affected by prominence-related mechanisms, radiative losses from the
prominence plasma being responsible for the attenuation of this
solution in the range of typically observed wavelengths; (2) the
prominence thermal conduction is only efficient for very small
wavelengths outside the observed range; (3) the fast mode is affected
by both prominence and coronal mechanisms; (4) the damping of the
external slow mode is entirely governed by coronal-related damping
mechanisms, mainly the coronal thermal conduction, which is the dominant
mechanism in the range of typically observed wavelengths.

\subsubsection{Damping of MHD Waves in a Partially Ionized Prominence
Plasma}
\label{sec:4.3.3}

Prominences are partially ionized plasmas since hydrogen lines are
observed (section 8.1 and 2.3 of Paper I). This fact was already considered by \citet{Mercier1977} when
they studied, from a theoretical point of view, the diffusion of
neutral atoms due to gravity. A few years ago, \citet{Gilbert2002}
studied the diffusion of neutral atoms in a
partially ionized prominence plasma, concluding that the loss
time scale is much longer for hydrogen than for helium.  Recently,
\citet{Gilbert2007} have investigated the temporal and spatial
variations of the relative abundance of helium with respect to
hydrogen in a sample of filaments. They have found that a majority of
filaments show a deficit of helium in the top part while in the bottom
part there is an excess. This seems to be due to the large loss
time scale for neutral helium with respect to neutral hydrogen.

\begin{figure}[t]
\centering
{\resizebox{10cm}{!}{\includegraphics{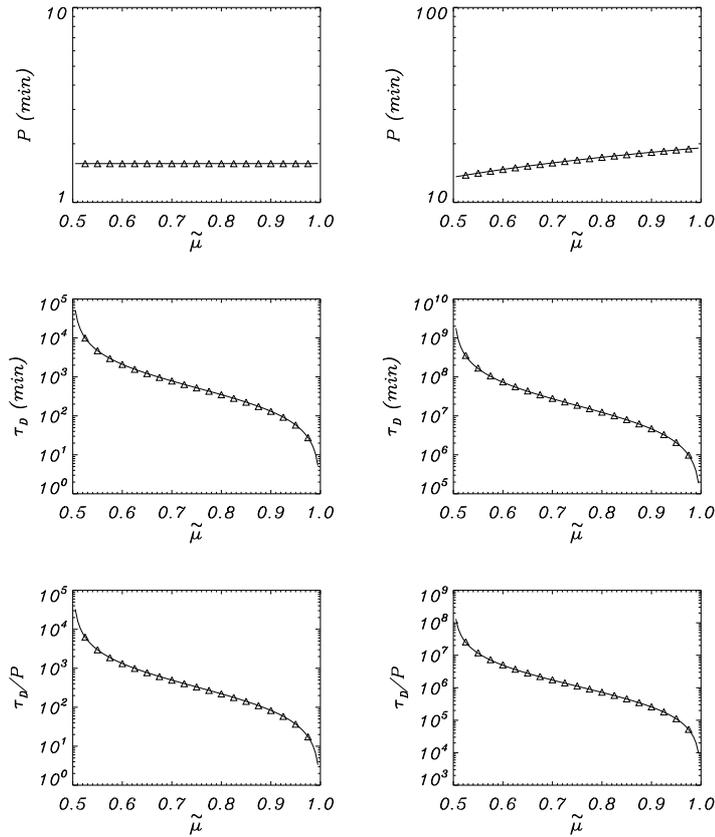}}}
\caption{
Period, damping time and $\dap$ versus the ionization fraction for the
fast wave (left) and the slow wave (right). The parameter values used
are $T_0=8\,000\ \mathrm{K}$, $\rho_0=5\times10^{-14}\
\mathrm{g/cm}^3$ and $B_{0}=10\ \mathrm{G}$, for which the Alfv\'en
speed is 126 km s$^{-1}$, the sound speed ranges from 10.5 to 14.9 km
s$^{-1}$ and the plasma $\beta$ varies between 0.008 and 0.017. In
addition, $k_x x_0=\pi/2$ and $k_zx_0=0.1$. The solid lines represent
the numerical solutions of the dispersion relation while the triangles
represent the results obtained with approximate expressions. Adapted
from \citet{Forteza2007}.}
\label{fig:jlb_fig5}
\end{figure}

The consideration of prominence plasmas as partially ionized is
extremely important for the physics of prominences, and the effects on
MHD waves in prominences need to be taken into account. In particular,
the frictional damping of magneto-acoustic waves in a partially ionized
plasma is much stronger than in a fully ionized plasma because the
presence of neutrals causes the Joule dissipation to increase as a
result of electron-neutral and ion-neutral collisions \citep[][]
{Khodachenko2004}. A comparative study of the role of ion-neutral
damping of MHD waves and their damping due to viscosity and thermal
conductivity was made by \citet{Khodachenko2004, Khodachenko2006},
finding that collisional damping is dominant.

The effects of partial ionization on fast and slow waves in an
unbounded prominence plasma have been studied by \citet{Forteza2007}.
They have considered a partially ionized hydrogen plasma but have not
included the effects of particle ionization and recombination, and
they have assumed a strong thermal coupling between the species,
so that electrons, ions, and neutrals have the same
temperature ($T_\mathrm{e} = T_\mathrm{i} = T_\mathrm{n} = T$).
The separate governing equations for the three species can be easily
substituted by a set of one-fluid equations for the whole partially
ionized plasma, where isotropic pressure has been assumed and gravity,
viscosity, heat conduction, and non-adiabatic effects have been
neglected.
A uniform plasma with density $\rho_0$ and pressure $p_0$ permeated by
a magnetic field $\vec{B}_0 = B_{0}\hat{x}=\mbox{const}$. has been
considered, and linear magneto-acoustic waves have been studied.  The
results shown in Fig.~\ref{fig:jlb_fig5} suggest that ion-neutral
collisions are more important for fast waves (for which $\dap$ is
between 1 and $10^{5}$) than for slow waves (for which $\dap$ varies
between $10^{4}$ and $10^{8}$).  Therefore, one can conclude that the
effects arising from the partial ionization of the plasma are
unimportant for slow waves, while fast waves can be damped
efficiently for moderate values of the ionization fraction,
$\tilde{\mu}\sim 1$, i.e., almost neutral plasmas.

Another interesting feature is that when plotting the period or the damping
time of the fast wave versus the wavenumber, at a certain critical
wavenumber ($k_\mathrm{c}$) the fast wave disappears. The reason is
that the critical wavenumber depends on the ionization fraction
through $\etac$ (Cowling's magnetic diffusivity).  Hence, in a
partially ionized plasma the fast mode only exists as a damped
propagating wave for wavenumbers below the critical value,
$k_\mathrm{c}$.  For wavenumbers greater than this critical value we
have a damped disturbance instead of a propagating wave.

\begin{figure}[t]
\centering\resizebox{\hsize}{!}{\includegraphics{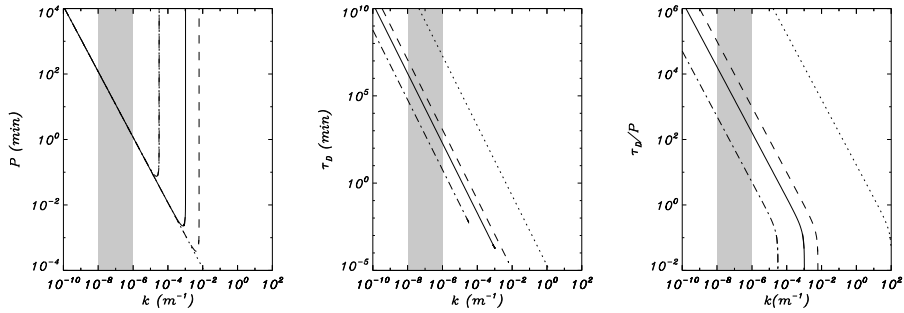}}
\caption{
Period, damping time and ratio of the damping time to the period for
the Alfv\'en wave in a partially ionized plasma with $\mutilde=0.5$
(dotted), $\mutilde=0.6$ (dashed), $\mutilde=0.8$ (solid) and
$\mutilde=0.99$ (dash-dotted). Adapted from \citet{Forteza2008}.}
\label{fig:jlb_fig6}
\end{figure}

Apart from magneto-acoustic waves, the time damping of Alfv\'{e}n waves
due to partial ionization effects has also been studied \citep[][]
{Forteza2008}. Figure~\ref{fig:jlb_fig6} shows the results obtained for the
period, the damping time and the ratio of the damping time to the
period. In this Figure the solution for a fully ionized plasma
($\mutilde=0.5$) with magnetic resistivity \citep[][]{Ferraro1961,
Kendall1964} is also shown.
The Alfv\'en wave behavior is similar to that of the fast wave in the
adiabatic partially ionized case.  When ion-neutral collisions become
the dominant mechanism, fast and Alfv\'en waves have similar period
and damping time and, as for the fast wave, the ratio of the damping
time to the period decreases when going to almost neutral plasmas.
A critical wavenumber ($k_\mathrm{c}^\mathrm{a}$) also appears for the
Alfv\'en waves.  This quantity depends on the ionization fraction and
on the propagation angle.  Usually, $k_\mathrm{c}^\mathrm{a}$ is larger
than $k_\mathrm{c}$ for fast waves, and both critical wavenumbers
become equal for parallel propagation.

\subsubsection{Damping of Prominence Thread Oscillations by Resonant
Absorption}
\label{sec:4.3.4}

\citet{Arregui2008} studied the resonant absorption of transverse kink
waves in filament threads. They considered a straight, cylindrically
symmetric flux tube of mean radius $a$, and neglected the effects of
gravity. The inhomogeneous filament thread occupies the full length of
the tube, and is modeled as a density enhancement with a non-uniform
radial distribution of density $\rho(r)$ across the structure. The
authors use the zero-$\beta$ approximation, so that slow waves are
absent, and they consider perturbations with azimuthal mode number
$m=1$, representing fast kink waves with transverse displacement of
the tube. For fast kink waves to be damped by resonant absorption, the
Alfv\'en speed must vary across the structure. The authors assume that
there is a transition layer with thickness $l$ between the interior of
the filament and the surrounding corona. The ratio $l/a$ provides a
measure of the transverse inhomogeneity length scale, and can vary in
between $l/a=0$ (homogeneous tube) and $l/a=2$ (fully non-uniform
tube). The global $m=1$ kink mode is resonantly coupled to local
Alfv\'en waves. This coupling causes excitation of localized
Alfv\'enic oscillations and damping of the global kink mode.

\begin{figure*}
\centering{\resizebox{11cm}{!}{\includegraphics{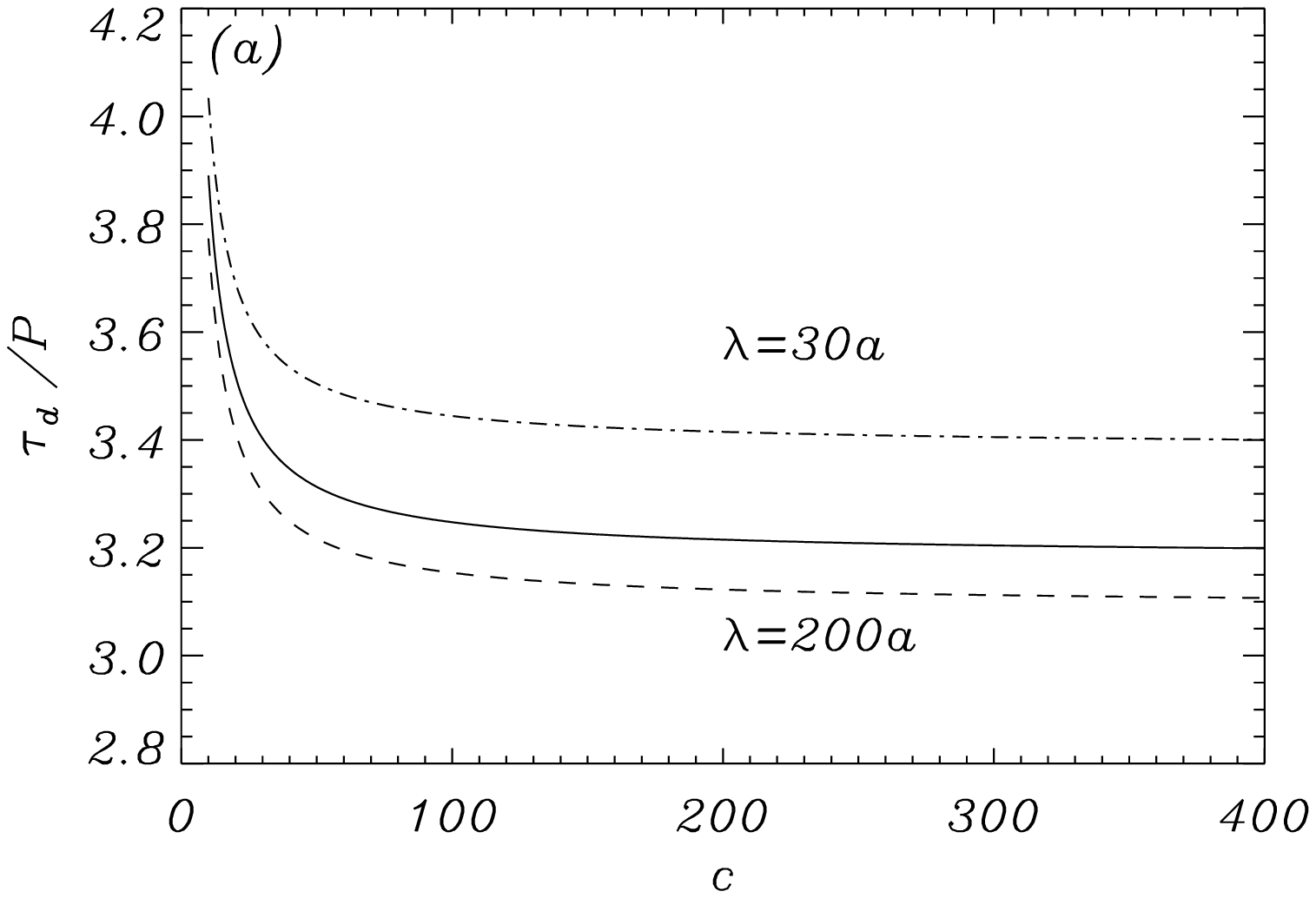}
                               \includegraphics{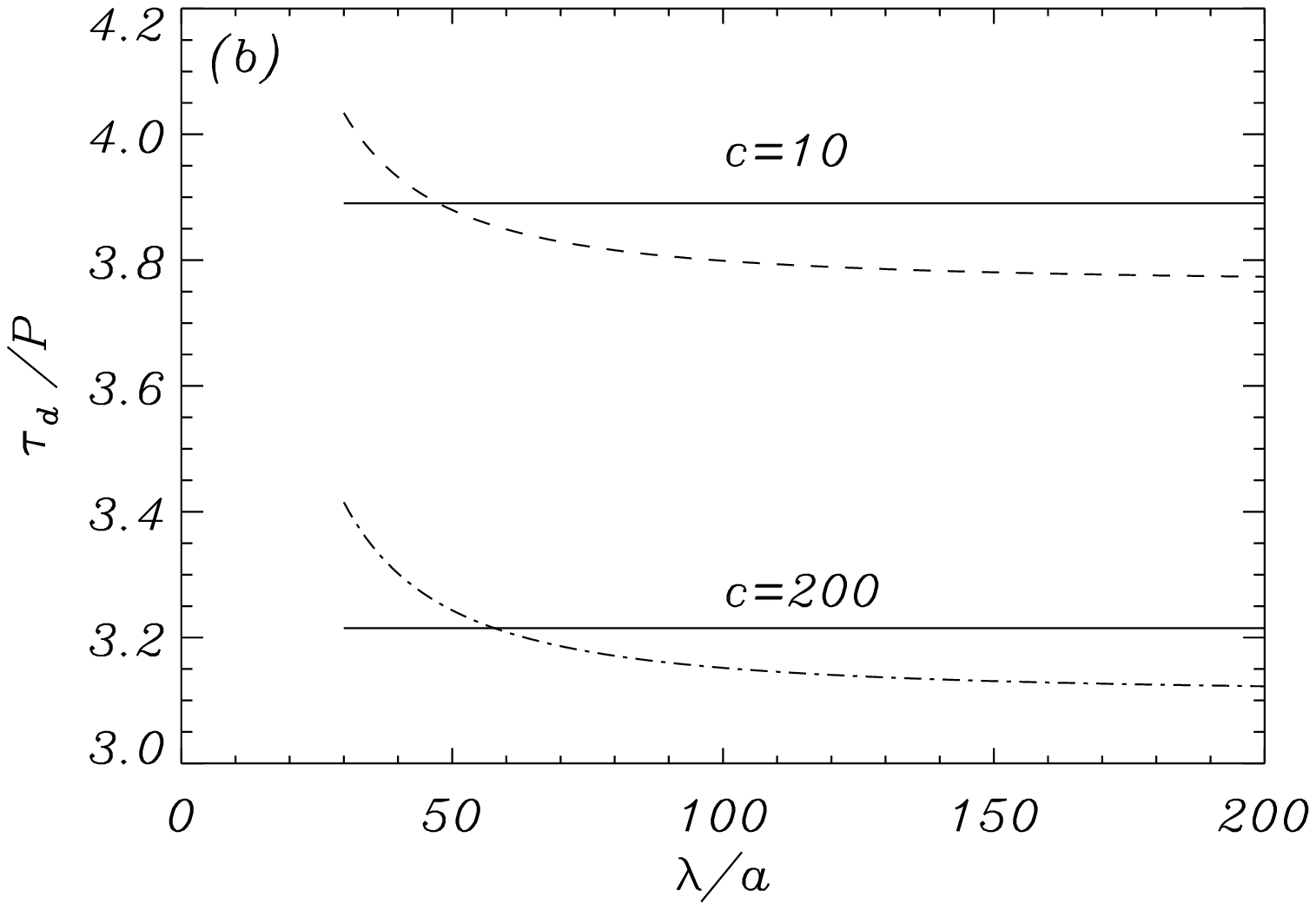}}}
\centering{\resizebox{11cm}{!}{\includegraphics{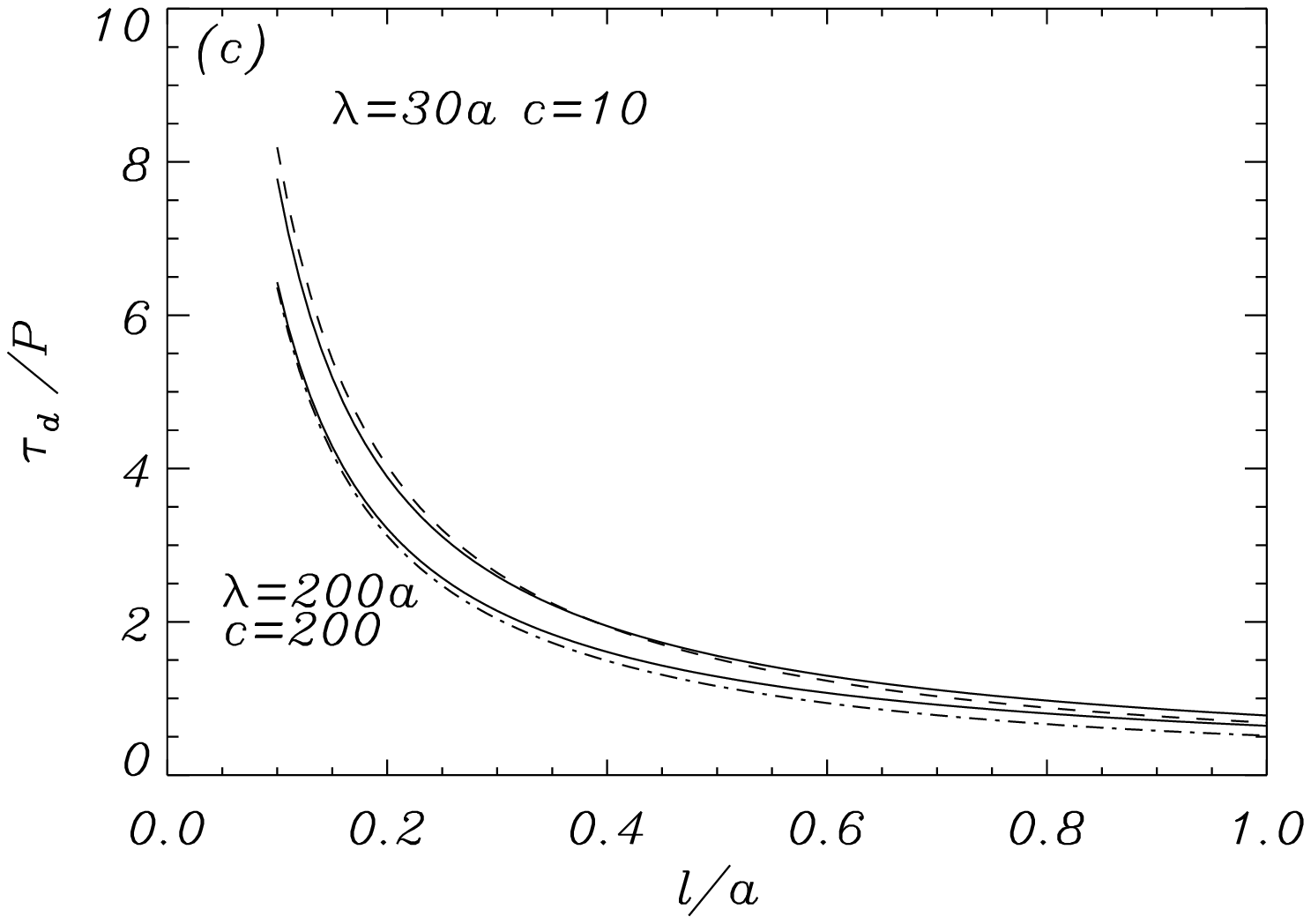}
                               \includegraphics{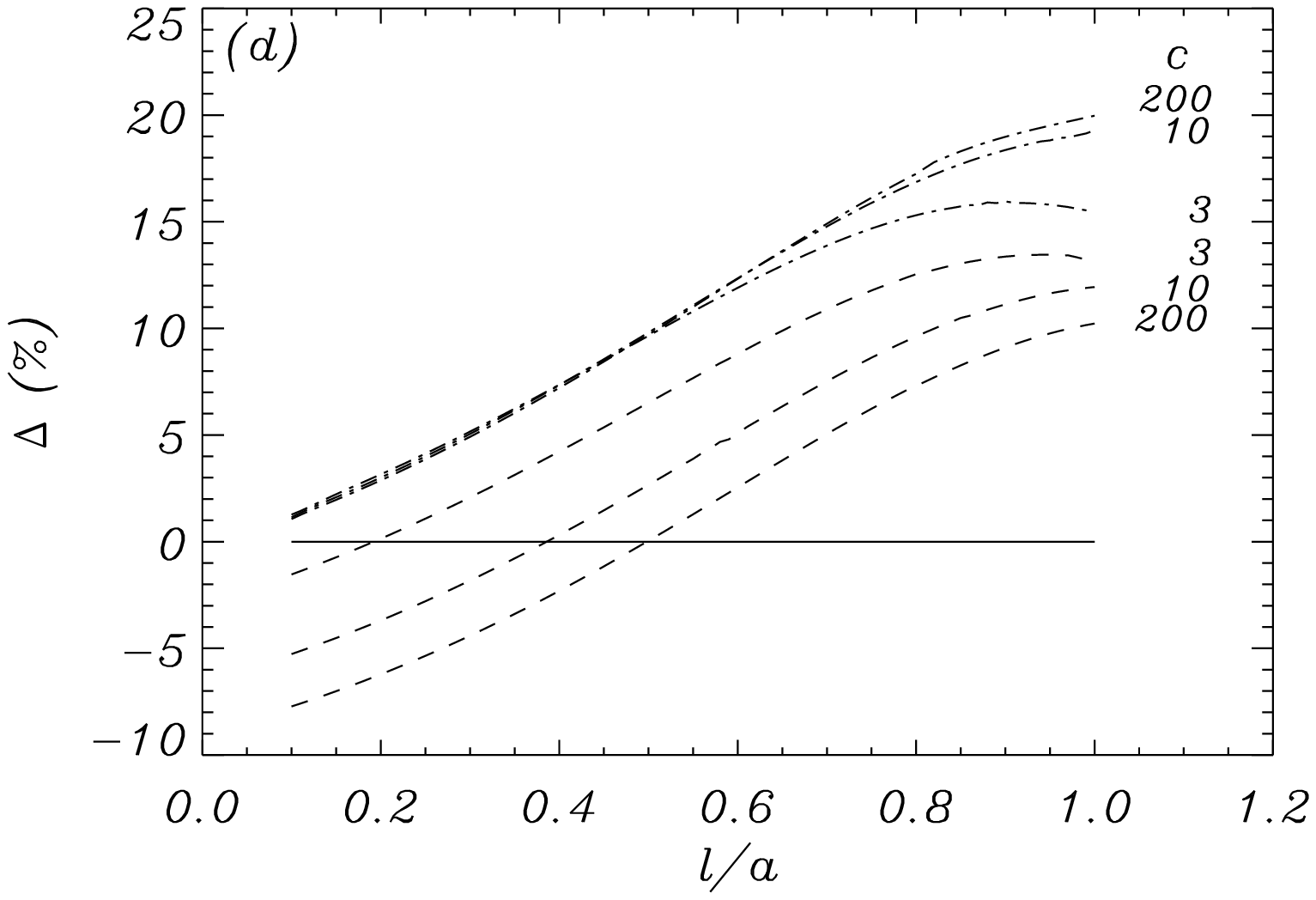}}}
\caption{
Ratio of damping time $\tau_d$ and wave period $P$ for fast kink waves
in filament threads with $a=100$ km.  In all plots solid curves
correspond to analytic solutions given by equation
(\ref{eq:dampingrate}) with $F=2/\pi$, and other curves correspond to
numerical solutions.
{\em (a)}: Damping ratio as function of density contrast $c$ with
$l/a=0.2$ and for two wavelengths.
{\em (b)}: Damping ratio as function of wavelength with $l/a=0.2$ and
for two density contrasts.
{\em (c)}: Damping ratio as function of transverse inhomogeneity
length scale $l/a$ for two combinations of wavelength and density
contrast.
{\em (d)}: Percentage difference $\Delta$ of numerical and analytical
values of the damping ratio, for $\lambda=30a$ (dashed lines) and
$\lambda=200a$ (dash-dotted lines), and for different values of
density contrast (see labels). Adapted from \citet{Arregui2008}.}
\label{fig:jlb_fig7}
\end{figure*}

Combining long-wavelength and thin boundary ($l/a \ll 1$)
approximations, analytic expressions for the damping time over the
period can be written as \citep[\eg][]{Hollweg1988, Sakurai1991,
Goossens1992, Goossens1995, Ruderman2002}:
\begin{equation}
\frac{\displaystyle \tau_{d}}{\displaystyle P} = F \;\; \frac
{\displaystyle a} {\displaystyle l}\;\; \frac{\displaystyle c +
1}{\displaystyle c - 1},
\label{eq:dampingrate}
\end{equation}
where $c$ is the density contrast of the thread, and $F$ is a
numerical factor that depends on the details of the density variation
in the transition layer.  \citet{Arregui2008} obtained numerical
solutions of the MHD wave equations for the $m=1$ modes.
Figure~\ref{fig:jlb_fig7} shows that analytical and numerical
solutions display the same qualitative behavior with density contrast
$c$, wavelength $\lambda$, and transverse inhomogeneity length scale
$l/a$. For low values of the density contrast, the ratio $\tau_d/P$
rapidly decreases with increasing thread density
(Fig.~\ref{fig:jlb_fig7}a).  Interestingly, for the larger contrasts
typically found in filament threads ($c \sim 100$), $\tau_d/P$ is
nearly independent of $c$ and $\lambda$ (Fig.~\ref{fig:jlb_fig7}b),
but rapidly decreases with increasing $l/a$
(Fig.~\ref{fig:jlb_fig7}c).

The above results suggest that resonant absorption is a very efficient
mechanism for the attenuation of fast kink waves in filament threads.
It is not yet clear how this affects of overall dynamics of prominence
plasmas.  One possibility is that the resonant absorption causes
significant heating of the prominence plasma, resulting in
chromospheric evaporation (section 3). This may alter the density profile of the
prominence thread, which changes the Alfv\'en wave resonance condition
\citep[\eg][]{Ofman1998}. Future modeling of the dynamics of
prominence plasmas should include the effects of resonant absorption.

\subsubsection{The Effects of Material Flows}
\label{sec:4.3.5}

\begin{figure}[t]
\centering{\resizebox{10cm}{!}{\includegraphics{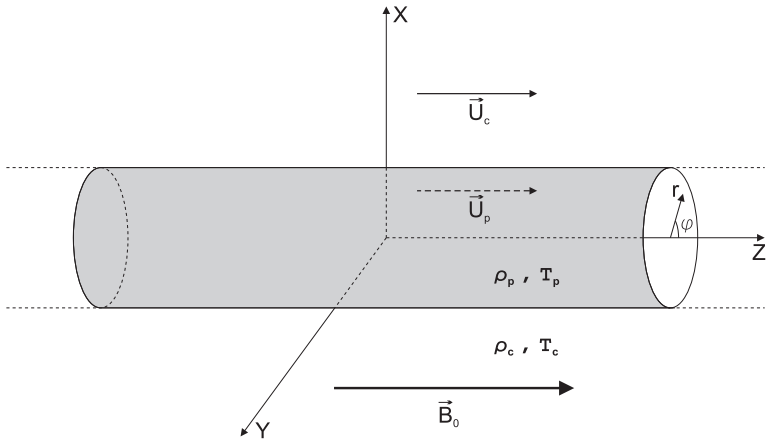}}}
\caption{
Sketch of the equilibrium. Adapted from \citet{Soler2008b}.}
\label{fig:jlb_fig8}
\end{figure}

As discussed in section \ref{sec:3}, flows are a ubiquitous feature in
prominences and filaments, and are routinely observed in H$\alpha$,
UV and EUV lines(see section 4, Paper I). \citet{Soler2008b} investigated the effects of both
mass flow and non-adiabatic processes on the oscillations of an
individual prominence thread.  The thread is modeled as an infinite
homogeneous cylinder with radius $a$, density $\rho_{\rm p}$,
temperature $T_{\rm p}$), and parallel flow velocity $U_{\rm p} \ge
0$. The cylinder is embedded in an unbounded homogeneous corona with
density $\rho_{\rm c}$ and temperature $T_{\rm c}$. The magnetic field
is $B_0$ everywhere and the total pressure is assumed to be continuous
across the interface between the flux tube and the external medium.
For simplicity, the hot coronal part of the magnetic tube that
contains the thread is not taken into account, and gravity is
neglected. The equilibrium configuration is shown in Figure
\ref{fig:jlb_fig8}.
\begin{figure}[t]
\centering{\resizebox{12cm}{!}{\includegraphics{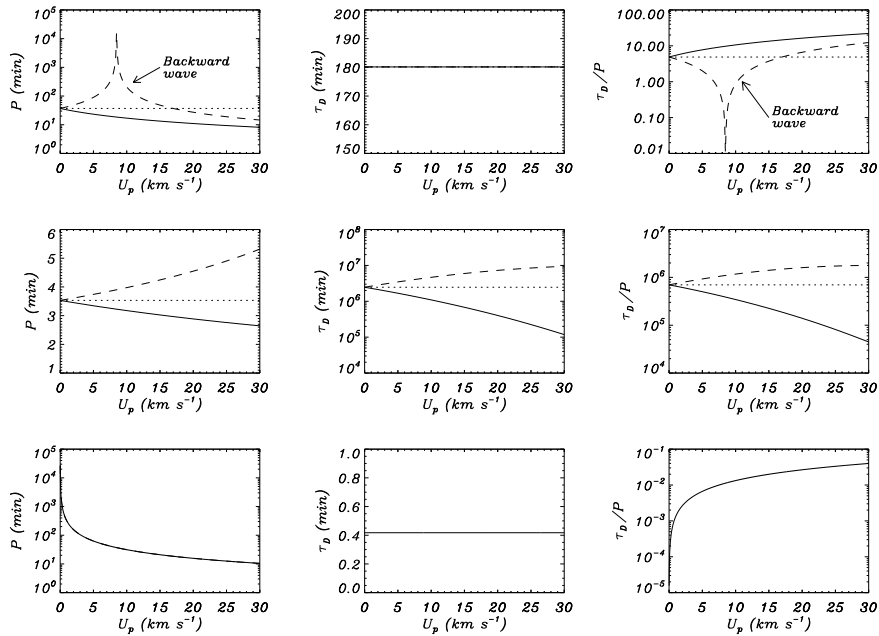}}}
\caption{
Period (left), damping time (center), and ratio of the damping time to
the period (right) versus the flow velocity for the fundamental
oscillatory modes with $k_z a = 10^{-2}$.  Upper, mid, and lower
panels correspond to the slow, fast kink, and thermal modes,
respectively.  Different line styles represent: parallel waves (solid
line), anti-parallel waves (dashed line), and solutions in the absence
of flow (dotted line).  Adapted from \citet{Soler2008b}.}
\label{fig:jlb_fig9}
\end{figure}

Figure~\ref{fig:jlb_fig9} shows the dependence of the period, damping
time, and their ratio as a function of the flow velocity for the slow,
fast and thermal modes \citep[for an explanation of the thermal mode,
see][]{Carbonell2009}.  The longitudinal wavenumber has been fixed to
$k_z a = 10^{-2}$, which corresponds to a value within the observed
range of wavelengths.  The flow velocities in the range 0 - 30 \kms
are considered, which corresponds to the observed flow speeds in
quiescent prominences. The anti-parallel slow wave becomes a backward
wave for $U_{\rm p} \approx 8.5$ \kms, which corresponds to the
non-adiabatic sound speed, causing the period of this solution to grow
dramatically near such flow velocity.  However, the period of both
parallel and anti-parallel fast kink waves is only slightly modified
with respect to the solution in the absence of flow, and the thermal
wave now has a finite period, which is comparable to that of the
parallel slow mode. The damping time of slow and thermal modes is
independent of flow velocity, but the attenuation of the fast kink
mode is affected by the flow. The larger the flow velocity, the more
attenuated the parallel fast kink wave, whereas the opposite occurs
for the anti-parallel solution. This behavior is due to weak coupling
of the fast modes to external slow modes \citep[for details
see][]{Soler2008b}.

\citet{Terradas2008} modeled the transverse oscillations of flowing
prominence threads as observed by \citet{Okamoto2007} with SOT. The kink
oscillations of a flux tube containing a flowing dense part, which represents 
the prominence material, were studied from both analytical 
and numerical points of view. The results determined that
there is almost no difference between the oscillation periods when
steady {\it vs.} flowing threads are considered. Also, the resulting period
matches that of a kink mode. In addition, to obtain information about 
the Alfv\'en speed in oscillating threads, a seismological analysis as 
described in \ref{sec:4.4} was performed.

\subsection{Prominence Seismology}
\label{sec:4.4}

The main goal of prominence seismology is to infer the internal
structure and properties of solar prominences from the study of their
oscillations. One of the most searched for parameters is 
the prominence magnetic field strength and attempts to determine it have 
been made. For instance, using the
significant periods obtained from the Fourier analysis of Doppler
velocity time series belonging to an active region filament,
\citet{Regnier2001} applied the theoretical model of
\citet{Joarder1993} to determine the magnetic field and the angle of
the magnetic field with the long axis of the observed filaments. They
used the periods to identify the modes involved in the oscillations
and after this identification they obtained $18^\circ \pm 2.5^\circ$
for the angle, assuming a prominence temperature of 8000 K and an
analytical relationship between the magnetic field strength and the
density. \citet{Pouget2006} followed the same procedure in the case
of filaments observed with CDS. They identified six modes (slow, fast
and Alfv\'en) involved in the oscillations and, using the same
theoretical model, they obtained values between $19^\circ$ and
$35^\circ$ for the angle and between 10 and 35 G for the magnetic
field strength. All these determinations must be considered with care
because the identification of the modes based on the periods is very
uncertain; apart from the periods, the velocity polarization is of
paramount importance for the mode identification in slab models.

\begin{figure}[t]
\centering
\resizebox{\hsize}{!}{\includegraphics{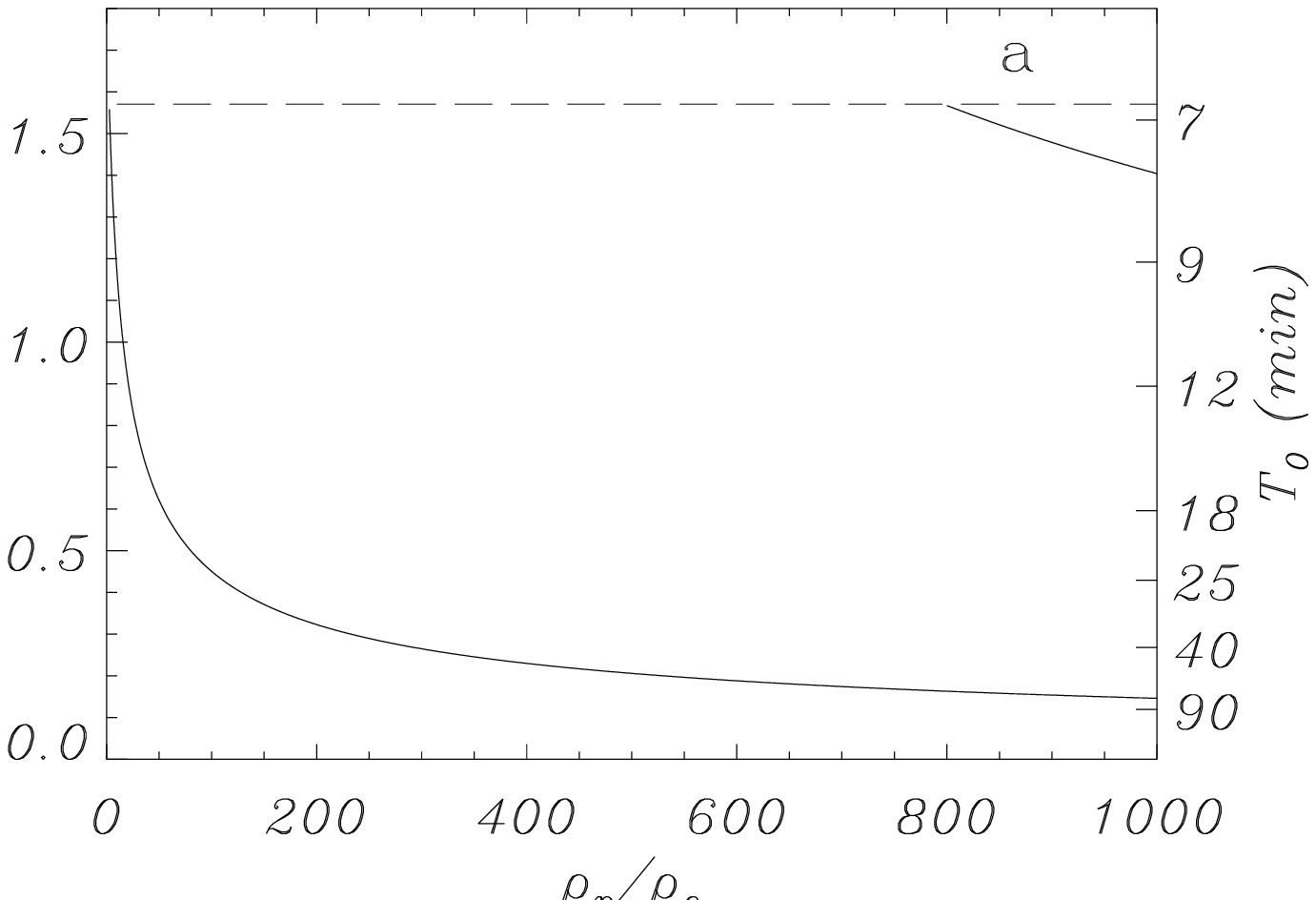} \hspace{2cm}
                      \includegraphics{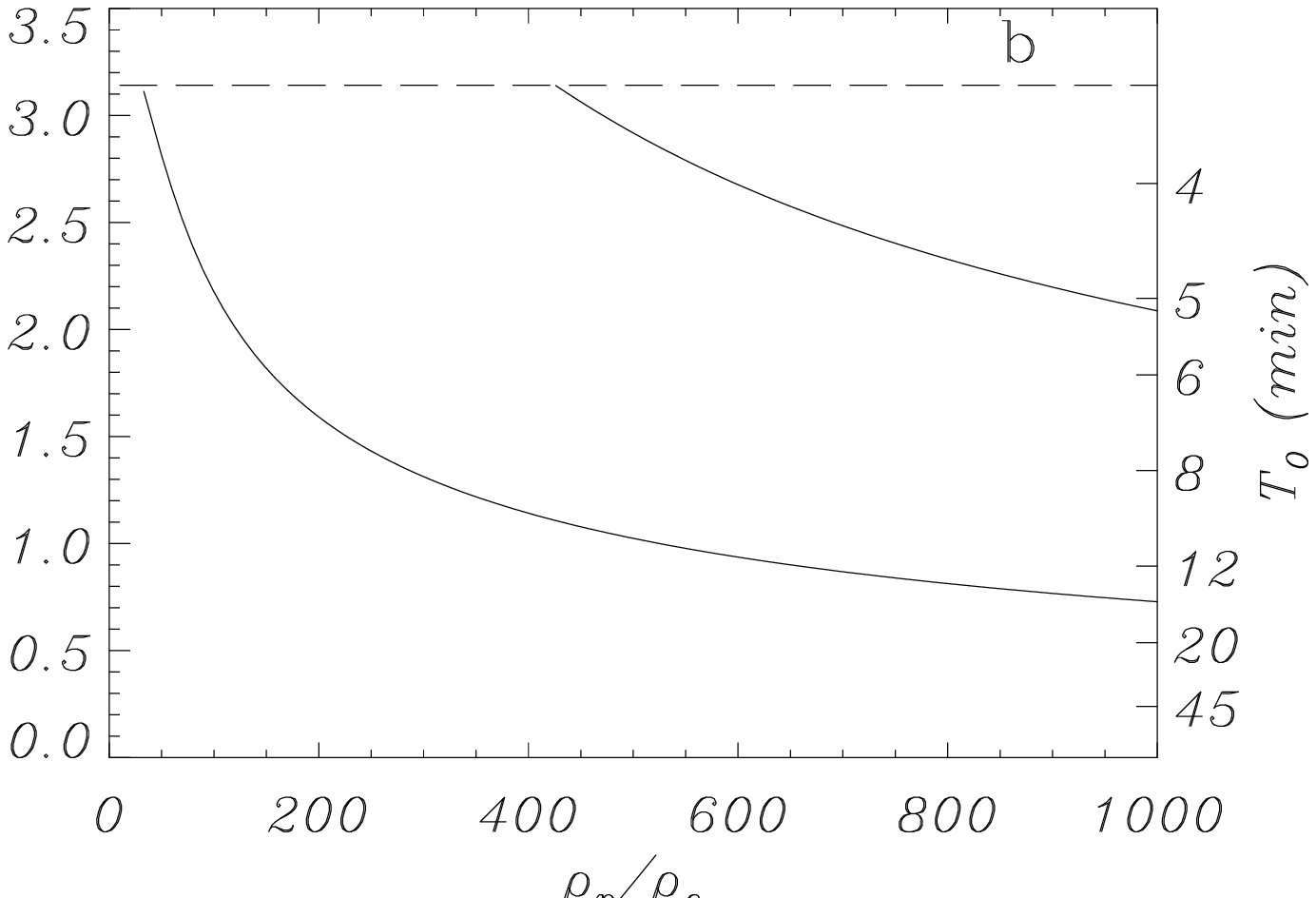}}
\resizebox{\hsize}{!}{\includegraphics{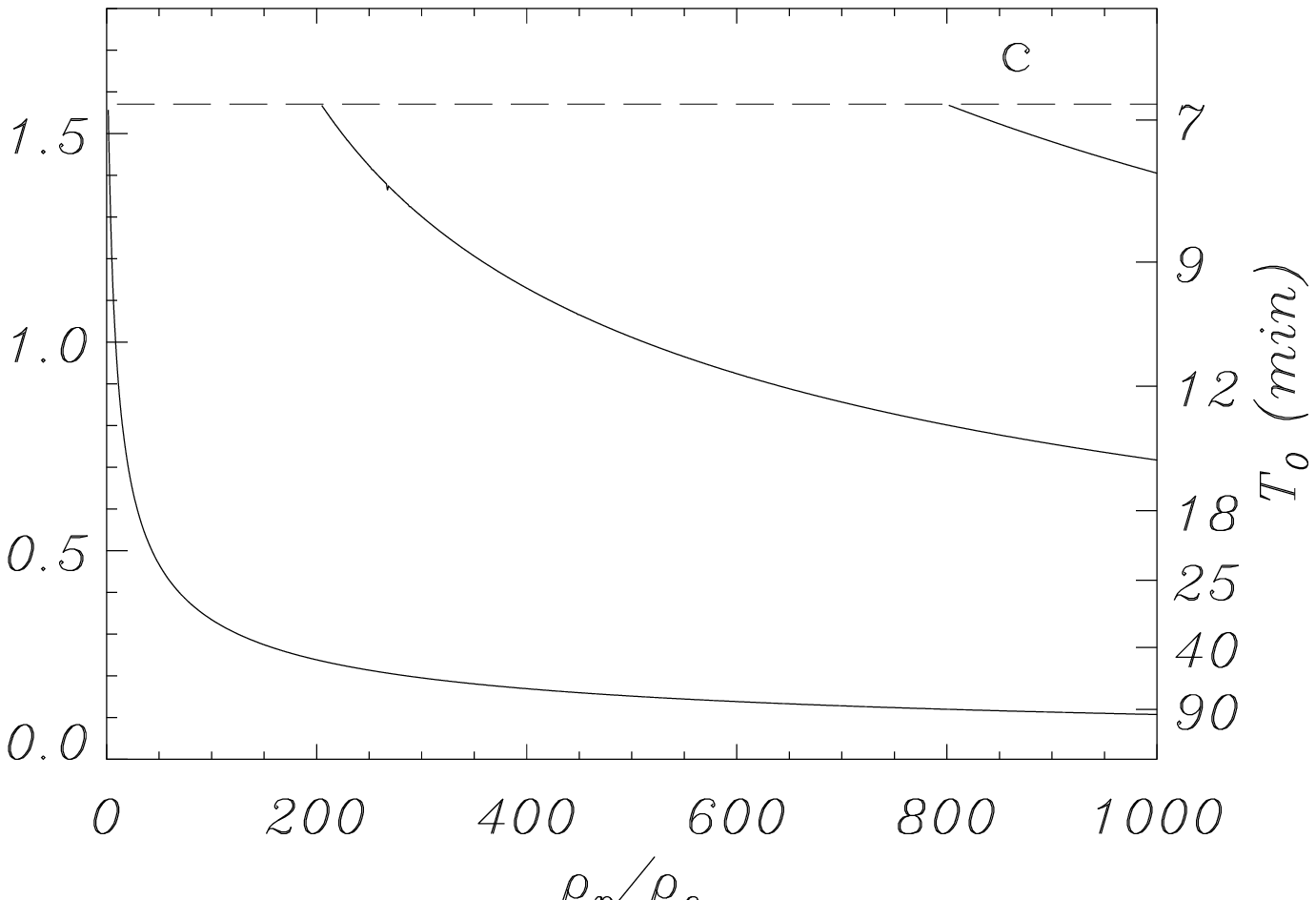} \hspace{2cm}
                      \includegraphics{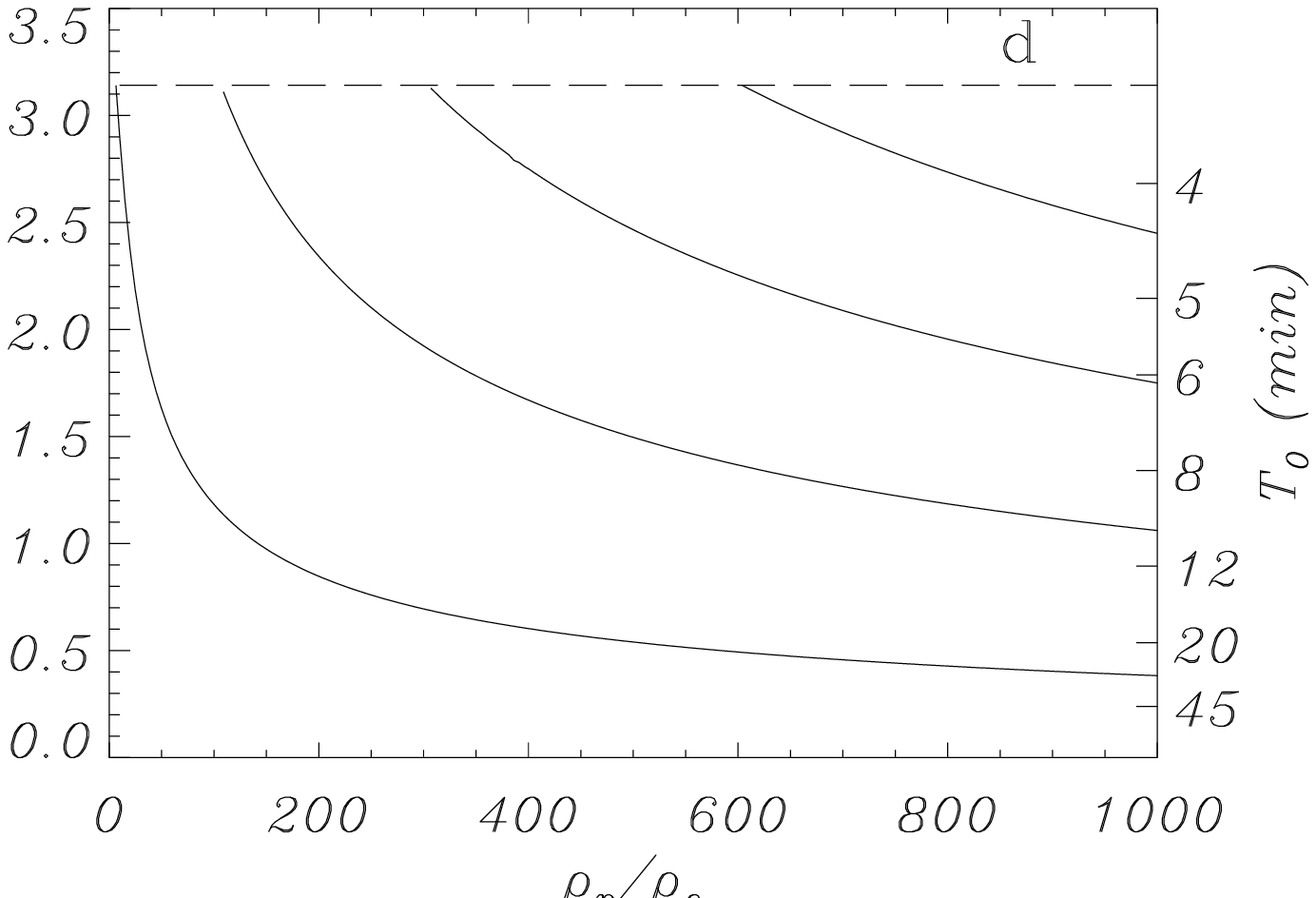}}
\vspace{2mm}
\caption{
Frequency of (a) and (c) the kink even modes and (b)
and (d) the kink odd modes against $\rho_{p}/\rho_{c}$ for the
parameters $\rho_{e}/\rho_{c}=0.6$ and $b/L=0.001$. In (a) and
(b) $W/L=0.1$, while in (c) and (d) $W/L=0.2$. The
right axis provides with the period $T_{0}$ obtained after assuming
the background magnetic field strength, the coronal density and the
half-length of field lines are $B_0=5$ G, $\rho_{c}=8.37\times
10^{-13}$ kg/m$^{3}$ and $L=5\times 10^4$ km. The horizontal dashed
line in each plot gives the cut-off frequency. Adapted from
\citet{Diaz2002}.}
\label{fig:jlb_rho_dep}
\end{figure}

As discussed in section \ref{sec:4.3.1}, \citet{Diaz2002} considered
a cylindrical model of a prominence thread. According to this model,
the cool prominence thread has a radius $b$, and its length $2W$ is
less than the length $2L$ of the flux tube on which the thread is located
(see Fig.~\ref{fig:jlb_fig1}). The density in the prominence is $\rho_p$,
the density in the coronal part of the flux tube is $\rho_e$, and the
density in the surrounding corona is $\rho_c$. An important result of
\citet{Diaz2002} is that the dimensionless oscillation frequency
$\omega L/c_{Ac}$ depends mainly on the ratios $\rho_{p}/\rho_{c}$ and
$W/L$, and is insensitive to the ratios $b/L$ and $\rho_{e}/\rho_{c}$. 
Here $c_{Ac}$ is the coronal Alfven speed. Fig.~\ref{fig:jlb_rho_dep}
shows the dimensionless frequency of some modes for fixed values of fibril
length $W/L$, thickness $b/L$, and $\rho_{e}/\rho_{c}$. Using such diagrams,
we can perform a seismological analysis if some of the involved quantities
are provided by observations. The availability of this observational
information would allow us to determine the numerical values of the rest of
prominence physical parameters. These examples point out that in order to determine 
the magnetic field strength, and apart of the oscillatory period, other physical properties of the prominence, such as the density, must be known or assumed.
\begin{figure}[t]
\centering{\resizebox{6cm}{!}{\includegraphics{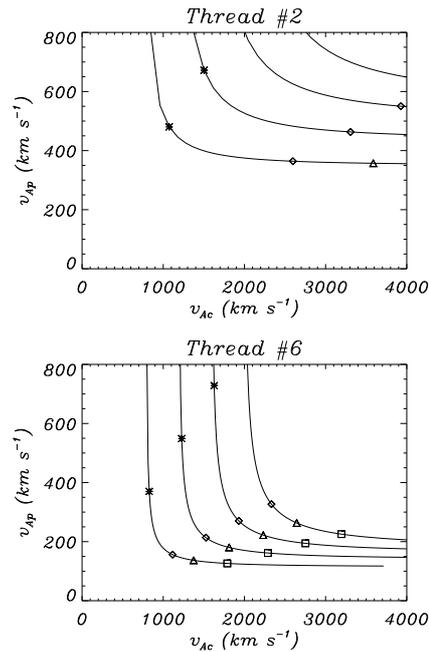}}}
\caption{
Dependence of the Alfv\'en velocity in the thread as a function of the
coronal Alfv\'en velocity for two of the threads studied by
\citet{Okamoto2007}.  In each panel, from bottom to top, the curves
correspond to a length of magnetic field lines of 100,000 km, 150,000
km, 200,000 km, and 250,000 km, respectively. Asterisks, diamonds,
triangles, and squares correspond to density ratios of the thread to
the coronal gas $\rho_{p}/\rho_{c }\simeq 5,50,100,200$. Adapted
from \citet{Terradas2008}.}
\label{fig:jlb_fig14}
\end{figure}

In the theoretical model used by  \citet{Terradas2008} (section \ref{sec:4.3.5}), prominence
and coronal Alfv\'en speeds are related through the ratio between
prominence and coronal densities. Therefore, once this ratio is fixed
and the coronal Alfv\'en speed is determined from the dispersion
relation for the kink mode, the Alfv\'en speed in prominence threads can be determined. 
Figure~\ref{fig:jlb_fig14} shows the dependence of the Alfv\'en velocity in the thread as a function of the
coronal Alfv\'en velocity, for two different threads, when
different density ratios and lengths of the magnetic field lines are considered.
It can be seen that for a certain value of the coronal Alfv\'en speed
the prominence Alfv\'en speed stabilizes. Using this analysis, lower
bounds on the Alfv\'en speeds in different prominence threads can
be obtained.

For the threads considered and taking the length of the magnetic
field lines to be 100,000 km, the lower bounds of the Alfven speed are
between 120 - 350 \kms. These values are consistent with strong magnetic
fields (50 G) and large densities ($1 - 8  \times 10^{11}$ $\rm cm^{-3}$),
or weaker magnetic fields (10 G) and lower densities
($0.4 - 3 \times 10^{10}$ $\rm cm^{-3}$). These densities are within the
ranges inferred from spectroscopic techniques (see sections 2.2 and 3.3 of
Paper I). Further assumptions about the density, appropriate for the considered 
model, but not based upon simultaneous observations (which are not available), must be made 
for a numerical determination of the magnetic field.

Finally, an important conclusion in \citet{Arregui2008} (section \ref{sec:4.3.4}) 
is that in equation~(\ref{eq:dampingrate}) the
damping rate becomes independent of density contrast for large values
of this parameter.  This fact has several seismological implications:
First, the determination of the density contrast in the case of
prominences is not as crucial as in other problems where the density
contrast is low; second, by assuming a density contrast $c$ of infinity
in equation~(\ref{eq:dampingrate}) an estimate of the transverse
inhomogeneity length scale $l/a$ can be determined. Using $\tau_{d}/P
= 4$, we find $l/a \sim 0.15$, i.e. high-density threads are compatible
with thin inhomogeneous layers. Furthermore, knowing the period and the wavelength, 
the Alfv\'en speed in threads can be also obtained. To make further progress and to obtain 
the magnetic field strength the prominence density is required.

In summary, the determination of the magnetic field strength is not 
straightforward. Depending on the theoretical model assumed to interpret 
the oscillations, additional information  on numerical values of parameters
such as prominence and coronal densities, geometrical dimensions, etc.,
need to be observationally acquired before the magnetic field strength can be 
determined.

\subsubsection{Concluding Remarks}
\label{sec:4.5}

The study and understanding of small-amplitude prominence oscillations
is a challenging task from the observational and theoretical point of
view.  The number of different physical effects (non-adiabaticity,
flows, partial ionization, etc.) involved suggests that they must be
explored in a systematic manner in order to obtain a full
understanding of the properties of these oscillations.  From a
theoretical point of view, the main question could be: What do we need
in order to foster theoretical understanding of prominence
oscillations?  Of course, a simple answer would be better observations
having high spatial and time resolution.  However, this would not be
enough because our current knowledge of the internal structure and
physical properties of solar prominences is not complete. For example, the
magnetic field structure which supports and shields the prominence is
not fully understood. Without a clear understanding of the magnetic
structure it becomes difficult to advance our knowledge of prominence
oscillations.
Furthermore, are the flows real or are there ionization or excitation
fronts changing the physical conditions and mimicking flows?  What is
the mechanism responsible for prominence heating?  Partial ionization
seems to be a key feature in prominences, influencing not only
magneto-hydrodynamic waves but also the manner in which prominence
plasma is tied to magnetic field lines and, right now, is starting to
be considered in some theoretical studies.

From the above considerations, one can easily conclude that from the
theoretical and observational point of view we are far from a complete
understanding of prominence oscillations, and this conclusion is
correct.  In the near future, the joint use of two-dimensional,
high-resolution observations, modern data analysis techniques,
and complex theoretical models, incorporating as much physics as
possible and using state of the art numerical simulations, should help
us to make further progress in this field.


\section{Formation and Large-Scale Patterns of Filament Channel and Filaments}
\label{sec:5}

\subsection{Global Patterns and Formation Locations}
\label{sec:5.1}

Solar filaments (a.k.a. prominences) form over a wide range of
latitudes on the Sun. Their locations spread everywhere, from the
active belts to the polar crown. Poleward transport of magnetic flux
across the solar surface during the solar cycle is accompanied by a
poleward migration of the preferred locations of filament formation
\citep{McI2002, Min1998, Amb2002, Mour1994}. Although filaments may
form at many locations on the Sun, they always form above Polarity
Inversion Lines (PILs), which divide regions of positive and negative
flux in the photosphere. As discussed in Section~\ref{sec:2.1.2}, a
necessary condition for the formation of a filament is the presence of
a filament channel at the height of the chromosphere \citep[\eg][]
{Gaiz1998}. Filament channels are the unique sites of filament formation.
Channels are more fundamental than the filaments that form within them, as not 
every channel contains a filament and a single channel may survive a succession 
of filament formations and eruptions. The structure of a filament channel is 
illustrated in Figure~\ref{fig:mackay_fig1}.  A key feature of a filament channel,
is that it is a region of dominant horizontal field where the field on either side 
points in the same direction. As a result, filament channels are interpret as 
locations of strong magnetic shear and highly non-potential magnetic fields.
The channels are believed to be a low atmosphere signature of a larger horizontal 
non-potential field that extends into the coronal volume 
(see Section~\ref{sec:2.1.2}). Presently it is unclear why channels and
their non-potential fields build up along PILs, so understanding filament
channel formation is key to understanding the evolution of magnetic fields
on the Sun and their relationship to eruptive phenomena. By observing and interpreting
their formation and evolution, we may examine directly the buildup to, and initiation of
geoeffective space weather. In the discussion below
we consider the global properties and formation locations as deduced from
H$\alpha$ observations. 

\subsubsection{Global Classification Schemes}
\label{sec:5.1.1}

While filaments form at many locations on the Sun, very few studies
have considered the exact nature or history of the PILs above which
they form. Those studies that have considered this, are mainly
restricted to studying large-scale, stable filaments and neglect smaller
unstable filaments forming in the centers of activity
complexes. Understanding the type of magnetic environment in which
filaments form is key to understanding the magnetic interactions
required for their formation.

Over the years many classification schemes for filaments and prominences
have been developed \citep{D'Az48}. These schemes have referred to different features, including: 
the dynamics of the material in the prominence, whether it is stable or eruptive, or finally
with respect to the distribution of the magnetic flux below the filament. For
a discussion see \citet{Tan-Han1995}. In this paper we use a classification
scheme of filaments in terms of their spatial location on the Sun   
\citep{Engvold1998}. In later sections when we discuss the possible mechanisms of 
filament formation this classification scheme will prove useful in illustrating that
different mechanisms may form different types of filaments. As given in the
introduction the types of filament were described as: ARF
(Active Region Filament), IF (Intermediate Filament), and QF (Quiescent
Filament). An ARF is one which forms in the centers of active regions or activity complexes. In
contrast, IFs form between active regions and decaying regions of
unipolar plage. Finally QFs form in regions of weak background
fields. Observations tend to show that IFs and QFs are larger, much more
stable structures with longer lifetimes (weeks to months) compared
to ARFs, which are generally unstable with a lifetime of only a few
hours to days.

\begin{figure}[t]
\centering\includegraphics[scale=0.25]{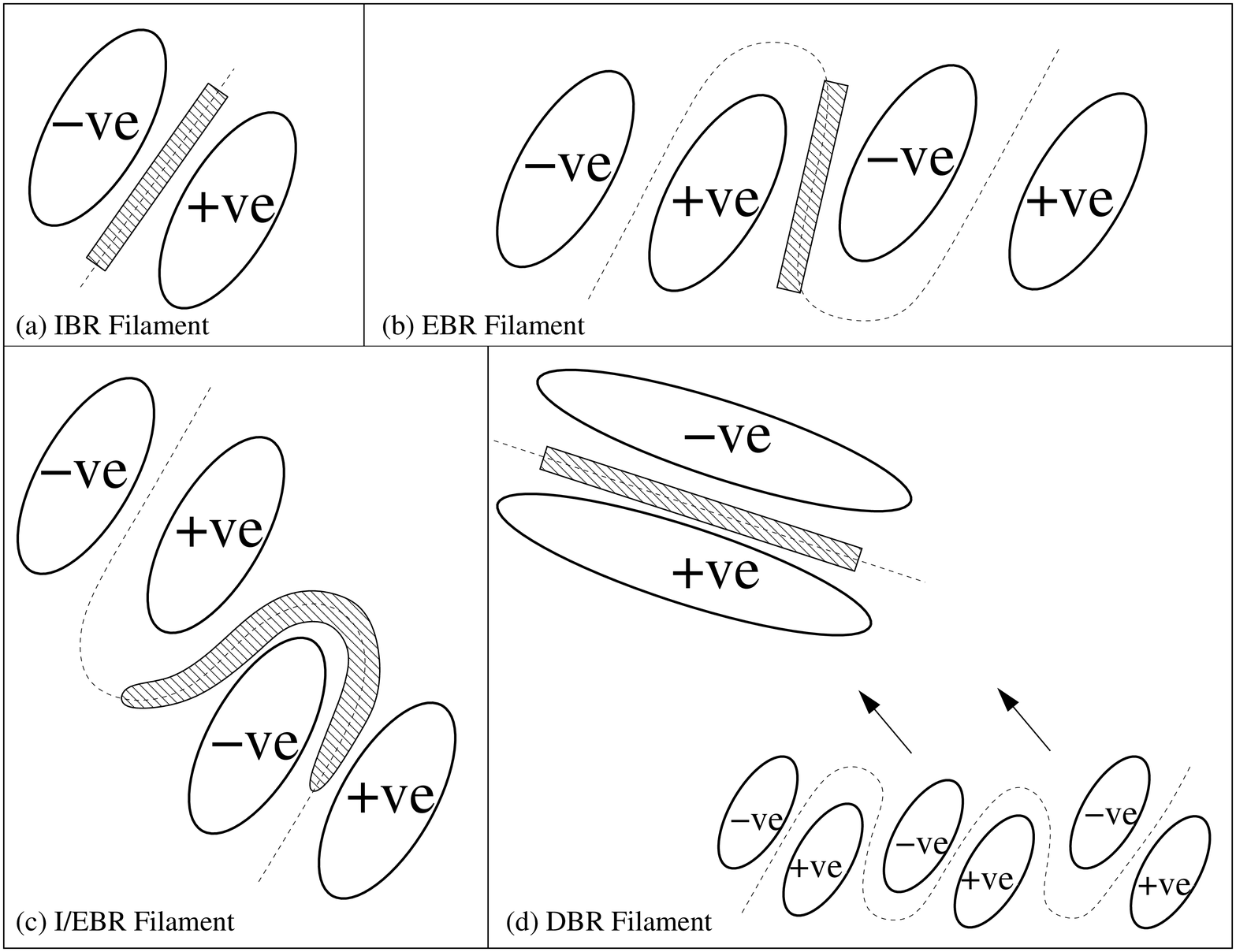}
\caption{Classification scheme for solar filaments based on those of
\citet{Tang1987} and \citet{Tan-Han1995} with two new categories (c
and d) introduced by \citet{Mackay2008}. (a) Filaments which form
above the internal PIL of a single bipole are classified as IBR. (b)
Those forming on the external PIL between bipoles or between bipoles
and unipolar regions of flux are classified as EBR. (c) Filaments that
lie both above the internal PIL within a bipole and the external PIL
outside the bipole are classified as I/EBR. (d) Filaments that form in
diffuse bipolar distributions created through multiple flux emergences
(such that the diffuse region can no longer be associated with any single
bipole emergence) are classified as DBR. This category is expected to
lie only at high latitudes. This figure is taken from Figure 2 of
\citet{Mackay2008}.}
\label{fig:mackay_fig2}
\end{figure}

\begin{figure}[t]
\centering\includegraphics[scale=0.5]{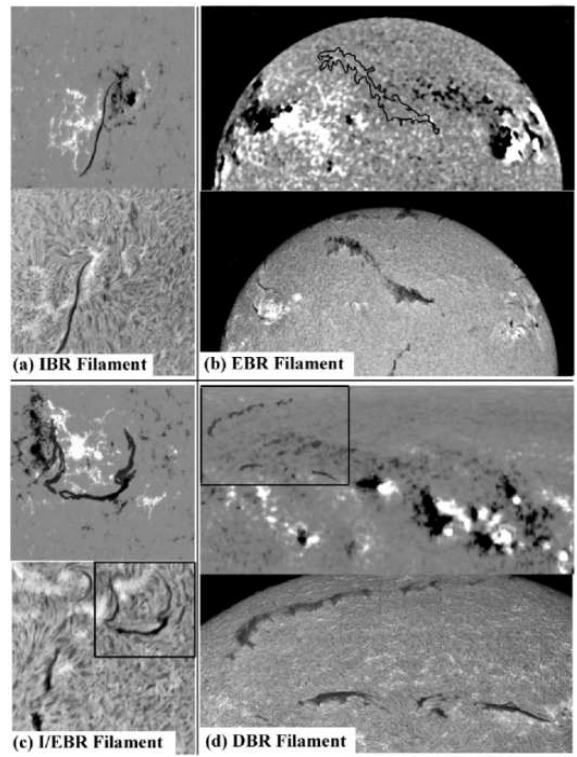}
\caption{Examples of the four categories of filaments shown in
Figure~\ref{fig:mackay_fig2} \citep[from][]{Mackay2008}. In each
of the panels (a)-(d), the bottom plot is an H$\alpha$ image from the
ORSO, while the top image shows the radial magnetic field derived from
either (a)-(c) a full-disk magnetogram or (d) a synoptic magnetogram
from Kitt Peak. Outlines of the H$\alpha$ filaments are superimposed
on each of the magnetograms. The dates of the observations are (a)
26$^{th}$ June 1979, (b) 6$^{th}$ May 1979, (c) 27$^{th}$ September
1979 and (d) 14$^{th}$ July 1979. For panels (c) and (d) the areas
enclosed by the boxes denote the corresponding area of (c) the
magnetogram and (d) the H$\alpha$ image. In panel (d) (top image) the
low latitude activity complexes which will extend poleward over time
and interact to produce diffuse regions of flux at high latitudes can
be clearly seen.}
\label{fig:mackay_fig3}
\end{figure}

While the classification scheme of \citet{Engvold1998} provides a
useful distinction between filaments forming inside and outside active
regions, to understand the role that magnetic fields play in the
formation, structure and evolution of filaments it is important to
understand the exact type of magnetic configurations wherein filaments
form. One early classification scheme \citep{Tang1987} splits them
into two categories based on the nature of the PIL above
which the filament lies.  The first category is one in which the
filament forms above a PIL lying within a single bipolar unit of flux,
and is classified as an ``Internal Bipolar Region Filament'' (see
Figures~\ref{fig:mackay_fig2}a and \ref{fig:mackay_fig3}a). In the
second, the filament forms above a PIL which lies between two
separate magnetic bipoles and is called an ``External Bipolar Region
Filament'' (Figures~\ref{fig:mackay_fig2}b and
\ref{fig:mackay_fig3}b). \cite{Tan-Han1995} describes these two
categories as Type A and Type B, respectively. Early observations by
\citet{Tang1987} showed that, when filaments are classified into these
two types, over 60$\%$ of filaments form external to bipolar regions.

In a more recent study \citet{Mackay2008} reconsidered where long,
stable solar filaments form. Their study followed the history and
evolution of the PILs underneath filaments using a wide range of
data. The data included H$\alpha$ synoptic maps, large-scale H$\alpha$
images from the Ottawa River Solar Observatory (ORSO), Kitt Peak (KP)
full disk and synoptic magnetograms, KP He~10830 synoptic images, and
results from magnetic flux transport simulations \citep{Yeates2007}.
Through following the history and evolution of the
PILs above which filaments form, the authors determined the origin of
flux regions at high latitudes from their initial bipolar source
regions at active latitudes. As with previous studies the data and
techniques employed were preferentially directed towards studying
large, stable filaments, so they did not include smaller unstable
filaments such as those found in the centers of activity complexes.

To distinguish the different bipole interactions that could lead to
the formation of filaments, \citet{Mackay2008} introduced two
additional categories for filaments compared to those defined by
\citet{Tang1987}: ``Internal/External Bipolar Regions Filaments''
(I/EBR) and ``Diffuse Bipolar Region Filaments'' (DBR). The I/EBR
filaments are defined as filaments that lie above both the
internal PIL of a bipole and the external PIL surrounding the bipole
(Figures~\ref{fig:mackay_fig2}c and \ref{fig:mackay_fig3}c), and
therefore could not be classified into the scheme proposed by
\citet{Tang1987}. In contrast, the DBR filaments are located in
essentially a bipolar distribution of flux, but where the polarities
defining the bipole did not emerge together. The formation of the
bipolar distribution was the result of many flux emergences,
coalescences and cancellations such that the polarities on either side
of the filament could not be attributed to a single bipole emergence
(Figures~\ref{fig:mackay_fig2}d and \ref{fig:mackay_fig3}d). On
comparing the classification scheme of \cite{Engvold1998} with those
of \citet{Tang1987} and \citet{Mackay2008}, ARFs could be of either
Internal Bipolar, External Bipolar or Internal/External Bipole filament
categories. IFs are always of External Bipolar type and QFs either
Internal, External, Internal/External or Diffuse Bipolar Region type. Therefore in principle
active region and quiescent filaments may form in very different magnetic 
environments.  By using the four categories (of IBR, EBR, I/EBR and DBR)  
and extending the work of \citet{Tang1987} to four distinct
phases of the solar cycle (two before and two after cycle maximum) the
authors were able to distinguish more clearly the different types of
bipole interactions leading to the formation of Intermediate
and Quiescent filaments. 

Of the 603 filaments studied by \cite{Mackay2008}, 92$\%$ formed at
locations requiring multiple bipole interactions (the breakdown
comprised of 62$\%$ EBR, 17$\%$ DBR and 13$\%$ I/EBR). Only 7$\%$
formed within a single bipole.  These results show that large-scale
filaments, namely those of the IF and QF type preferentially form at 
sites of multiple bipole interactions. Very few of them occur within a single 
bipole. Furthermore, by considering four distinct phases of the
solar cycle, \citet{Mackay2008} showed that only EBR filaments exhibit
any form of solar cycle dependence, with the other three types
remaining essentially constant \citep[see Figure 3 of][]{Mackay2008}.
The dependence showed that the number of EBR filaments varied in phase 
with the solar cycle with more at cycle maximum than minimum.
Such a variation indicates that the formation of EBR filaments must be 
strongly related to the amount of flux emergence in the solar cycle.

\subsubsection{The Hemispheric Pattern of Solar Filament Channels and
Filaments}
\label{sec:5.1.2}

While the basic properties of solar filaments have long been known,
filament channels and filaments have been classified more recently 
in terms of their chirality \citep[][]{Martin1994}.
As discussed in section \ref{sec:2.1.2}, this chirality may take one of
two forms: dextral or sinistral. Dextral/sinistral filament channels and
filaments have an axial magnetic field that points to the right/left when 
the main axis of the filament is viewed from the positive polarity side of 
the PIL (see Figure~\ref{fig:mackay_fig4}). The chirality of filament channels 
may be deduced from high resolution H$\alpha$ images combined with
magnetograms, whereas the chirality of filaments may be determined
from that of the channel, direct magnetic field measurements or from the
relationship of filaments to their barbs (dextral filaments $\sim$ right bearing
barbs, sinistral filaments $\sim$ left bearing barbs). In general due to the
lack of high resolution H$\alpha$ data and direct measurements of magnetic fields
within prominences, filaments are mostly classified using their relationship to barbs 
\citep{Pevtsov2003,Yeates2007}.
\citet{lop06} and
\citet{Martin2008} used the chirality rules to resolve the 180-degree
ambiguity in photospheric vector-field measurements. In force-free
field models \citep[e.g.,][] {Aulanier1998, Mackay1999, Balle2000,
Mackay2005} this chirality is directly related to the dominant sign
of magnetic helicity that is contained within the filament and
filament channel. A dextral
filament will contain dominantly negative helicity, while a sinistral
filament positive helicity. Hence, filaments and their channels may be regarded 
as indicators of sheared non-potential fields within the solar corona.
Their transport across the solar surface is therefore an indication of
the large-scale transport of magnetic helicity across the Sun
\citep{Yeates2008b} a key feature in explaining many eruptive 
phenomena. A surprising feature of the chirality of
filaments is that it displays an unusual large-scale hemispheric
pattern: dextral/sinistral filaments dominate in the northern/southern
hemispheres respectively \citep{Martin1994, Zirker1997, Pevtsov2003,
Yeates2007}. This
pattern is unusual as it is exactly opposite to that expected from
differential rotation acting on a North-South coronal arcade. Although
dextral/sinistral filaments dominate in the northern/southern
hemisphere, observations show that exceptions to this pattern do
occur. Therefore any model which tries to explain the formation of
filaments and their dominant axial magnetic fields must explain not
only the origin of this hemispheric pattern but also why
exceptions arise. The origin of this hemispheric pattern
will be discussed in Section~\ref{sec:5.5}.

It is clear from the above discussion that solar filaments form or are
found in a wide range of magnetic environments on the Sun, ranging
from the rapidly evolving activity complexes to the slowly evolving
poleward streams of flux that extend out of the active latitudes
towards the poles. To explain the formation of these filaments, 
observational studies and a wide range of theoretical models have 
been produced. The review will now consider observational 
case studies of the formation of filaments (Section~\ref{sec:5.2}). After
discussing these, models of filaments formation will
be discussed in Section~\ref{sec:5.3}. The observations will then be used to
clarify which models of filament formation are applicable to which filament 
formation locations (Section~\ref{sec:5.6}).

\subsection{Observations of Filament Channel and Filament Formation}
\label{sec:5.2}

To understand the magnetic environment and interactions leading to the
formation of filament channels and filaments, it is useful to discuss test
cases. To date, very few examples of filament channel formation have ever
been observed, so the exact formation mechanism remains debatable.
Six recent publications present detailed case studies. Four cases
show the formation of filament channels through surface effects that
reconfigure pre-existing coronal fields, while in the latter two examples
flux emergence of horizontal flux ropes is deduced by the authors to play 
a critical role.
Thus from interpreting the observations there appear to be two
opposing views on how filament channels and filaments form. In this discussion we
will consider the key observational features and determine whether
the two views may be reconciled.

\subsubsection{Evidence of the Reconfiguration of Pre-Existing Coronal
Fields in the Formation of Filament Channels}
\label{sec:5.2.1}

Observations reported by \cite{Gaiz1997} and \cite{Gaiz2001} show that
surface motions acting on pre-existing coronal fields play a critical
role in the formation of filament channels and filaments. In the first
case, an Intermediate Filament (IF) forms over a short period of a few days,
while in the second a Quiescent Filament (QF) forms over
a period of months. In both cases the filaments form on PILs external
to any single bipole and in the classification scheme of
Section~\ref{sec:5.1.1} would be classed as External Bipolar
Regions Filaments. Although the two cases occur over very different
time and length scales there are a number of important similarities.

Both cases begin with the emergence of a significant amount of
magnetic flux in the form of an activity complex. Importantly however,
no filaments form during the process of flux emergence. In fact, for
the large scale QF the filament forms approximately 27 days after
major flux emergence subsides. In both cases a necessary condition for
the formation of the filament channels was flux convergence and cancellation
at a PIL between separate bipolar regions. Such convergence and
cancellation of flux was also shown to be important for filament
formation in the papers by \cite{Gaiz2002} and \cite{Martin1998}.
Finally and most importantly, in each case a significant amount of
magnetic shear was seen to build up in the activity complexes as
they emerged. The redistribution of this non-potential field through surface
motions towards the PIL produces a preferred direction of the coronal
field above the PIL and plays a critical role in the formation of the
filament channels, the necessary ingredient for filament formation
\citep{Mackay2003}.
\begin{figure}[t]
\centering\includegraphics[scale=0.45]{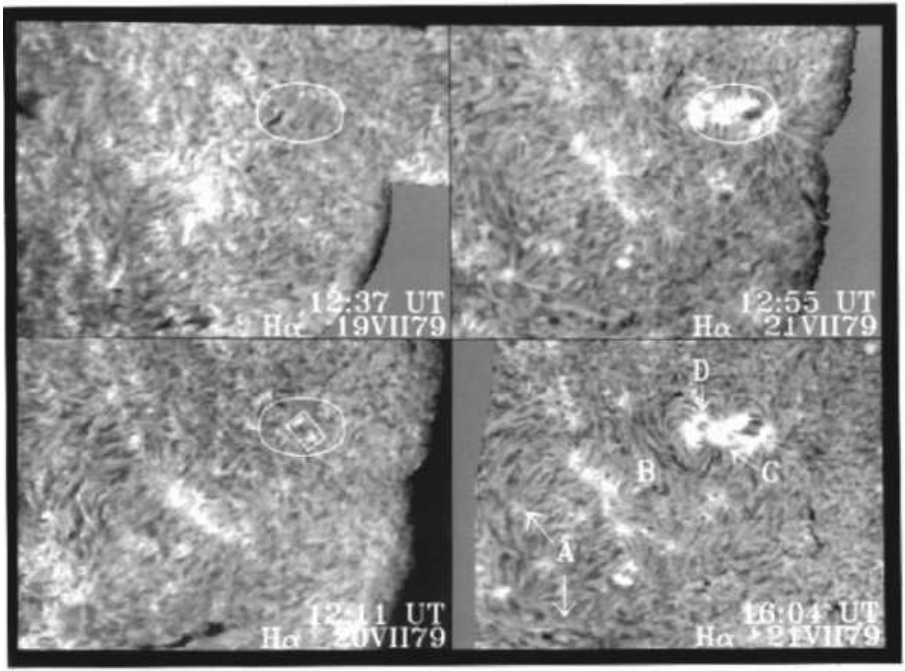}
\caption{H$\alpha$ images from \cite{Gaiz1997} (Figure 2) of the early
stages in the formation of an Intermediate Filament between an old
remnant region (bright plages A and B in bottom right images) and an
emerging activity complex (inside oval). The H$\alpha$ images
correspond to 19$^{th}$-21$^{st}$ July 1979 whereas the filament did
not form until the 25$^{th}$ July 1979.}
\label{fig:mackay_fig5}
\end{figure}

\begin{figure}[t]
\centering\includegraphics[scale=0.45]{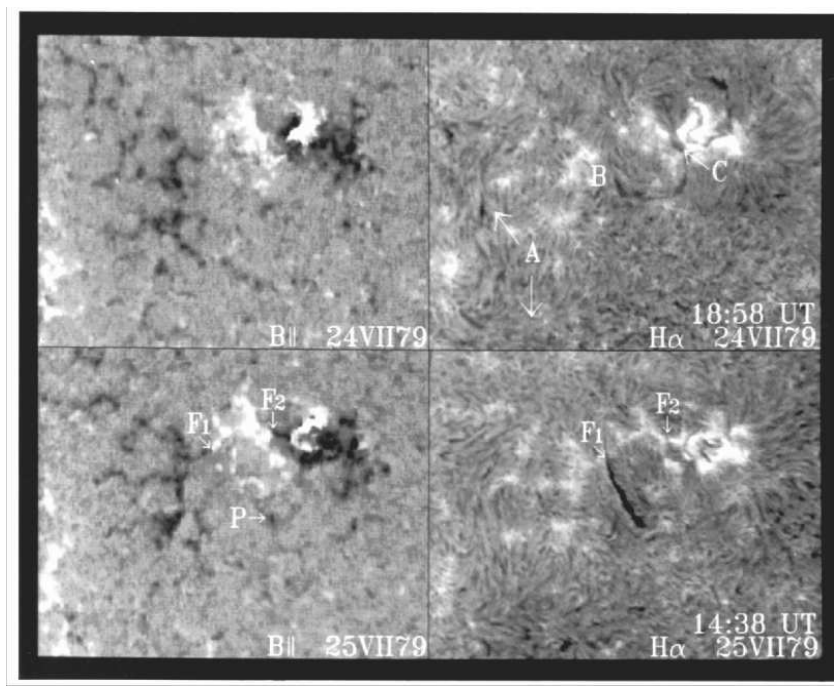}
\caption{H$\alpha$ image (right) and magnetogram (left) from
\citet{Gaiz1997} showing the final stages of the
formation of an Intermediate Filament on the 24$^{th}$ and 25$^{st}$
July 1979.  For the magnetogram images white represents positive and
black negative flux. The Intermediate Filament forms on the 25$^{th}$
July after flux convergence and cancellation occurs at F1.}
\label{fig:mackay_fig6}
\end{figure}

Figures~\ref{fig:mackay_fig5} and \ref{fig:mackay_fig6} illustrate the
main stages in the formation of a filament channel and IF over a period of five days
between the 20$^{th}$ -25$^{th}$ July 1979 \citep[see Figures 2 and 4
in][]{Gaiz1997}. The formation of this southern-hemisphere IF involves
the interaction of two distinct magnetic flux distributions, an old
remnant region (M$^c$Math 16159) and a new emerging region (M$^c$Math
16166). In the H$\alpha$ image of Figure~\ref{fig:mackay_fig5} (top
left) the bright North-South plage outlines the old remnant region;
the oval denotes the location where new magnetic flux will emerge on
the following day.  A key feature of this image is that the
chromosphere is free of any strong patterns of magnetic fields within
the oval.  New magnetic flux first emerges inside the rectangle within
the oval on the 20$^{th}$ July (bottom left) and then more strongly
between the 20$^{th}$ and 21$^{st}$ (top right). Over this time there
is very little change in the magnetic field of the old remnant
region. Magnetic field observations show that the activity complex
\citep{Gaiz1983, Ben2005} is made up of two or more sunspot
pairs. Significantly, no filament forms near or around the activity
complex during this period of rapid flux emergence.  The key
development in the formation of the filament channel occurs between
the two right hand panels taken just three hours apart on the
21$^{st}$ July (at the location denoted by B in lower right panel).
Over this period a band of co-aligned fibrils form at
the tail end of the new activity complex, between it and the old
remnant region. These co-aligned fibrils indicate a magnetic field at
this location with a dominant horizontal component, i.e. that a
filament channel has formed. According to a model by \cite{Mackay1997}
this pattern of co-aligned fibrils can only be explained by the
extended non-potential magnetic field of the activity complex in which
the field contains a large amount of positive helicity (correct sign
for the southern hemisphere).

No filament forms as magnetic flux continues to emerge within the
activity complex. The emergence ceases on the 23$^{rd}$ July,
thereafter the activity complex expands and flux begins to disperse
into the filament channel.
In Figure~\ref{fig:mackay_fig6} the distribution of magnetic
flux (left column) and corresponding H$\alpha$ images (right column)
can be seen for the 24$^{th}$ and 25$^{th}$ July.  In the magnetogram
images, white (black) represents positive (negative) flux.  After
major flux emergence ceases on the 23$^{rd}$ the trailing positive
polarity of the activity complex disperses or diffuses out. This
dispersion causes a convergence of flux between the old and new
regions. On 25$^{th}$ July, five days after the complex started to
emerge, cancellation of flux occurs at the point F1 (bottom
right). The filament forms after this cancellation and passes through
the location of flux cancellation. It was a stable structure which
survived for a full solar rotation and can clearly be seen to lie on a
PIL which is external to any one bipolar region. Subsequent modeling
by \cite{Mackay1997} showed that the resulting magnetic structure of
the filament could only be explained by the interaction of the
combined fields of both the old and new magnetic distributions. Both
fields were highly non-potential, again with a significant amount of
positive helicity which must have originated during the creation of
the new activity complex. It is clear from the observations that
reconfiguration of the previously emerged fields played a critical
role in the formation of the filament channel.

\cite{Gaiz2001} described a similar process of filament channel and
filament formation, but
this time for a QF which is nearly 1$R_\odot$ in length. The process
of formation once again begins with the emergence of new flux, but this
time in the form of two neighboring activity complexes in the northern
hemisphere.  Fibril alignment at the chromosphere once again shows
that the extended fields of the activity complexes are highly
non-potential with negative helicity. As with the previous case,
the filament only forms after major flux emergence ceases and the
activity complexes converge and partially cancel with one another.
In contrast to the filament in the previous case, which took 5 days
to form, for the large-scale case the QF takes nearly one full solar
rotation (27 days) to appear.

In both cases described above no stable filaments form
during the periods of the highest rates of flux emergence, and the
authors concluded that surface motions acting on pre-existing coronal
fields play a critical role in the formation of stable filaments
through the interaction of multiple bipoles. This result is consistent
with the classification of filaments given in Section \ref{sec:5.1.1}
where the majority of filaments are found to lie in magnetic
configurations that involve more than one bipole. A key role of these
surface motions is to redistribute the helicity which is seen to
emerge in the early stages to form the filament channel
\citep{Gaiz1997, Gaiz2001, Mackay2003, Mackay2005, Mackay2006}.
At a later stage a filament may appear when mass is deposited into the channel.
Convergence and cancellation of flux have also
been shown to be important for filament channel formation according to the
observations reported by \cite{Martin1998} who reviews several clear
examples, and \cite{Gaiz2002}, who shows that early in the solar
cycle a unipolar region of flux has to extend 180$^\circ$ around the
Sun to interact and cancel with an opposite polarity region before a
filament can form on that PIL. For these cases the redistribution of flux,
after emergence, is inferred to be a key process in the formation of the
filament channel. In the more recent paper by \citet{Gaiz2008},
three different arrangements of interacting nests are considered for
the formation of filament channels and filaments as flux disperses across 
the Sun. In each case channels and quiescent filaments form on the boundaries
of these activity nests.

\citet{Schmieder2004} studied the formation of a filament in the
complex center of a decaying active region formed out of smaller 
individual components, using multi-wavelength observations obtained
during a ``Joint Observing Programme'' between ground-based
instruments in the Canary Islands (the SVST and the MSDP on the VTT)
and the TRACE satellite. They followed the evolution of three
individual filament segments denoted F1, F2 and F3 over several days,
and found that F1 and F2 gently merged into a single structure, as
observed by a gradual filling in H$\alpha$ of the gap between them.
This merging was associated with mild EUV brightenings and with
small H$\alpha$ Doppler shifts at the merging point. While EUV
brightenings are a good indicator of magnetic reconnection, the flows
revealed that the merging first took place by dynamic exchanges
between the two progenitors, until they formed a single long stable
filament. Two days later segments F2 and F3 came into contact and
produced a confined flare, as evidenced by EUV post-flare loops
\citep[][]{Deng2002}. To determine the directions of the axial fields
in the three filament segments, \citet[][]{Schmieder2004} used the
chirality rules for chromospheric fibrils and magnetic field polarity,
the skew of the overlying coronal arcades, and the sense of twist in
neighboring sunspots. It was then confirmed that when two filaments
interact, magnetic reconnection takes place and leads to a merging
when their chiralities are of the same sign, but leads to a flare
when the chiralities are opposite \citep[also see][]
{Malherbe1989, Martin1998, Rust2001, vanb04, DeVore2005}. It was also
inferred that magnetic helicity must slowly accumulate prior to filament
merging, as seen by the rotation of a small twisted sunspot close to the
merging point. Finally, it was suggested that magnetic reconnection first
accelerates plasma between both progenitor filaments, and that it may
later result in a change of topology which can sustain stable plasma
all along the new filament.

More recent observations by \cite{Wang2007} have supported the work of
\citet{Gaiz1997,Gaiz2001}.  \cite{Wang2007} describe
examples of the formation of filament channels and filaments through
comparing BBSO H$\alpha$ images with MDI normal component
magnetograms. The authors describe how fibrils which are initially
normal to the PIL rotate to lie parallel to the PIL over a period of 1
to 2 days and in doing so form a filament channel.  Through studying
the evolution of the magnetic fields the authors deduce that flux
cancellation as a result of supergranular convection plays a key role
in the formation of the filament channels. They argue that this
cancellation process between opposite polarity elements removes the
normal component of the field but leaves the component parallel to the
PIL which builds up gradually to form the axial field of the filament
channel. In contrast to \citet{Gaiz1997,Gaiz2001} they do not observe
any significant helicity resulting from the emergence of the active
regions. For the two clearest examples of filament formation, 15th
January 2002 and October 7th 2002 (see \cite{Wang2007} Figures 2 and
5) the filaments form between bipolar regions of flux and background
fields and therefore would be classified as IF and Exterior Bipolar
Region Filaments. For the third case which is not so clear, the
filament partially lies on both the internal and also the external
portions of the PIL so would be an I/EBR filament or an AR filament.

On comparing the results of \cite{Gaiz1997} and \cite{Wang2007}, while
there are many similarities, there are also some differences in time scale.
The clearest is the time difference required to form the
filament channel. For \cite{Gaiz1997} the formation of the filament
channel occurs over a 3 hour period and is attributed to the extended
non-potential field of the activity complex containing a large amount
of helicity. Cancellation of flux could not produce such a strongly
sheared field over such a short period of time. In contrast,
\cite{Wang2007} do not report any strong patterns of fibrils
associated with helicity emerging in the active regions but rather
form the filament channel over a period of 1 to 2 days in a much
slower process of cancellation.  Therefore there appear to be two
complementary methods of forming a filament channel over different
time scales.

\subsubsection{Evidence of Emerging Horizontal Flux Tubes in Filament
Formation}
\label{sec:5.2.2}

It is clear from the above observations that surface effects play a
critical role in forming the studied IFs and QFs (which are long stable
structures). \cite{Lites1997} describe a different process for forming
short, unstable active-region filaments. In \cite{Lites1997} the emergence of a
$\delta$-spot is traced through vector magnetic field measurements
using Advanced Stokes Polarimetry. Magnetic field vectors along part
of the PIL within the emerging $\delta$-spot show a concave up or
dipped magnetic structure (see Figure 1 of \cite{Lites1997}; also see
\citet{lites05}). A small active region filament forms at this location.
The filament was however unstable with a lifetime of only 2 days.
\citet{Lites1997} suggest photospheric material is dragged up into the
corona through the levitation process, as a horizontal flux rope emerges
(\citet{Rust1994}, also see section \ref{sec:3}).

A more recent example of the effect of evolving magnetic fields on the
structure and stability of an active region filament is described by 
\citet{Okamoto2008} and \citet{Okamoto2009}. In two papers, the authors present
observations of a time series of vector magnetic fields taken by SOT
underneath a pre-existing filament. The vector magnetic field measurements show 
a PIL with dominant horizontal 
field along it. This horizontal field probably represents that of the 
filament channel of the pre-existing filament. Over a period of 1.5 days 
the horizontal field vector changes from normal to inverse polarity and 
a dominant blue shift is observed. During this period the filament alters its 
appearance from a single structure, to a fragmented one and back again. Before 
returning to a single structure, brightenings are observed along the filament
fragments in the Ca II H line.

From the observations the authors deduce two possible scenarios. 
In the first scenario they interprete the observations in terms of
an emerging horizontal flux rope which fully 
emerges into the corona and occupies the position of the pre-existing filament. 
With this scenario the mass of the prominence originates from below the 
photosphere. The second scenario interprets the brightenings in Ca II H as evidence 
for reconnection between the pre-existing filament and a new flux rope that emerges
free of mass. The reconnection then produces a single structure along the PIL. A 
difficulty with both scenarios is that no simulations of magnetic flux 
emergence have been able to emerge a horizontal flux rope through the photosphere. 
  
In contrast too that put forward by the authors, a third possibility also exists.
This fits the theoretical models discussed 
in Section~\ref{sec:3.2}. As the top part of a flux rope emerges, a likely outcome is 
the emergence of sheared arcades. A coronal flux rope may then be formed out of
these arcades through the process of reconnection. This reconnection may lift cool 
material into the corona, as has been discussed in Section~\ref{sec:3}. If the axial 
component of the emerging arcade lies in the same direction as that of the pre-existing 
filament channel, the new and old flux systems may join to produce a single structure 
(see Section~\ref{sec:5.4}). To consider which, if any of these three scenarios are correct, 
new high resolution magnetic field observations at multiple levels in the solar atmosphere 
(e.g. photosphere, chromosphere and corona) are required. 

\subsubsection{Summary of Observations}
\label{sec:5.2.3}

The observations described in Section~\ref{sec:5.2.1} and \ref{sec:5.2.2} 
provide evidence for filament formation
arising from surface motions that reconfigure already existing coronal
fields or, emerging flux tubes. So can the two methods
be reconciled? The important distinction between these cases is the
type and location of filaments formed in each case. For the first
four cases surface motions play an important role in forming long
stable Quiescent or Intermediate filaments which are External Bipolar Region
Filaments, the dominant type of
large-scale filament found at all latitudes on the Sun.  In contrast,
flux tubes emerging in a $\delta$-spot forms an Active Region or Internal Bipolar
Region Filament which is unstable, lasting merely two days.

While it is difficult to draw general conclusions from just six
specific observations, they indicate that two different mechanisms might
form filaments in different magnetic environments on the Sun.
Thus large stable filaments of the IF and QF type (External
or Diffuse Bipolar Region) may require surface motions to gradually
reconfigure pre-existing coronal fields, while small, short-lived
ARFs (Internal Bipolar Region) may form due to flux emergence. To
determine whether different mechanisms do produce different types of
filaments at different locations on the Sun, the formation of
filaments over a wide range of latitudes needs to be considered in
detail. Observational programs required to do this will be discussed
in Section~\ref{sec:5.7}. 

\begin{table}[!t]
\caption{Surface Models of Filament Formation}
\label{table:mackay_table1}
\smallskip
{\small
\begin{tabular}{ll}
\hline
\hline
Single Bipole & Multiple Bipoles  \\
\noalign{\smallskip}
\hline
\noalign{\smallskip}   
\cite{Balle1989}$^{1,3,4,10}$      & \cite{Kuperus1996}$^{1,3,4}$  \\ 
\cite{DeVore2000}$^{1,4}$          & \cite{Kuijpers1997}$^{3,4,8,10}$ \\
                                   & \cite{Mackay1998}$^{3,4,6,8,10}$ \\
                                   & \cite{Galsgaard1999}$^{3,4}$  \\ 
                                   & \cite{Balle2000}$^{1,4,10}$   \\  
                                   & \cite{Martens2001}$^{3,4,10}$ \\
                                   & \cite{Lionello2002}$^{8,10}$  \\
                                   & \cite{DeVore2005}$^{1,3,4}$   \\
                                   & \cite{Mackay2005}$^{1,4,8,10}$ \\
                                   & \cite{Welsch2005}$^{3,4,8,10}$ \\
                                   & \cite{Litvinenko2005}$^{3,4,8,10}$ \\ 
                                   & \cite{Yeates2008a}$^{1,4,8,10}$ \\
\hline
\end{tabular}
}
\end{table}

\begin{table}[!t]
\caption{Sub-Surface Models of Filament Formation}
\label{table:mackay_table3}
\smallskip
{\small
\begin{tabular}{ll}
\hline
\hline
Single Bipole & Multiple Bipoles  \\
\noalign{\smallskip}
\hline
\noalign{\smallskip}   
\cite{Low1994}$^7$                 &  \cite{Balle1990}$^{2,3,4,7}$  \\ 
\cite{Rust1994}$^{7,9}$            &  \cite{Priest1996}$^{2,3,4,6}$  \\
\cite{Gibson2004}$^{7,9}$          &  \cite{Oliver1999}$^{2,3,4,6}$  \\
\cite{Low1995}$^{7,9}$             &                                 \\ 
\cite{Fan2004}$^{7,9}$             &                                 \\ 
\cite{Fan2006}$^{7,9}$             &                                 \\ 
\cite{Gibson2006}$^{7,9}$          &                                 \\ 
\cite{mag06}$^{7,9}$               &                                 \\ 
\cite{mag08}$^{7,9}$               &                                 \\  
\citep{fan09}$6,9$                 &                                 \\
\hline
\end{tabular}
}
\end{table}

\subsection{Theoretical Models of Filament Formation}
\label{sec:5.3}

Over the years many models have been constructed, each employing a
variety of mechanisms in order to describe the formation of
filaments. These models vary from descriptive papers to full numerical
MHD simulations and consider two main problems. First, how to obtain
the correct dipped magnetic field configuration with dominant axial
magnetic field that follows the hemispheric pattern, and secondly, the
origin of the dense plasma. While the second question relates more to
thermodynamics (\cite{Karpen2001}, section \ref{sec:3}), this section
which relates filaments to their underlying magnetic polarities is
relevant to the first group of models. It is widely accepted that 
magnetic flux ropes are a suitable configuration to represent solar
filaments; the main area of debate is how exactly these flux ropes may
form. Therefore, the various models in that group may be broadly split
into two distinct sub-groups: those employing surface effects to
reconfigure coronal fields (Table~\ref{table:mackay_table1}) and those
employing subsurface effects (Table~\ref{table:mackay_table3}). This
split naturally arises from the discussion of the observations in
Section~\ref{sec:5.2}. At the present time only those employing
surface mechanisms can be directly compared to observations.  In these
tables the surface/subsurface models have also been subdivided into
those acting in single or multiple bipolar configurations in account
of the observations discussed in Section~\ref{sec:5.1.1}. The list 
contains models which consider the physical processes and mechanisms that 
may produce the 3D magnetic structure of filaments. It should only be 
regarded as representative and not exhaustive. Where the same authors 
publish multiple papers based on the same mechanism, generally only the 
first paper outlining the process is listed. Due to this, readers are
recommend to search for other such papers in the literature. For each 
of the entries in the table the numbers
attached correspond to the various mechanisms that the models employ,
as listed in Table~\ref{table:mackay_table2}. From
Table~\ref{table:mackay_table1} it is clear that surface models rely
on a variety of mechanisms combined together, while subsurface models
generally rely on the emergence of twisted flux ropes where the
filament forms in the dips of the flux rope or U-loop.

Surface mechanisms include: differential rotation; shear flows along a
PIL (differential rotation is just a weak shear flow); and converging
flows onto a PIL. Recent helioseismic observations by
\cite{Hindman2006} show that underneath a well developed filament
strong shear flows may be observed. However this was after the
filament had formed and not during the formation process. Photospheric
converging or shearing flows may be detected by local correlation tracking
\citep[cf.][]{Magara1999, Rondi2007, Roudier2008}. For some
surface models diffusion of flux towards a PIL with subsequent
cancellation plays the role of the converging flow.  To produce an
axial magnetic field consistent with observations, these flows
generally have to occur in a specific order. In contrast to these
surface motions, subsurface shear flows have also been employed. In
both sets of models magnetic reconnection is generally required to
reconfigure the fields; the reconnection may occur either above or
below the surface.

\begin{table}[!t]
\caption{Mechanisms of Filament Formation}
\label{table:mackay_table2}
\smallskip
{\small
\begin{tabular}{llll}
\hline
\multicolumn{2}{c}{\bf{Surface Mechanisms}} &
\multicolumn{2}{c}{\bf{Subsurface Mechanisms}} \\
\noalign{\smallskip}
\hline
\noalign{\smallskip}
\multicolumn{2}{l}{(1) Differential Rotation (shear flows)} &
\multicolumn{2}{l}{(2) Subsurface Motions}   \\
\multicolumn{2}{l}{(3) Converging Flows                   } &  \\
\multicolumn{2}{l}{(4) Magnetic Reconnection (atmosphere) } &
\multicolumn{2}{l}{(5) Magnetic Reconnection (subsurface)}     \\
\multicolumn{2}{l}{(6) Flux Emergence (bipoles)           } &
\multicolumn{2}{l}{(7) Flux Emergence (U-loops)}  \\
\multicolumn{2}{l}{(8) Magnetic Helicity                  } &
\multicolumn{2}{l}{(9) Magnetic Helicity}     \\
\multicolumn{2}{l}{(10) Flux Cancellation/Diffusion       } &  \\
\hline
\end{tabular}
}
\end{table}

Another feature common to both sets of models is flux emergence, but
it is used in very different ways. For surface models, magnetic
bipoles which emerge either untwisted or twisted are advected across
the solar surface and reconfigured with other pre-existing coronal
fields as discussed in the observations of
Section~\ref{sec:5.2.1}. A key element in recent papers describing
filament formation is that these bipoles are non-potential and include
an initial magnetic helicity \citep{Mackay2005,Yeates2008a}.  In
contrast, flux emergence for subsurface models is presumed to occur in
the form of twisted U-loops (Section~\ref{sec:5.2.2}).

Whilst it is impractical to describe each of the models listed in
Tables~\ref{table:mackay_table1} and~\ref{table:mackay_table3} in
detail, key elements may be considered from a few selected cases. 
The cases chosen are picked solely for illustrative purposes. The
key feature of any sub-surface model is described in the papers by
\cite{Low1994} and \cite{Rust1994}. For these models a filament is
formed by a horizontal twisted magnetic flux tube in the convective
layer of the Sun. Due to magnetic buoyancy the tube rises through the
convective layer and emerges into the photosphere and corona, dragging
cool dense material with it, to produce the mass of the prominence. Not
every part of the tube has to rise at the same time; subsequent rising
parts could explain canceling magnetic features
\cite[see Figure 13 in][]{Rust1994}. Such a feature has been
considered in the numerical simulations of \citet{Gibson2006} and
\citet{mag08}.  In \citet{mag08} the flux rope is forced to rise into
the solar atmosphere by imposed velocity fields. In other flux emergence
simulations where the authors use only buoyancy and magnetic buoyancy
instabilities, it is found that the axis of the flux rope does not
rise through the photosphere \citep[][]{Moreno2004, arch04, arch08,
Murray2006, gals07}. Although the axis and U-loops of the emerging 
tube do not rise to coronal heights, the process of flux emergence may still 
produce a coronal flux rope with dips. A flux rope may form through the 
reconnection of emerged sheared field lines that lie above the emerging
tubes axis \citep{man04,mag06,arch08b,fan09}.

\begin{figure}[!t]
\centering\includegraphics[scale=0.42]{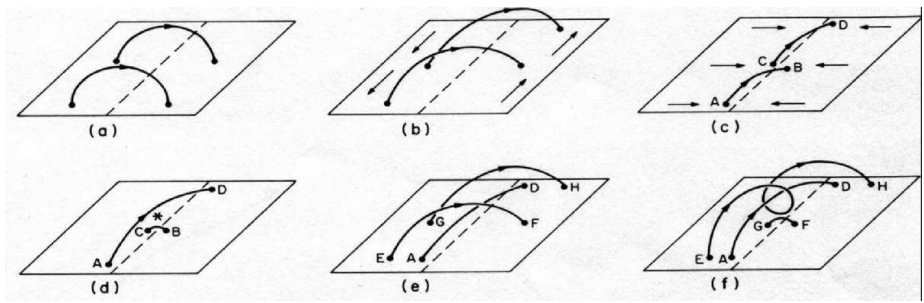}
\caption{Example of the formation of a filament's axial magnetic field
through shearing motions, convergence, and reconnection as put forward
by \cite{Balle1989}.}
\label{fig:mackay_fig7}
\end{figure}

In contrast, one of the first surface models, by \citet{Balle1989},
considers shearing motions acting on a coronal arcade in a bipolar
configuration. The footpoints of the arcade are sheared in such a way
that their separation increases and an axial field component is
produced along the PIL (see Figures~\ref{fig:mackay_fig7}a and
\ref{fig:mackay_fig7}b). In principle this shear could be supplied
by solar differential rotation or by other shear flows on the Sun.
Next convergence, or diffusion of the flux towards the PIL, brings the
foot points together where they may reconnect to produce, a long axial
field line along the PIL and also a small loop which submerges through
the surface (Figures~\ref{fig:mackay_fig7}c and
\ref{fig:mackay_fig7}d). Subsequent repetition of this process creates
dipped magnetic field lines consistent with the topology required for
filaments (Figures~\ref{fig:mackay_fig7}e and \ref{fig:mackay_fig7}f).

In an extension to the \cite{Balle1989} model, \cite{Martens2001} put
forward a ``head-to-tail" linkage model for the formation of filaments
in a multiple bipolar configuration.  A key element of their model was
once again flux convergence and cancellation, which acts as the driver
for reconnection between initially unconnected magnetic sources 
\citep[also see][]{Kuperus1996, Kuijpers1997}). In this scenario
\cite{Martens2001} describe how the filament channel and filament may
be produced by the interaction of two bipoles, one older and more
diffuse lying at a slightly higher latitude. As long as these sources
satisfy Hale's Polarity Law and Joy's Law, then convergence and
cancellation as a consequence of differential rotation, of the
following polarity of the lower latitude bipole and lead polarity of
the high latitude bipole, could result in a strongly sheared field
line along the PIL. As with \cite{Balle1989} successive repetitions of
this process would build up helical field structures. In addition, if
multiple bipoles are involved over a range of latitudes large
filaments extending over a full solar radius could be produced. While
Martens and Zwaan considered the interaction of multiple bipoles in a
conceptual model, \cite{Mackay2000,Mackay2001} carried out numerical
simulations of a similar process.

\cite{Galsgaard1999} consider a similar scenario with 3D numerical MHD
simulations without the processes of flux cancellation and submergence.
The interaction of two bipolar regions of flux, one which is older and
more spread out than the other, is considered. The numerical
simulations are based on the magnetic configurations discussed by
\cite{Gaiz1997}. In the model the two bipoles are initially connected
to one another and the authors demonstrate how a current sheet may
form in the coronal volume between the two bipoles as a result of flux
convergence but no cancellation or submergence. Subsequent reconnections resulting
from the convergence, then lift cool dense plasma over a number of
pressure scale heights where it is able to form a region of high
density plasma overlying the PIL. As part of this reconnection process
dipped and sometimes helical field lines are formed with an upward
tension force. Although the authors carried out the simulation with a
simple isothermal atmosphere it remains one of the few full MHD
simulations to consider both the origin of mass and of the shear in the 
magnetic field.

An alternative method of forming a similar magnetic structure was
proposed by \cite{DeVore2000} using a single bipolar configuration
\cite[also see][]{Antiochos1994}. In this model, a bipolar magnetic
field distribution is subjected to a strong shearing motion parallel
to the PIL, however, no converging flow is applied. Once the
footpoints of the field lines are sheared a distance comparable to the
bipole width, an untwisted dipped magnetic configuration forms. The
authors show that through further shearing of the dipped field lines
the initially untwisted field may form a helical structure similar to
that of \cite{Balle1989} through a two stage reconnection process.
Therefore, in contrast to \cite{Balle1989} and \cite{Martens2001},
\cite{DeVore2000} do not rely on convergence and cancellation of flux
to produce the helical field.

From the discussion above it is clear that a wide range of theoretical 
models exist to explain the 3D magnetic structure of solar filaments.
These models employ a variety of mechanisms. As will be discussed in
Section~\ref{sec:5.6} at the present time none of these models may be ruled 
out. However, by combining the observations discussed in Sections~\ref{sec:5.1}
and ~\ref{sec:5.2} it may be argued that some models 
are more relevant than others for the formation of large stable filaments
(Quiescent and Intermediate) compared to Active Region filaments. A full discussion 
along with the presented hypothesis will be carried out in  Section~\ref{sec:5.6}.

\subsection{Models of Filament Merging}
\label{sec:5.4}

\begin{figure}
\includegraphics[width=4.5in]{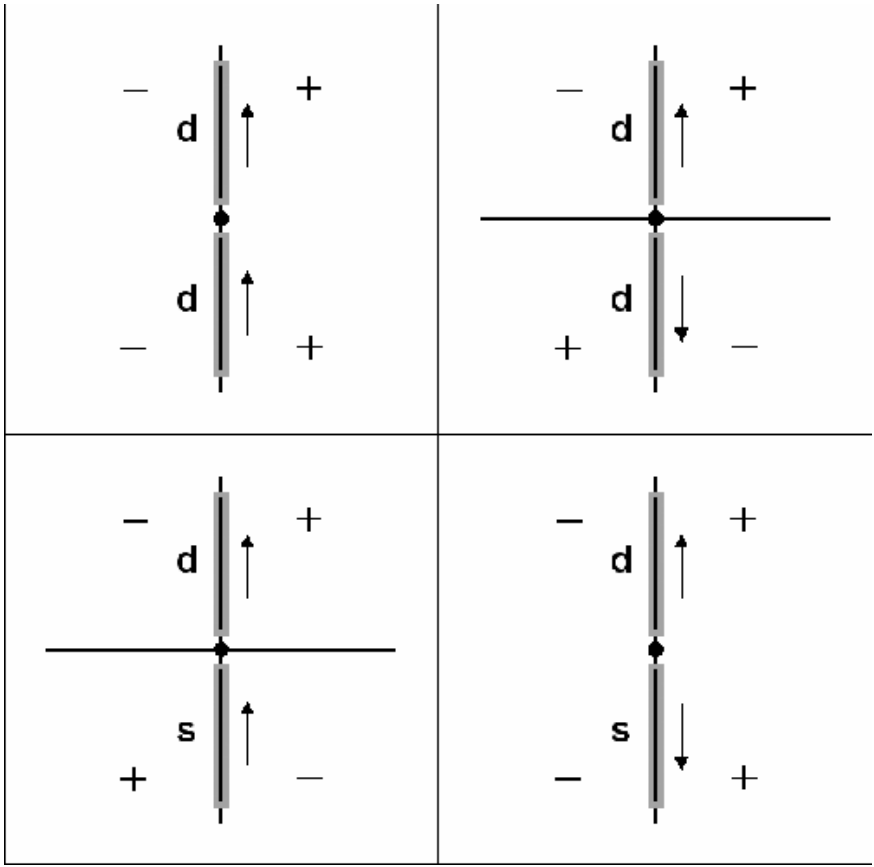}
\caption{Schematic diagram of the four possible configurations of
interacting prominences: identical (top) or opposite (bottom)
chiralities, and aligned (left) or opposed (right) axial magnetic
fields. Black lines are the polarity inversion lines of the vertical
field, whose direction is denoted by $+$ (upward) and $-$ (downward);
shaded gray rectangles are the prominences, whose chiralities are
indicated by {\bf d} (dextral) and {\bf s} (sinistral) and whose
axial field directions are shown by the arrows; and filled black
circles mark the prominence interaction regions.}
\label{fig:devore_fig1}
\end{figure}

Numerical MHD simulations of the formation and interaction between
filament segments were conducted by \citet{DeVore2005}. The footpoint
motions of the sheared arcade model (section \ref{sec:2.2.2}) were
applied to two adjacent, initially current-free, magnetic dipoles. As
shown in Fig.~\ref{fig:devore_fig1}, four possible combinations of
chiralities (identical or opposite) and axial magnetic fields (aligned
or opposed) between the participating filaments were considered. These
four simulations exhibited substantially different degrees of linkage.
 
When a single polarity inversion line is shared by the two bipoles,
then identical chiralities necessarily imply aligned axial fields
(Fig.~\ref{fig:devore_fig1}, left). In this case, magnetic
reconnection between field lines from both bipoles links the two
initial prominence segments. Both acoustic and Alfv\'en waves
propagate along these newly reconnected field lines, and should be
capable of driving existing plasma condensations from one progenitor
to another. As the shear increases, the volume between the original
segments becomes filled with reconnected field lines, so that they
gradually merge into a single filament. This multi-step merging
mechanism couples photospheric shear, coronal reconnection, and
relaxation to form a longer structure containing numerous dipped field
lines where plasma can collect most easily \citep{Aulanier2006}.
Furthermore, the case of identical chiralities and aligned axial
fields successfully reproduces the observations of filament merging
\citep[e.g.,][]{Martin1994, Rust2001, Schmieder2004, vanb04}. A second
model configuration (Fig.~\ref{fig:devore_fig1}, upper right) also
contains a shared PIL, but opposite helicities and axial fields
between the two segments. No merging ensues because little
reconnection occurs between the filament segments, as is consistent
with observations \citep[e.g.,][] {Martin1994, Rust2001, Deng2002,
Schmieder2004}.

When the initial topology instead is quadrupolar, so that the system
contains two orthogonal PILs (Fig.~\ref{fig:devore_fig1}, lower
panels), then the converse relation holds between chirality and
axial-field alignment. Reconnections that form linking field lines now
occur only between filament segments with opposite chiralities
(Fig.~\ref{fig:devore_fig1}, lower right), while reconnection between
same-chirality segments only results in footpoint exchanges
(Fig.~\ref{fig:devore_fig1}, lower left).  The results for quadrupolar
topologies have not yet been verified with solar data because
multipolar filament interactions are rarely observed. However, these
key predictions present an important objective for future
observational campaigns.

\subsection{Origin of the Hemispheric Pattern of Filaments}
\label{sec:5.5}

Any model which tries to explain the origin of a filament's magnetic
field must also explain why this magnetic field exhibits a hemispheric
pattern.  The first attempt through detailed numerical simulations was
carried out by \cite{Balle1998} who considered whether the surface
flux transport effects of differential rotation, meridional flows and
supergranular diffusion could in fact create the observed axial fields
in filaments. By using observed magnetic flux distributions and
initial coronal fields which were potential, they simulated the
evolution of both the photospheric and coronal fields. They found that
the above-mentioned surface effects when acting on potential fields
create approximately equal numbers of dextral and sinistral channels
in each hemisphere, in contradiction with observations. However, these
authors did not take into account the force balance of the coronal
plasma, and only considered statistical relationships between filament
chirality and latitude.

In the more recent study of \cite{Mackay2005} the authors
re-consider the origin of the hemispheric pattern through combined
flux transport and magneto-frictional relaxation simulations. In these
simulations as the flux transport effects shear up the photospheric
field, the coronal field responds to these motions by relaxing to a
non-linear force-free field equilibrium. In the simulations the
authors consider an idealized setup of two initially isolated
interacting bipoles. They perform a parameter study to determine the
type of chirality formed along the PIL between the two bipoles, as the
bipoles initial tilt angle (Joy's law) and helicity are
varied. Therefore in contrast to the study of \cite{Balle1998} they
did not use initial potential fields. Through the simulations the
authors demonstrate that surface diffusion can play a key role in
canceling flux between the bipoles and building up an axial field
through transporting helicity and sheared fields from the inner parts
of the bipoles to the outer edges. The authors demonstrate that the
hemispheric pattern of filaments may be explained through the
observational properties of newly emerging bipoles such as, (1) their
dominant tilt angles (-10:30$^\circ$, \cite{Wang1989}), and (2) the
dominant helicity that they emerge with in each hemisphere
\citep{Pevtsov1995}. In addition to this, a key feature of the
simulations was that for the first time the occurrence of exceptions
to the hemispheric pattern as a result of large positive bipole tilt
angles and minority helicity could be quantified.

\begin{figure}[!t]
\centering\includegraphics[scale=0.5]{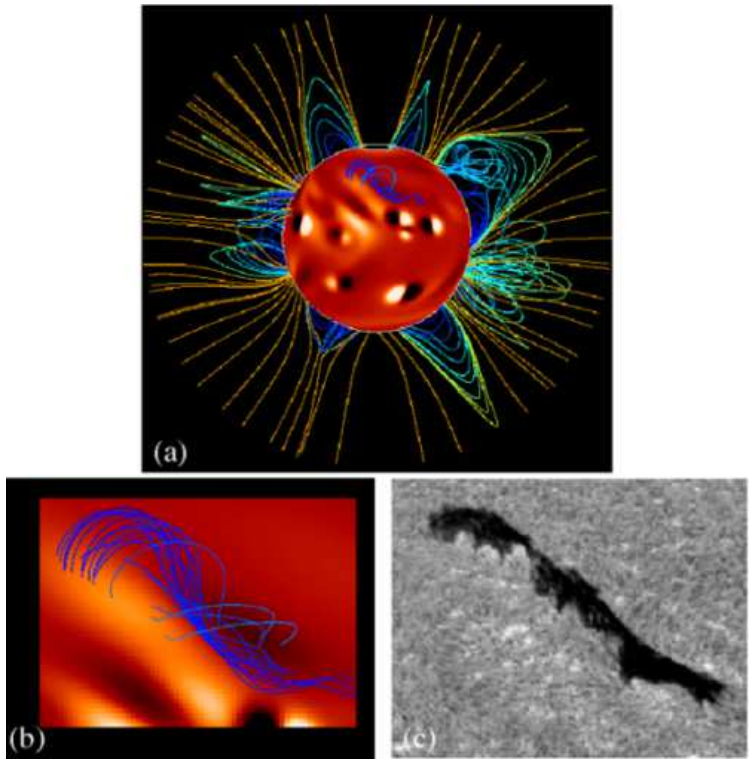}
\caption{Example of the comparison of theory and observations
performed by \cite{Yeates2008a}.  (a) Example of the magnetic field
distribution in the global simulation after 109 days of evolution,
showing highly twisted flux ropes, weakly sheared arcades and near
potential open fields. On the grey-scale image white/black represents
positive/negative flux. (b) Close up view of a dextral flux rope lying
above a PIL within the simulation. (c) BBSO H$\alpha$ image of the
dextral filament observed at this location. }
\label{fig:mackay_extra2}
\end{figure}

The results of \cite{Mackay2005} have been tested through a direct
comparison between theory and observations by
\cite{Yeates2007,Yeates2008a}.  In this comparison
\cite{Yeates2007,Yeates2008a} develop a new technique to model the
long term, global evolution of the Suns magnetic field based on actual
magnetogram observations. To carry out the comparison
\citet{Yeates2007, Yeates2008a} first consider H$\alpha$ observations
from BBSO over a 6 month period and determine the location and
chirality of 109 filaments \citep{Yeates2007} relative to the
underlying magnetic flux. It should be noted that all of the filaments
in the study lie below 65$^\circ$ latitude and no polar crown filaments were
included since their chirality could not be determined. To determine the chirality 
for such filaments direct magnetic field measurements would be required.  In the 
second stage they use combined
magnetic flux transport and magneto-frictional relaxation simulations,
where the simulations are based on actual photospheric magnetic
distributions found on the sun. They show that by including the
process of flux emergence in the flux transport simulations they can
reproduce to a high accuracy KP synoptic magnetograms from one
Carrington Rotation to the next. Over the 6 month simulation to
maintain the accuracy, 119 bipoles were emerged with their properties
determined from observations.  When the bipoles were emerged they were
initially isolated from the surrounding fields and could contain either
positive or negative helicity. Their subsequent interaction with
neighboring field regions and transport across the solar surface could
then be followed. A key difference of these simulations from previous
studies, was that they were run for the whole six month period without
ever resetting the surface field back to that found in observations or
the coronal field to potential. Therefore the simulations were able to
consider long term helicity transport across the solar surface from
low to high latitudes.

Using this technique \cite{Yeates2008a} carried out a direct
one-to-one comparison of the chirality produced by the model with the
observed chirality of the filaments at the exact location that the
filaments were observed. An example of this can be seen in
Figure~\ref{fig:mackay_extra2}.  In Figure~\ref{fig:mackay_extra2}a
the global field distribution can be seen after 108 days of
evolution. The photospheric field distribution is given by the
grey-scale image (white is positive flux, black is negative flux),
while the lines denote the field lines of the non-linear force-free
coronal field.

The coronal field can be seen to be made up of highly twisted flux
ropes, slightly sheared coronal arcades and near potential open field
lines. A zoomed in area of the central portion of this image can be
seen in Figure~\ref{fig:mackay_extra2}b. In this figure a simulated
flux rope structure can be seen where the axial field is of dextral
chirality. For comparison the H$\alpha$ filament that formed at this
location can be seen in Figure~\ref{fig:mackay_extra2}c.
Through studying the barbs and applying a
statistical technique, the filament was determined to be of dextral
chirality so the chirality formed in the simulation matches that of
the filament.

Through varying the sign and amount of helicity emerging within the
bipoles, \cite{Yeates2008a} (see their Figure 5b) show that by
emerging dominantly negative helicity in the northern hemisphere and
positive in the southern, a 96$\%$ agreement can be found between the
observations and simulations where the agreement is equally good for
minority chirality filaments as well as for dominant chirality
filaments.  A key feature of the simulations is that a better
agreement between the observations and simulations is found the longer
the simulations are run.  This indicates that the Sun has a long term
memory of the transport of helicity from low to high latitudes. The
reason for this high agreement is described in the paper of
\cite{Yeates2009} where seven different mechanisms are involved in
producing the observed chiralities. The results demonstrate that the
combined effects of differential rotation, meridional flow,
supergranular diffusion and emerging twisted active regions are
sufficient to produce the observed hemispheric pattern of filaments.

\subsection{Discussion of Filament Classification Schemes,
Observations of Formation and Theoretical Models}
\label{sec:5.6}

So far we have discussed a number of properties of
solar filaments.  These properties have ranged from observations and
locations of filament formation, to the wide variety of theoretical
models used to explain them. In this sub-section we turn our
attention to tying all of these observations together, by forming a unifying hypothesis,
to quantify where and at what locations the mechanisms and models discussed in 
Section~\ref{sec:5.3} are appropriate. We hope that this hypothesis will stimulate 
new observational studies to test it. We will first discuss the formation
of large-stable filaments, namely those of the Intermediate and Quiescent filament
type. After this we will consider Active Region filaments. A distinction is made
as present observational evidence suggests that different mechanisms may apply.

\subsubsection{Intermediate and Quiescent Filament Formation}

From the observations of large-scale stable filaments discussed in
Section~\ref{sec:5.1.1} it can be seen that the majority of IF and QF
(92\%) form in magnetic configurations involving multiple
bipole interactions. Very few (7\%) form in single
bipolar configurations along the internal PIL of the bipole.
From this it is clear that it is rare to find either IF or QF in such single
bipole configurations. Therefore, while none of the models listed in
Table~\ref{table:mackay_table1} or \ref{table:mackay_table3} can be
ruled out, it is clear that those involving multiple bipole interactions are 
the most appropriate.

The question now turns to whether the IF and QF are formed
due to surface motions acting on pre-existing coronal fields as
\cite{Gaiz1997}, \cite{Gaiz2001} and \cite{Wang2007} argue
(Section~\ref{sec:5.2.1}) or whether they are due to sub-surface
processes as argued by \cite{Lites1997} and \cite{Okamoto2009}.

The answer to this may be alluded from the observations of
\cite{Mackay2008} who showed (see their Figure 4) that only External
Bipolar Region Filaments (EBR), those forming between two neighboring
bipoles show any form of solar cycle dependence. Their numbers
were found to increase in phase with the solar cycle, peaking around solar
maximum and with more present in the declining phase of the cycle than in the
rising phase. In contrast, Internal Bipolar Region Filaments (IBR) those
forming in a single bipole, clearly showed no solar cycle dependence, with their 
number remaining at a low constant value throughout
the cycle. If flux emergence in the form of flux ropes as used by
subsurface models is key to the formation of IF and QF, then one
would expect that IBR Filaments should show a strong solar cycle
dependence. In addition they should be the dominant type. As more bipoles 
emerge on the Sun there is a much greater chance of having emerging flux ropes, 
hence more Internal Bipolar Region Filaments. As they are not the dominant type
and exhibit no solar cycle dependence, another process other than flux rope emergence
must be acting to form IBR filaments. An alternative explanation for
the formation of IBR filaments, which from the observations discussed in this review
generally occurs outside activity complexes, is that they may result
from strong surface shearing motions or helicity injection and flux
cancellation. Evidence for such shearing
motions, at least in the later development of a filament, has recently
been published by \cite{Hindman2006}. This fits in with the single
bipole model of \cite{DeVore2000} where shear flows play a key role in
the formation of the filament. In contrast, the alternate process of injection of 
helicity followed by cancellation has been described by \citet{Wang2007}.

If flux rope emergence is unlikely to form IBR filaments, then it is
extremely unlikely that such a process could apply to External Bipolar
Region Filaments (EBR), the dominant type which tend to form between distinct 
bipoles after flux emergence has ended (see Section~\ref{sec:5.2.1}).
Again, another process must be acting, one that is closely related to
the amount of magnetic flux on the Sun.

The obvious choice deduced from the observations of \cite{Gaiz1997},
\cite{Gaiz2001} and \cite{Wang2007} is the convergence between
individual bipoles resulting in flux cancellation and reconnection,
since the rates of convergence increase due to the
increased rate of flux emergence during periods of high activity.
During these periods of high activity the most widespread source of
convergent flows would be the natural expansion of all bipolar pairs
as soon as they emerge. If these processes are key in explaining the
formation of filaments this would lead to an increase in the number of
filaments forming between individual magnetic bipoles as activity
increases. In addition, one expects more EBR Filaments after cycle
maximum than after cycle minimum; after maximum there is still a
significant amount of magnetic flux on the Sun so that convergence and
cancellation or magnetic reconnection can still take place.

Therefore, it may be argued that convergence leading to subsequent
cancellation and reconnection (i.e., items 3, 4 and 10 in
Table~\ref{table:mackay_table2}) are the mechanisms that result
in the formation of the majority of large stable filaments found on
the Sun, and that flux rope emergence does not play a major role for these
filaments.
The models in Table~\ref{table:mackay_table1} which include these
mechanisms appear to be the most appropriate. At the present time no further
distinction can be made between these models. While this argument is
put forward for EBR Filaments, it also applies to Diffuse Bipolar
Region Filaments (DBR) because the flux patterns in which they form
are a natural consequence of flux convergence and cancellation
occurring over long periods of time. 

Finally, for Internal/External Bipolar Region Filaments again
convergence and cancellation/reconnection may apply but this
time strong shear flows or helicity injection may again be applicable. While the above
discussion applies to Intermediate and Quiescent filaments, in the next sub-section
we consider Active Region filaments. 

\subsubsection{Active Region Filaments}

We now turn our attention to Active Region filaments to consider their
formation mechanism. Observations by \citet{Lites1997} and \citet{Okamoto2009}
suggest that small-scale unstable active region filaments may be formed as the result
of flux rope emergence dragging cool dense photospheric plasma into the corona.
While this remains a possibility, most numerical simulations of emerging flux ropes fail to
lift the axis and cool material of the original flux tube into the corona. Therefore is 
remains unclear whether such a process may occur. In contrast, many 
authors have shown that during the process of flux emergence, after the top of the flux 
rope has emerged, magnetic reconnection or helicity injection  
\citep{man04,mag06,arch08b,fan09} may reconfigure the emerged coronal arcade to produce
a secondary coronal flux rope. During the formation of the secondary flux rope the reconnection
may then lift cool dense material to coronal heights. Therefore while emerging flux appears
to be important for the formation of active region filaments a key element may still
be atmospheric reconnection of pre-emerged fields. To resolve this issue many new observational studies are
required.

From the argument presented above it appears that different formation
mechanisms may apply to different types of filaments. Quiescent filaments
and Intermediate filaments which mainly fall into the Exterior and Diffuse bipolar
region types rely on surface effects acting on coronal fields. In contrast for
active region filaments a strong possibility is the emergence of flux ropes
or the formation of flux ropes during emergence as a result of coronal reconnection. Therefore it is useful
to distinguish between IF and QF compared to ARF as they may have a different formation
mechanism.

\subsubsection{Future Observation of Filament Channels}
\label{sec:5.7}

The formation, structure, and evolution of solar filaments is an
important part of our understanding of coronal physics and the
behavior of magnetic fields as they are transported across the solar
surface. New observational results show that surface motions acting on
non-potential magnetic fields may play an integral part in the
formation of large-scale stable filaments while flux rope emergence
may play a role for small, unstable active region filaments.  In
addition the large-scale hemispheric pattern of filaments may just be
the result of surface effects redistributing the helicities of new
emerging bipoles across the solar surface from low- to high-
latitudes.

A better understanding of the formation of prominences requires
multi-wavelength observations of prominences situated over a wide
range of latitudes, from the active region belts up to the polar
crowns. It is imperative to determine whether different formation
mechanisms occur at different latitudes on the Sun.  To distinguish
this, spectral lines from $H\alpha$ to X-rays along with magnetic
information are needed to provide full coverage of the wavelength
ranges associated with the formation and structure of filaments.
Before ordering up this large menu of observations, we must be able
to use hindsight as to where and when a long-lived filament might
form. Therefore, maintenance of existing synoptic data sets is a
vital part of advanced studies of prominence formation.

\section{Open Issues}
\label{sec:6}

The present review is focused on four aspects of solar prominences,
namely, their magnetic structure, the dynamics of prominence plasma,
prominence oscillations, and the formation and evolution of filament
channels. Other aspects, such as the physical properties of prominence
plasmas, spectroscopic methods used to measure these properties and
non-LTE models are discussed in Paper I.  In the following we outline
several outstanding issues, and the observations and modeling required
to resolve these issues. The ordering reflects the 
order of topics considered in the review and no attempt is made to 
prioritise them:
\begin{enumerate}
\item
Why do different prominences have such different morphology (e.g.,
horizontal vs.~vertical threads), and why do prominences/filaments
look so different in different wavelengths? Current models do not
readily explain such differences. What physical mechanism 
causes the observed thin thread structures and subsequently
the gaps between individual threads?  To resolve these issues, images
with high spatial resolution of filaments and prominences obtained
simultaneously are required at different wavelengths, including H$\alpha$, He I
10830 {\AA} and He~II 304 {\AA}. These observations should track the
same filament as it rotates across the solar disk and above the limb.
Three-dimensional (3D) models of prominence threads should be
developed (see Paper I).
\item
What is the 3D magnetic structure of different types of prominences?
In particular, what is the magnetic field orientation in filament
barbs, and what is the relationship between the barbs and the evolving
photospheric magnetic fields (e.g., parasitic polarities)? What is the
effect of the evolving magnetic carpet on prominence magnetic
structure? Are non-erupting filaments outside active regions suspended 
in detached flux ropes ? Answering such questions will require 
spectro-polarimetric measurements of the magnetic fields in prominences at 
multiple heights including both measurements in prominences (using the 
Zeeman and Hanle effects) and in the photosphere/chromosphere below filaments. 
Techniques should be developed for modeling the evolution
of filament magnetic structure in response to time-dependent
photospheric boundary conditions, including the motions of discrete
magnetic elements, flux emergence, and flux cancellation.
\item
To what extent can vector field extrapolations from the photosphere or
chromosphere be used to deduce magnetic fields within filaments?
Answering this question requires photospheric and/or chromospheric
vector field measurements, reliable NLFFF modeling to extrapolate the
magnetic fields to larger heights, and verification that extrapolation
results match the observed filament channel structures and magnetic
fields.
\item
How does the prominence plasma collect and evolve? What
physical mechanisms are responsible for prominence plasma formation and
dynamics? Which mechanisms best explain the observations of plasma 
distributions and flows? As different magnetic field models exhibit
a different field structure (e.g. sheared arcade vs flux rope). What
effect do these differences have on the distribution and dynamics of the plasma
produced by the different plasma formation mechanisms. 
This requires simultaneous, high cadence, high
spatial resolution observations of many spectral lines covering the
full range of temperatures present in the solar atmosphere.  An
important goal is to obtain proper motion and Doppler shift
information.  Tests of the proposed mechanisms (\S3) requires 3D MHD
modeling, including relevant energy release, deposition, transport and
loss processes.
\item
What is the relationship between the observed signatures of the
oscillations (spectral line shift, width, and intensity variation) and
the physical properties of the MHD waves (velocity, magnetic field,
temperature and density perturbations)?  How are prominence
oscillations excited and damped, and how does this affect the plasma
energetics? Can prominence seismology be developed as a powerful
diagnostic tool, too for example deduce the distribution and strength of the
magnetic field in prominences? Solving this problem requires 2D observations of
Doppler shifts with high spatial and temporal resolution, as well as
independent estimates of key physical properties (e.g. densities and
temperatures). Interpretation of these observations will require
multi-dimensional modeling of wave propagation and damping in
prominences, eventually incorporating wave excitation processes.
\item
How are prominences heated, e.g., by waves (see point 5), shocks or other
mechanisms? The temperatures in prominences are higher than the
radiative equilibrium temperature of about 5000 K. We need high
cadence observations in several spectral lines, and accurate modeling
of the ionization state of the plasma and of radiative transport in
optically thick spectral lines (see Paper I).
\item
How are filament channels formed at all latitudes on the Sun?
Currently, few observations of filament channel formation are
available. What is the helicity distribution within activity
complexes and how does it relate to filament channels? How is 
this helicity dispersed across the solar
surface?  This requires low cadence (hours) global
observations, including the far side of the Sun over long periods of
time (years): photospheric LOS magnetic field, H$\alpha$, He I 10830
{\AA}, He II 304 {\AA}. Techniques should be developed to simulate the
global-scale, long-term evolution of the coronal magnetic field,
incorporating the observed distribution of emerged magnetic flux and
helicity.
\item
 How and at what locations along the PILs  do filament channels form
within active regions and activity complexes? Answering this question requires high-cadence,
high-resolution observations of active regions and activity complexes
in H$\alpha$, EUV lines, and LOS magnetic field. To elucidate the role
of subsurface processes (\S5.3 and \S5.6), the emergence of twisted
flux ropes should be modeled self-consistently.
\item
What is the magnetic structure of polar crown prominences? Do they have
the same chirality as mid-latitude prominences in the same hemisphere?
If they exhibit a hemispheric pattern what is the cause of it ?
This requires spatially resolved, synoptic measurements of prominence
magnetic fields using the Hanle effect. Such observations have not
been carried out since the 1980's.  Do sub-surface processes play a
role in the formation of polar crown prominences?  To answer this
question requires out-of-ecliptic observations of photospheric
magnetic fields (LOS component) and local helio-seismology to
determine subsurface flows near the polar crowns.
\item
What can filaments and prominences tell us about the distribution
of magnetic helicity across the solar surface and subsequently the solar dynamo?
Attacking this fundamental issue requires the development of 3D dynamo
models that include the distribution and transport of magnetic
helicity. The results of such dynamo models should be compared with
observations of the helicity in active regions and filament channels
on the quiet Sun.
\end{enumerate} 

These issues can be addressed by combining spectral and imaging
observations from ground- and space-based instruments. Some of the
goals may be achieved with currently available resources, but others
will require new missions. Planners of future space missions are
encouraged to take the above observational requirements into account.

\begin{acknowledgements}
The authors would like to thank ISSI (Bern) for support of the team
``Spectroscopy and Imaging of Quiescent and Eruptive Prominences from
Space.'' They also would like to thank their teammates for the nice
working atmosphere and the lively discussions. Financial support by
the European Commission through the European Solar Magnetism  Network
(HPRN-CT-2002-00313) and the SOLAIRE Network (MTRN-CT-2006-035484) is
gratefully acknowledged.
JK would like to thank S.~Antiochos, C.R.~DeVore and J.A.~Klimchuk for
their collaboration on the thermal non-equilibrium studies discussed
in section \ref{sec:3}.
JLB would like to acknowledge I.~Arregui, M.~Carbonell, A.~D\'{i}az,
P.~Forteza, R.~Oliver, R.~Soler and J.~Terradas, colleagues from the
Mallorca Solar Physics Group, for their contributions to the research
reported in section \ref{sec:4}. JLB would also like to acknowledge
the financial support received from the Spanish Ministery of Science
and Innovation under grant AYA2006-07637, FEDER funds, and from the
Conselleria de Economia, Hisenda i Innovaci\'{o} under grant
PCTIB-2005GC3-03.
DHM would like to thank the UK STFC for their financial support.

This paper is based on data from many solar observatories.
Figure 6 is based on observations from the
TH\'{E}MIS, which was built by the INSU/CNRS (France) and the CNR
(Italy), and is installed at the International Observatory of the
Canary Islands (Tenerife, Spain), which is operated by the Instituto
de Astrof\'{i}sica de Canarias. Figures 23, 24 and 25 used data from
the Ottawa River Solar Observatory of the National Research Council
(Canada), and from the National Solar Observatory (NSO), which is
supported by the National Science Foundation (NSF) and by NASA. SOHO
is a mission operated by ESA and NASA. Hinode is a Japanese mission developed
and launched by ISAS/JAXA, collaborating with NAOJ as a domestic partner, NASA 
and STFC(UK) as international partners.
\end{acknowledgements}


\end{document}